\documentclass[iicol,default]{sn-jnl}

  \RequirePackage{flushend}
  \RequirePackage[numbers,sort&compress]{natbib}

  \usepackage{hyphenat}
  \usepackage[utf8]{inputenc}
  \usepackage[english]{babel}
  \usepackage{upgreek}
  \usepackage{times}
  \usepackage{stackrel}
  \usepackage{stmaryrd}
  \usepackage{xcolor}
  \usepackage{amssymb}
  \usepackage{amsmath}
  \usepackage{mathtools}
  \usepackage{graphicx}
  \usepackage{tabularx}                              
  \usepackage{cleveref}                              

  \DeclareMathAlphabet\mathbfcal{OMS}{cmsy}{b}{n}

  \newcommand{\kR}{\mathcal{R}}
  \newcommand{\kP}{\mathcal{P}}                      
  \newcommand{\wRe}{\widetilde{\Re}}
  \newcommand{\bnabla}{\boldsymbol{\nabla}}

  \newcommand{\sig}{\Upsigma} 
  \newcommand{\sigSBL}{\Upsigma_{\text{SBL}}} 
  \newcommand{\sigLBL}{\Upsigma_{\text{LBL}}} 
  \newcommand{\sigEff}{\Upsigma_{\text{eff}}}

  \newcommand{\VVVV}{V_{{\beta}j}V_{{\alpha}i}V^*_{{\beta}i}V^*_{{\alpha}j}} 

  \def\v{\text{v}}
  
  \def\ToyModel{$\pi$-$\mu$}
  \def\TM{\pi\text{-}\mu}
  
  \def\Zeta{\mathfrak{T}}

  \renewcommand{\vec}{\boldsymbol}                   
 
  \DeclareMathOperator\Tr{Tr}
  \DeclareMathOperator\diag{diag}
  \DeclareMathOperator\erf{erf}
  \DeclareMathOperator\Ierf{Ierf}
  \DeclareMathOperator\erfc{erfc} 

  \def\approxprop{%
    \def\p{%
      \setbox0=\vbox{\hbox{$\propto$}}%
      \ht0=0.6ex \box0 }%
      \def\s{\vbox{\hbox{$\sim$}}}%
      \mathrel{\raisebox{0.7ex}{%
      \mbox{$\underset{\s}{\p}$}%
    }}%
}

\begin{document}

\selectlanguage{english}

\title[ ]{Virtual neutrino propagation at short baselines}

\author*[1]{\fnm{Vadim~A.~Naumov}}\email{vnaumov@theor.jinr.ru}            
\equalcont{These authors contributed equally to this work.}
\author[2]{\fnm{Dmitry~S.~Shkirmanov}}\email{dmitry@shkirmanov.com}        
\equalcont{These authors contributed equally to this work.}

\affil[1]{\orgdiv{Bogoliubov Laboratory of Theoretical Physics},
           \orgname{Joint Institute for Nuclear Research},
           \orgaddress{\street{Joliot-Curie}, \city{Dubna},
           \postcode{141980}, \state{Moscow Region}, \country{Russian Federation}}}

\affil[2]{\orgdiv{Dzelepov Laboratory of Nuclear Problems},
          \orgname{Joint Institute for Nuclear Research},
          \orgaddress{\street{Joliot-Curie}, \city{Dubna},
          \postcode{141980}, \state{Moscow Region}, \country{Russian Federation}}}

\abstract{
Within a covariant perturbative field-theoretical approach, the wave-packet modified neutrino propagator is expressed
as an asymptotic expansion in powers of dimensionless Lorentz- and rotation-invariant variables.
The expansion is valid at high energies and short but macroscopic space-time distances between the vertices
of the proper Feynman macrodiagram.
In terms of duality between the propagator and the effective neutrino wave packet, at short times and distances,
neutrinos are deeply virtual and move quasi-classically.
In the lowest-order approximation, this leads to the classical inverse-square dependence of the modulus squared
flavor transition amplitude and related neutrino-induced event rate from distance $L$ between the source and detector,
and the above-mentioned asymptotics results in the corrections to the classical behavior represented by powers of $L^2$.
This is very different from the long-baseline regime, where similar corrections are given by an asymptotic expansion
in inverse powers of $L^2$.
However, in both short- and long-baseline regimes, the main corrections lead to a decrease in number of neutrino events.
}

\maketitle

\section{Introduction}
\label{sec:Introduction}

A considerable amount of work has been devoted to the quantum-field theoretical (QFT) analysis of spatial evolution of mixed massive neutrinos
\cite{Kobzarev:1981ra,Giunti:1993se,Blasone:1995zc,Grimus:1996av,Giunti:1997sk,Campagne:1997fu,Kiers:1997pe,Zralek:1998rp,Ioannisian:1998ch,%
      Blasone:1998hf,Grimus:1998uh,Grimus:1999ra,Cardall:1999bz,Cardall:1999ze,Beuthe:2000er,Stockinger:2000sk,Beuthe:2002ej,%
      Giunti:2002xg,Beuthe:2001rc,Garbutt:2003ih,Asahara:2004mh,Dolgov:2005nb,Nishi:2005dc,Kopp:2009fa,Akhmedov:2009rb,Keister:2009qn,%
      Naumov:2009zza,Naumov:2010um,Akhmedov:2010ms,Akhmedov:2010ua,Bernardini:2010zba,Akhmedov:2012uu,Akhmedov:2012mk,%
	  Kobach:2017osm,Egorov:2019vqv,Grimus:2019hlq,Naumov:2020yyv,Ciuffoli:2020flo,Smaldone:2021mii,Cheng:2022lys}.      
These studies were, in large part, initiated by a harsh critique of the quantum-mechanical (QM) theory of plane-wave neutrino oscillations,
burdened with numerous inconsistencies and paradoxes.
A comprehensive discussion of this and related issues, as well as further references, can be found  
in books~\cite{Giunti:2007ry,Xing:2011zza}. 
The QFT-based computations consistently reproduce, as an approximation, the standard QM formula for the neutrino flavor transition probability,
and usually predict deviations from it (often different for different authors and nonequivalent approaches), thereby providing proper
interpretation of the QM result, determining the scope of its applicability, and indicating possible directions of the experimental
search for new related phenomena, such as 
decoherence and dispersion effects~\cite{Chan:2015mca,An:2016pvi,Wong:2017jbh,Cheng:2020jje,DeGouvea:2020hfl,deGouvea:2021uvg,JUNO:2021ydg}
and inverse-square law violation~\cite{Naumov:2013bea,Naumov:2015hba,Naumov:2017pgt,Naumov:2021vds}.

The above-cited works are mainly concerned with studying the neutrino oscillations at long baselines, where the transition probability
is large enough to be detected experimentally. 
The short baseline oscillation regime was discussed in Ref.~\cite{Ioannisian:1998ch} within a solvable QFT model, and later
in Ref.~\cite{Beuthe:2001rc} in a more general approach.
Here, there is an immediate need for clarification: By short and long baselines (SBL and LBL) we will mean the distances $L$ between the
neutrino production and detection points that satisfy the conditions $L \ll E_\nu/\sigSBL^2$ and $L \gg E_\nu/\sigLBL^2$ respectively,
where $E_\nu$ is the mean neutrino energy, and $\sigSBL$ and $\sigLBL$ are the representative scales of the neutrino energy uncertainty
related to the SBL and LBL regions.
Later in the paper we explain where these conditions come from and what might be interesting about the SBL region, 
(depending on the values of $L$, $E_\nu$, and $\sigSBL$) 
which may or may not be related to SBL experiments with accelerator or reactor (anti)neutrinos, or neutrinos from artificial radioactive sources.

Throughout the paper, we use the natural units ($\hbar=c=1$) and the Feynman metric. The Greek letters $\mu, \nu, \ldots$
are used for Lorentzian indices, while the Latin indices $k, l, n, \ldots$ enumerate the spatial components of vectors and tensors.
The letters $\alpha$ and $\beta$ are reserved for lepton flavors. 
The formalism under consideration does not depend on the number of flavors or number of the neutrino mass eigenfields,
but, unless otherwise noted, one may have in mind three generations of the Standard Model.

\section{Essentials of formalism}
\label{sec:General}

\begin{figure}[htb]
\centering\includegraphics[width=0.62\linewidth]{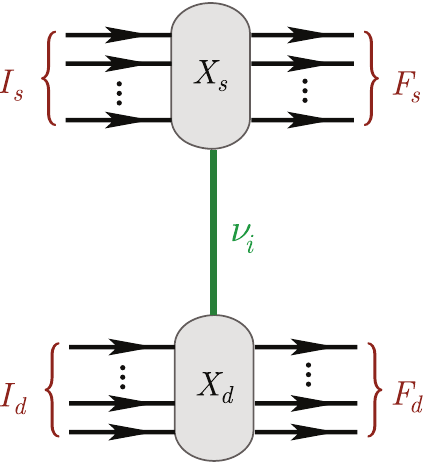}
\caption{A generic macrodiagram with virtual neutrino exchange.
         $I_s$ and $I_d$ denote the sets of initial wave-packet states in the source and detector vertices $X_s$ and $X_d$,
         respectively, $F_s$ and $F_d$ denote the sets of the final states.
         The vertices $X_s$ and $X_d$, which may include arbitrary inner lines and loops, are assumed to be
         macroscopically separated in space and time.
         The inner line connecting $X_s$ and $X_d$ denotes the causal Green's function for
         the neutrino $\nu_i$ with definite mass $m_i$.
         \label{Macrograph_generic}
        }
\end{figure}

The subsequent discussion is based on the covariant (``diagrammatic'' ) QFT formalism developed in Refs.~\cite{Naumov:2009zza,Naumov:2010um}.
In this section, we shortly discuss the features of the formalism most essential for the aims of the present study, for more details, see Ref.~\cite{Naumov:2020yyv}.
The basic idea of the diagrammatic approach is that the flavor transition of neutrinos is the result of interference of ``macroscopic''
Feynman diagrams, such as the one shown in Fig.~\ref{Macrograph_generic}, perturbatively describing a lepton number violating process
in which the massive neutrinos are internal lines connecting the vertices of their production and absorption.
In other words, the neutrino mass eigenfields, $\nu_i$ ($i=1,2,3, \ldots$), are treated as virtual intermediate states.
The neutrinos with definite flavors, $\nu_\alpha$ ($\alpha=e,\mu,\tau$), do not participate in the formalism at all,
although can be used to compare the predictions with those from other approaches, in particular, with the standard QM
approach based on the concept of flavor mixing.

The external lines of the macrodiagrams correspond to asymptotically free incoming (``in'') and outgoing (``out'')
wave packets $\vert\vec{p}_a,s_a,x_a\rangle$ ($a{\in}I_{s,d}$) and $\vert\vec{p}_b,s_b,x_b\rangle$ ($b{\in}F_{s,d}$)
of the following generic form: 
\begin{equation}
\label{x_state}
\vert\vec{p},s,x\rangle = \int\frac{d\vec{k}\,e^{i(k-p)x}}{(2\pi)^32E_{\vec{k}}}\phi(\vec{k},\vec{p})\vert\vec{k},s\rangle
\end{equation}
representing the superposition of the one-particle Fock's states
$\vert\vec{k}, s\rangle=\sqrt{2E_{\vec{k}}}a_{\vec{k}s}^\dag\vert0\rangle$
with 3-momenta $\vec{k}$ and energy $E_{\vec{k}}=\sqrt{\vec{k}^2+m^2}$, which may also depend on discrete variables $s$,
such as the spin projection; here $a_{\vec{k}s}^\dag$ is the creation operator, $m$ is the mass of the particle,
$d\vec{k} \equiv d^3k$.
Wave-packet state \eqref{x_state} is characterized by the most probable 3-momentum $\vec{p}$ and the space-time coordinate
$x=(x_0,\vec{x})$ of its barycenter or, simply, center of symmetry, which is always implicitly assumed.
The function $\phi(\vec{k},\vec{p})$ is a model-dependent Lorentz scalar (``form factor''), which determines the shape
of the packet and generally depends on a set of parameters $\vec\sigma$ that characterize the uncertainty of the
momentum $\vec{p}$; $k=(k_0,\vec{k})$ and $p=(p_0,\vec{p})$ are the on-shell 4-momenta, i.e., $k^2=p^2=m^2$.
In the plane-wave limit (PWL) equivalent to the definite momenta approxiamtion, the form factor becomes proportional
to the $3d$ delta function,
\begin{equation*}
\phi\left(\vec{k},\vec{p}\right)\,\stackrel{\text{PWL}}{\longmapsto}\,(2\pi)^32E_{\vec{p}}\delta\left(\vec{k}-\vec{p}\right).
\end{equation*}
and, as a result, the coordinate dependence in Eq.~\eqref{x_state} disappears and the wave packet turns into the state
with definite momentum:
\begin{equation*}
\vert\vec{p},s,x\rangle\,\stackrel{\text{PWL}}{\longmapsto}\,\vert\vec{p},s\rangle.
\end{equation*}
From this point we will consider only very narrow in the momentum space external wave packets, for which
the form-factor function $\phi(\vec{k},\vec{p})$ is strongly peaked at the point $\vec{k}=\vec{p}$.
Such packets are very close to their plane-wave limit and in the configuration space are described by simple scalar function
\begin{equation}
\label{psi(p,x)}
\psi(\vec{p},x) = \int\frac{d\vec{k}e^{ikx}}{(2\pi)^32E_{\vec{k}}}\phi(\vec{k},\vec{p}).
\end{equation}

The blocks (vertices) $X_s$ and $X_d$ (hereinafter referred to as ``source'' and ``detector'') symbolically depict
the microscopic regions of the in and out particle interactions.
The spatial interval between $X_s$ and $X_d$ can be arbitrarily large.
Without risk of confusion, we will denote by the same symbols $X_s$ and $X_d$ the 4-vectors that determine
the effective space-time coordinates of the interaction regions;
the precise definition of these coordinates (called ``impact points'') depends on the form factor model and will be done later.
Macroscopic separation between the source and detector vertices means that the values of $\vert{X_d^0-X_s^0}\vert$
and $\vert{\vec{X}_d-\vec{X}_s}\vert$ are large compared to the microscopic space-time scales of the interaction regions
(determined by the dynamics) and to the micro- or even mesoscopic effective dimensions of the in and out wave packets;
this clarifies the meaning of the term ``macroscopic diagram''.

It is further assumed that the particle interactions are described by the Standard Model (SM) Lagrangian,
phenomenologically augmented by a Dirac or Majorana mass term with the number of the neutrino mass eigenfields $N\ge3$
and by corresponding kinetic terms. So the set of initial or final states (or both sets) at each vertex must contain
a charged lepton or (anti)neutrino.
To avoid Irrelevant complications, we rule out the possibility that the external states 
include gauge or Higgs bosons.

Wave packets as external lines of macrodiagrams lead to a modification of Feynman's rules.
For the narrow in and out packets, they are modified in a rather simple way \cite{Naumov:2009zza,Naumov:2010um}:
for each external line, the standard (plain-wave) factor must be multiplied by
\begin{subequations}
\label{FeynmanRules}
\begin{gather}
e^{-ip_a(x_a-x)}\psi_a  \left(\vec{p}_a,x_a-x\right) \enskip \text{for} \enskip a \in I_s{\oplus}I_d,
\intertext{and}
e^{+ip_b(x_b-x)}\psi_b^*\left(\vec{p}_b,x_b-x\right) \enskip \text{for} \enskip b \in F_s{\oplus}F_d,
\end{gather}
\end{subequations}
where each function $\psi_{\varkappa}\left(\vec{p}_{\varkappa},x\right)$ defined by Eq.~\eqref{psi(p,x)}
is generally specified by its own form factor $\phi_{\varkappa}(\vec{k},\vec{p}_{\varkappa})$, 
most probable 3-momentum $\vec{p}_{\varkappa}$, space-time coordinate $x_{\varkappa}$, and mass $m_{\varkappa}$ ($\varkappa=a,b$).
The internal lines (including loops) and vertex factors remain unchanged.
The additional factors \eqref{FeynmanRules} provide the following two common multipliers
in the integrand of the scattering amplitude \cite{Naumov:2009zza,Naumov:2010um}: 
\begin{gather}
\mathbb{V}_{s,d}(q) = \int d^4x e^{\pm iqx} \!
      \prod_{a\in{I_{s,d}}}e^{-ip_ax_a}\psi_a  \left(\boldsymbol{p}_a,x_a-x\right) \nonumber \\
\times\prod_{b\in{F_{s,d}}}e^{+ip_bx_b}\psi_b^*\left(\boldsymbol{p}_b,x_b-x\right).
\label{OverlapVolumes_def}
\end{gather}
These functions are called overlap integrals, because they characterize the space-time integrated degree of
overlap between the in and out wave-packet states in the source and detector vertices.
Obviously, in the plane-wave limit $\psi(\vec{p},x) \mapsto \exp(ipx)$ and integrals \eqref{OverlapVolumes_def} 
turn into the ordinary $4d$ delta-functions to within a multiplier: 
\begin{equation*}
\mathbb{V}_{s,d}(q)\,\stackrel{\text{PWL}}{\longmapsto}\,(2\pi)^4\delta\left(q \mp q_{s,d}\right),
\end{equation*}
where
\begin{equation*}
q_s = \sum_{a\in{I_s}}p_a-\!\sum_{b\in{F_s}}p_b
\enskip\text{and}\enskip
q_d = \sum_{a\in{I_d}}p_a-\!\sum_{b\in{F_d}}p_b
\end{equation*}
are the 4-momentum transfers in the source and detector vertices, respectively.

Our main interest in this paper is to study the space-time dependence of the count rate of neutrino-induced events.
The expression for the count rate or number of events is obtained after various averages of the modulus squared amplitude 
of the process under study and is primarily determined by behavior of the causal neutrino propagator
\begin{equation}
\label{eq_ThePropagator}
\int\frac{d^4{q}}{(2\pi)^4}\,\frac{\widetilde\delta_s(q-q_s)\widetilde\delta_d(q+q_d)(\hat{q}+m)e^{-i{qX}}}{{q}^2-m_i^2+i\epsilon}
\end{equation}
``dressed''  (through the overlap integrals) by the external wave packets.
Here $m_i$ is the neutrino mass, \[X=(X_0,\vec{X})=X_d-X_s,\] $\widetilde\delta_{s,d}\left(q \mp q_{s,d}\right)$
are the ``smeared'' $\delta$-functions which provide {approximate} energy-momentum conservation at the vertices of
the macrodiagram, and which in the plane-wave limit become the delta-functions, 
\begin{equation}
\label{delta_s,d_PWL}
\widetilde\delta_{s,d}\left(q \mp q_{s,d}\right)\,\stackrel{\text{PWL}}{\longmapsto}\,\delta\left(q \mp q_{s,d}\right);
\end{equation}
in this case, the energy and momentum are exactly conserved at the microscopic level,
while the uncertainty of coordinates of in and out particles tends to infinity
and the space-time description of the process shown in Fig.~\ref{Macrograph_generic} loses its physical meaning.
The smeared $\delta$-functions are defined by the external wave packets and therefore their explicit form depends
on the form factors $\phi(\vec{k},\vec{p})$, but the limit \eqref{delta_s,d_PWL} is model independent.

\subsection{Relativistic Gaussian packets}
\label{sec:RGP}

Below we will use the so-called relativistic Gaussian packets (RGP) proposed in Ref.~\cite{Naumov:2009zza}.
The RGP model corresponds to a very simple form factor $\phi(\vec{k},\vec{p})\propto\exp\left[-(kp)/2\sigma^2\right]$
and the function \eqref{psi(p,x)} describing RGP in the coordinate representation is
\begin{equation}
\label{psi_RGP_exact}
\psi\left(\vec{p},x\right) = \frac{K_1({\zeta}m^2/2\sigma^2)}{{\zeta}K_1(m^2/2\sigma^2)} \equiv \psi_G\left(\vec{p},x\right).
\end{equation}
Here $\vec{p}$ and $m$ are the most probable 3-momentum and mass, respectively, $x=(x_0,\vec{x})$ is the space-time coordinate
of the packet's center,%
\footnote{RGP is spherically symmetric in its own frame of reference (denoted by $\star$), which is defined by the condition 
          $\vec{p}_\star=0$.}
$\sigma$ is the momentum spread (just a constant assumed small compared to $m$), $K_1(z)$ is the modified Bessel function
of the third kind (or Macdonald function) of order 1, and
\begin{gather*}
\zeta = \sqrt{1-\frac{4\sigma^2}{m^2}\left[\sigma^2x^2+i(px)\right]}.
\end{gather*}
For our aims, it will suffice to restrict ourselves to the simplest approximate form of Eq.~\eqref{psi_RGP_exact} 
\begin{equation}
\label{psi_coord}
\psi(\vec{p},x) = \exp\left\{ipx-\frac{\sigma^2}{m^2}\left[(px)^2-m^2x^2\right]\right\}
\end{equation}
(called contracted RGP, or CRGP), which describes a non-difluent mode characterized by the condition that
in the intrinsic frame the function $\vert\psi(\vec{p},x)\vert=\vert\psi(\vec{0},x_\star)\vert$ is independent
of time $x^0_\star$, namely,
$
\vert\psi(\vec{0},x_\star)\vert = \exp\left(-\sigma^2\vec{x}_\star^2\right).
$
An analysis of the asymptotic expansion of $\ln\left[\psi_G(\vec{0},x_\star)\right]$ in powers of the small parameter $\sigma^2/(m^2\zeta)$,
taking into account the inequalities $\vert\zeta\vert\ge1$ and $\vert\arg\zeta\vert<\pi/2$, provides the necessary and sufficient conditions
of the non-difluent behavior:
\begin{eqnarray}
\label{TheRestrictions_psi}
\sigma^2(x^0_\star)^2             \ll \frac{m^2}{\sigma^2},
\enskip
\sigma^2\vert\vec{x}_\star\vert^2 \ll \frac{m^2}{\sigma^2}.
\end{eqnarray}
As is seen, the area of applicability of the CRGP approximation expands with decreasing $\sigma$.

\subsection{Process under study}
\label{sec:ProcessUnderStudy}

Many physical assumptions, limitations, and simplifications are the same for the SBL and LBL cases;
these will be mentioned here only briefly.
As in the LBL case \cite{Naumov:2009zza,Naumov:2010um,Naumov:2020yyv}, we will limit ourselves with studying the following process:
\begin{equation}
\label{Macroprocess_A}
I_s {\oplus} I_d \to F_s'+\ell_\alpha^+~{\oplus}~F_d'+\ell_\beta^-,
\end{equation}
where $\oplus$ sign indicates the macroscopic remoteness of the corresponding subsets of particles.
This process is interesting because it includes virtually all terrestial experiments with short baselines.
\begin{figure}[htb]
\centering\includegraphics[width=0.95\linewidth]{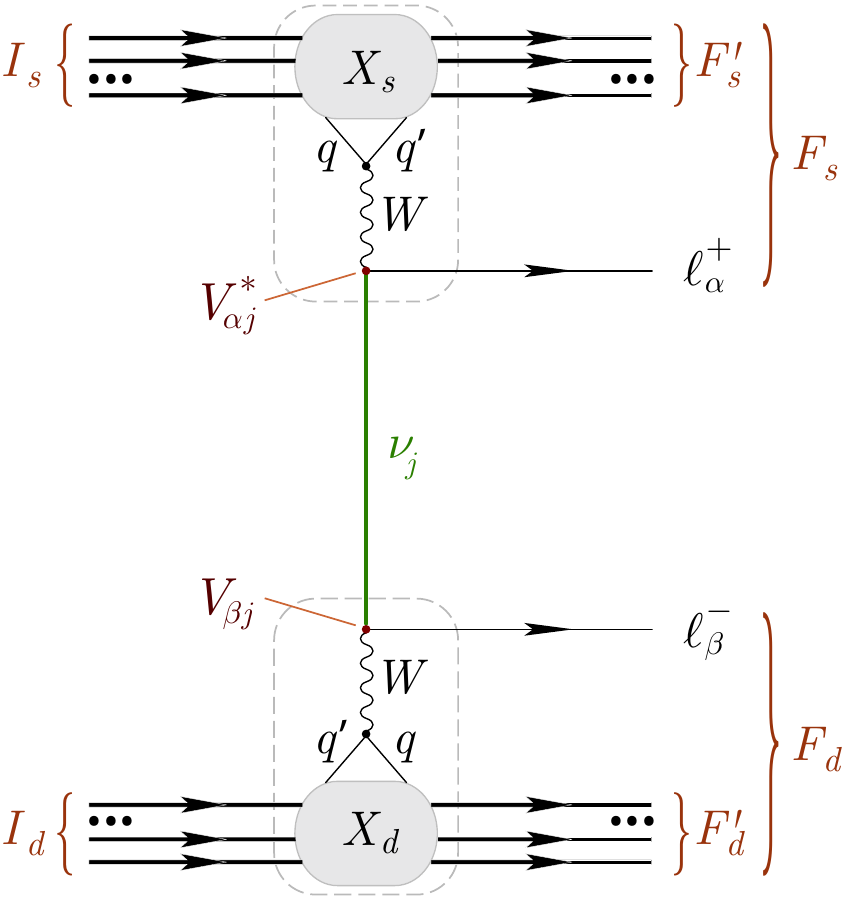}
\caption{A diagram of process \protect\eqref{Macroprocess_A}, see text for details.
         \label{fig:Macrograph_Class_A}
        }
\end{figure}
The macroscopic Feynman diagram corresponding to process \eqref{Macroprocess_A} is shown in Fig.~\ref{fig:Macrograph_Class_A};
it is a particular case of diagram \ref{Macrograph_generic}.
Initial and final states $I_{s,d}$ and $F_{s,d}'$ in Fig.~\ref{fig:Macrograph_Class_A} consist of hadronic states,
while $\ell_\alpha^+$ and $\ell_\beta^-$ represent the final-state leptons with the 4-momenta
$p_\alpha=\left(p_\alpha^0,\vec{p}_\alpha\right)$ and $p_\beta=\left(p_\beta^0,\vec{p}_\beta\right)$, respectively ($\alpha,\beta=e,\mu,\tau$).
Recall: it is assumed for definiteness that the subgraphs of the diagram containing lepton and quark lines, $W$-propagators,
the elementary vertices proportional to the elements $V_{{\alpha}j}^*$ and $V_{{\beta}j}$ of the neutrino mixing matrix, as well as
of the quark mixing matrix (not indicated in Fig.~\ref{Macrograph_generic}), are described by the Standard Model,
but this assumption is not at all a critical part of the formalism under consideration.
As above, $X_s$ and $X_d$ symbolically denote the neutrino source and detector blocks and, at the same time, the impact points
macroscopically separated in space and time.

According to Ref.~\cite{Naumov:2009zza}, the amplitude of the process \eqref{Macroprocess_A} can be written in the following form:
\begin{multline}
\label{Amplitude_2}
\mathcal{A}_{\beta\alpha} = \frac{g^4}{64\mathcal{N}}\mathcal{J}_d^{*\nu}\mathcal{J}_s^{\mu}\sum_j V_{{\beta}j}V^*_{{\alpha}j} \\
\times\overline{u}(\vec{p}_\beta)O^{\nu'}\mathbb{G}^j_{\nu\nu'\mu'\mu}\left(\{\vec{p}_{\varkappa},x_{\varkappa}\}\right) O^{\mu'}v(\vec{p}_\alpha).
\end{multline}
This result is based on a theorem on factorization of hadronic blocks of the macrodiagram \ref{fig:Macrograph_Class_A} proved in Ref.~\cite{Naumov:2020yyv}.
Let us concisely describe the ingredient of Eq.~\eqref{Amplitude_2}, necessary for a further presentation of the formalism.
Here $g$ is the $SU(2)$ electroweak gauge coupling, $\mathcal{N}$ is the normalization factor,
$\mathcal{J}_d^{\nu}$ and $\mathcal{J}_s^{\mu}$ are the standard QFT (c-number) hadronic currents (see Ref.~\cite{Naumov:2020yyv} for more details),
$\overline{u}(\vec{p}_\beta)$ and $v(\vec{p}_\alpha)$ are the usual Dirac bispinors, describing the final-state charged leptons
$\ell_\beta$ and $\ell_\alpha$, respectively, and  $O^{\nu}=\gamma^\nu(1-\gamma_5)$;
$\vec{p}_{\varkappa}$ and $x_{\varkappa}$ are, respectively, the most probable 3-momentum of the packet $\varkappa{\in}S{\oplus}D$ and space-time
position of the packet's center.

The tensor function $\mathbb{G}^j_{\nu\nu'\mu'\mu}$
accumulating the effects of all external wave packets has the form
\begin{multline}
\mathbb{G}^j_{\nu\nu'\mu'\mu} =
\frac{ie^{-\mathfrak{S}-i\Theta}}{16\sqrt{\vert\Re_s\vert\vert\Re_d\vert}}\int d^4q\,e^{-iqX}\varPhi(q) \\
\times\frac{(\hat{q}+m_j)}{q^2-m_j^2+i\varepsilon}\Delta_{\nu\nu'}(q-p_\beta)\Delta_{\mu\mu'}(q+p_\alpha).
\label{tenzor_function}
\end{multline}
The so-called overlap tensors $\Re_s$ and $\Re_d$ are given by
\begin{equation}
\label{Re_sd}
\Re_s^{\mu\nu} = \sum_{\varkappa{\in}S}T_{\varkappa}^{\mu\nu}
\enskip\text{and}\enskip
\Re_d^{\mu\nu} = \sum_{\varkappa{\in}D}T_{\varkappa}^{\mu\nu},
\end{equation}
respectively, where $S=I_s{\oplus}F_s$, $D=I_d{\oplus}F_d$,
\begin{equation*}
T_{\varkappa}^{\mu\nu} = \sigma_{\varkappa}^2\left(u_{\varkappa}^{\mu}u_{\varkappa}^{\nu}-g^{\mu\nu}\right),
\end{equation*}
$u_{\varkappa}=p_{\varkappa}/m_{\varkappa}=\Gamma_{\varkappa}(1,\vec{v}_{\varkappa})$ is the 4-velocity and $\sigma_{\varkappa}$
is the momentum spread of the packet $\varkappa$.
The tensors $\wRe_s$ and $\wRe_d$ are inverse to the overlap tensors, i.e.  
\begin{equation*}
\widetilde{\Re}_s^{\mu\lambda}\left(\Re_s\right)_{\lambda\nu}=\delta^{\mu}_{\nu}
\enskip\text{and}\enskip
\widetilde{\Re}_d^{\mu\lambda}\left(\Re_d\right)_{\lambda\nu}=\delta^{\mu}_{\nu}.
\end{equation*}
It is important that both $\Re_{s,d}$ and $\wRe_{s,d}$ are symmetric and positive definite.
Therefore, the determinants $\vert\Re_{s,d}\vert$ and $\vert\wRe_{s,d}\vert=1/\vert\Re_{s,d}\vert$ are positive and
the same is also true for the principal minors of $\Re_{s,d}$ and $\wRe_{s,d}$.

The functions $\Delta_{\nu\nu'}(q-p_\beta)$ and $\Delta_{\mu\mu'}(q+p_\alpha)$ are the $W$-boson propagators,
whose explicit form we will not need.
The functions $\varPhi(q)$, $\mathfrak{S}$, and $\Theta$ are defined as
\begin{gather} 
\label{Phi_just_gauss}
\varPhi(q) = \exp\left[-\frac{1}{4}F(q)\right], \\
\mathfrak{S} = \mathfrak{S}_s+\mathfrak{S}_d, \nonumber
\quad 
\Theta = q_dX_d+q_s X_s,
\intertext{where}
\begin{aligned}
\label{F(q)_def}
F(q) = &\  \wRe_s^{\mu\nu}\left(q-q_s\right)_\mu\left(q-q_s\right)_\nu \\
       &\ +\wRe_d^{\mu\nu}\left(q+q_d\right)_\mu\left(q+q_d\right)_\nu,
\end{aligned} \\
\label{mathfrak_S_sd_final_a}
\mathfrak{S}_{s,d} = \sum_{\varkappa{\in}S,D}T_{\varkappa}^{\mu\nu}
                     \left(x_{\varkappa}-X_{s,d}\right)_{\mu}
                     \left(x_{\varkappa}-X_{s,d}\right)_{\nu}, \nonumber
\end{gather}
and
\begin{subequations}
\label{X_sd}
\begin{align}
\label{X_sd_a}
X_{s,d}^{\mu} = &\ \widetilde{\Re}_{s,d}^{\mu\nu}\,\sum_{\varkappa{\in}S,D}T_{\varkappa\nu}^{\lambda}x_{\varkappa\lambda} \\
\label{X_sd_b}
              = &\ \widetilde{\Re}_{s,d}^{\mu\nu}\,\sum_{\varkappa{\in}S,D}\sigma_{\varkappa}^2
	               \left[(u_{\varkappa}x_{\varkappa})u_{\varkappa\nu}-x_{\varkappa\nu}\right].
\end{align}
\end{subequations}
From Eqs.~\eqref{Re_sd} and \eqref{X_sd_a} follow the identities
\begin{equation*}
\sum_{\varkappa{\in}S,D}T_{\varkappa}^{\mu\nu}\left(x_{\varkappa}-X_{s,d}\right)_\nu = 0,
\end{equation*}
from which we see that the impact points represent the weighted-mean space-time coordinates of the external wave packets.
Considering that $\mathfrak{S}_{s,d}\ge0$, the functions $\exp(-\mathfrak{S}_s)$ and $\exp(-\mathfrak{S}_d)$ 
are the geometric suppression factors conditioned by the overlapping degree of the in and out wave packets
in space and time: they suppress the amplitude when the packets do not intersect or intersect only partially near
the points $X_s$ and $X_d$ in the source and detector, respectively.
At the impact points $\mathfrak{S}_s=\mathfrak{S}_d=0$;
the closer are the world lines of the packets' centers to the impact points, 
the more probable is the interaction between the packets. 
The condition that the interaction regions in the source and detector vertices are macroscopically separated from each other is therefore 
equivalent to the macroscopic separation of the points $X_s$ and $X_d$. 
The world line configurations have no concern with the specific dynamics governed by the interaction Lagrangian, being uniquely defined by the initial
(final) coordinates, group velocities and effective dimensions of the asymptotically free in (out) wave packets. 

\section{Long baselines in short}
\label{sec:ISL-LBL}

An important tool for studying LBL asymptotics is the Grimus-Stokinger theorem~\cite{Grimus:1996av}
used in most abovecited analyses of neutrino oscillations based on QFT approaches with wave packets.
The theorem predicts that the spatial part of the ``dressed'' neutrino propagator \eqref{eq_ThePropagator}
behaves as $1/L$ at $L\equiv\vert\vec{X}\vert\to\infty$, which leads to the classical inverse-square law (ISL)
for the detected number of neutrino events.
In Refs.~\cite{Naumov:2013bea,Korenblit:2014uka} it was shown that the classical ISL is violated at finite $L$.
This is a consequence of an extension of the Grimus-Stockinger theorem, which allows one to compute corrections
to the long-distance asymptotics of the propagator \eqref{eq_ThePropagator}.
The theorem as formulated in Ref.~\cite{Naumov:2013bea} states that for any Schwartz function $\varPhi(\vec{q})$%
\footnote{That is, the function $\varPhi(\vec{q}) \in C^{\infty}(\mathbb{R}^3)$ which goes to zero faster than any inverse power
          of $\vert\vec{q}\vert$, as do all its derivatives. The (positive) function 
          in the integrand of the neutrino propagator~\eqref{eq_ThePropagator} belongs to this class; $\upkappa^2=q_0^2-m_i^2$.}
the integral
\begin{equation*}
\mathfrak{J}(\vec{X},\upkappa)=\int\frac{d\vec{q}}{(2\pi)^3}\frac{\varPhi(\vec{q})e^{i\vec{qX}}}{\vec{q}^2-\upkappa^2-i\epsilon}
\end{equation*}
may be represented by the asymptotic series
\begin{gather}
\label{ExGS}
\mathfrak{J}(\vec{X},\upkappa)
= \frac{\varPhi(\upkappa\vec{l})e^{-i{\upkappa}L}}{4{\pi}L}\Bigg[1+\sum_{a\ge1}\frac{\mathcal{D}_a(\upkappa)}{L^a}\Bigg],
\intertext{where}
\mathcal{D}_a(\upkappa)
         = \frac{(-i)^a}{\varPhi(\upkappa\vec{l})}\sum_{b=0}^{a}\sum_{c=0}^{\left[\frac{a-b}{2}\right]}
           \left(\frac{\upkappa}{4}\right)^bD_{abc}\varPhi(\vec{q})\Big\vert_{\vec{q}=\upkappa\vec{l}}, \nonumber \\
D_{abc}  = c_{abc}\left(\vec{l}\times\!\bnabla_{\vec{q}}\right)^{2(b+c)}
                       \left(\vec{l}\bnabla_{\vec{q}}\right)^{a-b-2c},                                  \nonumber                        
\end{gather}
$\vec{l} = \vec{X}/\vert\vec{X}\vert$, and $c_{abc}$ are recursively defined positive numbers
(given in Ref.~\cite{Naumov:2013bea} for $a\le6$).
Equivalent but recursive-free formulation of the theorem is given in Ref.~\cite{Korenblit:2014uka}.
The applicability of the leading (Grimus-Stokinger) approximation is defined by explicit form of $\varPhi(\vec{q})$,
but in general case it can be written as
\begin{equation*}
\sigLBL^2{L}/q_0 \gg 1, 
\end{equation*}
where $\sigLBL$ is an effective parameter dependent on the momenta, masses, and momentum spreads
($\sigma_\varkappa$ in the CRGP model) of the external (in and out) wave packets.
It is also shown in Ref.~\cite{Naumov:2013bea} that $\left\vert\mathfrak{J}(\vec{X},\upkappa)\right\vert^2$ is
given by a series in powers of $1/L^2$ and the lowest ($\propto 1/L^4$) ISL violating (ISLV) correction is negative in the physical region
$\kappa^2\ge0$. The explicit formula for $\sigLBL$ obtained within the CRGP model in the third order saddle-point asymptotic expansion
is given in Ref.~\cite{Naumov:2021vds}. It is far beyond the scope of this article to analyze the contribution of the non-physical domain.

Another important consequence of the theorem is that the virtual neutrinos are on-mass-shell, and
both energies and momenta of different eigenfields $\nu_i$ do not coincide.
This strongly contradicts the standard assumption of the plane-wave QM theory about the equality of
the momenta of the neutrino states with different masses, but, surprisingly, in the ultrarelativistic limit
leads to the same oscillation phases.


Omitting the rather lengthy calculations, let us write down the final expression for the number of events in
a detector located at a large distance from the source, obtained within the framework of the formalism described above:
\begin{multline}
\label{EventNumber}
N_{\beta\alpha}
= \frac{\tau_d}{V_{\mathcal{D}}V_{\mathcal{S}}}\int\limits_{V_{\mathcal{S}}}d\vec{x}\int\limits_{V_{\mathcal{D}}}d\vec{y}
\int d\mathfrak{F}_\nu \\
\times\int d\sigma_{{\nu}\mathcal{D}}\mathcal{P}_{\alpha\beta}\left(E_\nu,\vert\vec{y}-\vec{x}\vert\right).
\end{multline}
The detais of the calculations, conditions of applicability, and interpretation of this somewhat abstract approximate formula
can be found in Refs.~\cite{Naumov:2010um,Naumov:2020yyv}. Here we will only briefly explain its constituents.

The indices $\alpha$ and $\beta$ indicate that the final states must necessarily include the leptons $\ell_\alpha$
and $\ell_\beta$.
Parameter $\tau_d$ represents the detector exposure time, $V_{\mathcal{S}}$ and $V_{\mathcal{D}}$ are
the fiducial volumes of, respectively, the source ($\mathcal{S}$) and detector ($\mathcal{D}$) ``devices'';
the spacial integrations are performed over these volumes.
The differential form $d\sigma_{{\nu}\mathcal{D}}$ is defined such that the expression
\begin{equation*}
\label{CrossSection}
\frac{1}{V_{\mathcal{D}}}\int d\vec{y}\,d\sigma_{{\nu}\mathcal{D}}
\end{equation*}
represents  the differential cross section of neutrino scattering at the detector $\mathcal{D}$ \emph{as a whole}.
In the practically important case of neutrino scattering on single particles and under the usual assumption that the momentum
distributions of scatterers are very narrow (such as the Maxwellian distribution), the differential form $d\sigma_{{\nu}\mathcal{D}}$
becomes the elementary differential cross section for the given reaction multiplied by the number of scatterers in the volume $V_{\mathcal{D}}$.
The function $\mathcal{P}_{\alpha\beta}(E_\nu,L)$ in Eq.~\eqref{EventNumber}
is the QFT analogue of the quantum-mechanical (QM) flavor transition probability;
it is a function of neutrino energy $E_\nu$ and distance $L=\vert\vec{y}-\vec{x}\vert$ between the points
$\vec{x}\in\mathcal{S}$ and $\vec{y}\in\mathcal{D}$;
for any pair $(\vec{x},\vec{y})$, $L$ is assumed to be large compared to the dimensions of $\mathcal{S}$ and $\mathcal{D}$.
In the conventional ultrarelativistic approximation, the ``probability''
includes the standard QM oscillation phase factor and corrections for decoherence effects of different kinds.
These corrections will be discussed in Sect.~\ref{sec:CountRate}, when comparing the expressions for the number of events in
the LBL and SBL modes.

Finally, the differential form $d\mathfrak{F}_\nu$ is the differential neutrino flux in $\mathcal{D}$ defined such that the quantity
\begin{equation*}
\label{NeutrinoFlux}
\frac{d\vec{x}}{V_{\mathcal{S}}}\int\frac{d\mathfrak{F}_\nu}{dE_{\nu}}
\end{equation*}
is the number of neutrinos appearing per unit time and unit neutrino energy in the elementary $3d$ volume $d\vec{x}=d^3x$
around the point $\vec{x}$, moving within a unit solid angle in the direction from point $\vec{x}$ to point $\vec{y}$,
and crossing the unit area placed around $\vec{y}$ and perpendicular to $\vec{y}-\vec{x}$.
According to the extended Grimus-Stockinger theorem \eqref{ExGS},
\begin{equation}
\label{TheISLviolatinFactor}
d\mathfrak{F}_\nu \propto \frac{1}{\vert\vec{y}-\vec{x}\vert^2}
\Bigg[1+{\sum_{n\ge1}\frac{\mathfrak{C}_n}{\vert\vec{y}-\vec{x}\vert^{2n}}}\Bigg].
\end{equation}
As a consequence, ISL is broken.
The coefficients of the series \eqref{TheISLviolatinFactor} are discussed in Ref.~\cite{Naumov:2013bea}.
Recall that $\mathfrak{C}_1 < 0$, so the lowest-order ISLV correction reduces the number of neutrino events:
\begin{equation}
\label{ISLVforN}
N_{\beta\alpha} \approxprop \frac{1}{L^2}\left(1-\frac{\mathfrak{L}_0^2}{L^2}\right).
\end{equation}
Here $L$ is the distance between the source and detector (assumed large compared to their sizes)
and $\mathfrak{L}_0$ is an energy-dependent parameter of dimension of length:
\begin{equation*}
\mathfrak{L}_0 \sim \frac{E_\nu}{\sigLBL^2} \approx 
20\left(\frac{E_\nu}{1~\text{MeV}}\right)\left(\frac{\sigLBL}{1~\text{eV}}\right)^{-2}~\text{cm}.
\end{equation*}
Recent analysis \cite{Naumov:2021vds} shows that current reactor antineutrino data timidly hints that the ISLV effect
is already manifesting itself, although the accuracy of the data is still too low  and the uncertainty of the calculated
reactor antineutrino energy spectra is too large to draw confident conclusions.

\section{Short baseline asymptotics}
\label{sec:SBLasymptoticsOfAmplitude}

We are now ready to study the short-baseline (SBL) behavior of the amplitude \eqref{Amplitude_2}.
It will be shown that the SBL formula for the number of neutrino events leads, like Eq.~\eqref{EventNumber},
to an ISL behavior, which is also violated by higher-order corrections.

To simplify things a bit, we restrict ourselves to the conditions
\[\big\vert\left(q_s-p_\alpha\right)^2\big\vert \sim \big\vert\left(q_d-p_\beta\right)^2\big\vert \ll m_W^2.\]
Then the $W$ propagator $\Delta_{\mu\nu}$ can be approximated by $-ig_{\mu\nu}/m_W^2$.
Next, in a good approximation, we can replace $\hat{q}$ with $\hat{\kP}$, where the 4-vector $\kP$
is the stationary point of $\varPhi(q)$.
From Eqs.~\eqref{Phi_just_gauss} and \eqref{F(q)_def} it immediately follows that
\begin{equation}
\label{SaddlePoint}
\kP_\mu=\kR_{\mu\nu}Y^\nu, \quad
Y^{\mu} = \wRe_s^{\mu\nu}q_{s\nu}-\wRe_d^{\mu\nu}q_{d\nu},                         
\end{equation}
where $\kR$ is the tensor inverse of $\wRe$:%
\footnote{It may be expedient to recall that the tensors $\wRe_{s,d}$ are inverse of $\Re_{s,d}$;
          their explicit form can be found in Ref.~\cite{Naumov:2013bea}. Since these tensors are
          symmetric and positive definite, the same is true for the tensor $\kR$.}
\begin{equation}
\label{TheEquationForkR}
  \kR_{\mu\lambda}\wRe^{\lambda\nu} = \delta^{\nu}_\mu,
\quad
\widetilde{\Re} \equiv \widetilde{\Re}_s+\widetilde{\Re}_d.  
\end{equation}
In the exact energy-momentum conservation limit $\kP=q_s=-q_d$ and, due to the suppresion factor \eqref{Phi_just_gauss}
in the amplitude, $\kP \simeq q_s \simeq -q_d$ even otherwise. 
So, within the saddle-point approximation, it is natural to interpret $\kP$ as the 4-momentum of the virtual neutrinos.
It should be remarked that $\kP$ is independent of the neutrino masses and is determined only by the momenta, masses and momentum spreads
of the external wave packets (from both vertices of the macrodiagram), but it will be shown later that although
neutrino virtuality, $\kP^2-m_j^2$, can be rather large compared to $m_j^2$, 
zero virtuality is the most probable for each ($j$-th) contribution to the full amplitude \eqref{Amplitude_2},
which, of course, is a trivial consequence of the pole $q^2=m_j^2$ in the neutrino propagator \eqref{eq_ThePropagator}.
Another simple but nonetheless important observation: At no configuration of external momenta can all the neutrinos $\nu_j$
($j=1,2,3,...$) together be on their mass shell.
Thus, at a substantial neutrino mass hierarchy (which, as the experiments suggest, is the case in the real world), neutrinos at short baselines
are \emph{deeply virtual}. 
This is basically different from the propagation regime occurring at long baselines, when all the neutrinos
end up on their own mass shell and have different energies, 3-momenta, and velocities.

Since the Grimus-Stokinger theorem and its extension \eqref{ExGS} are inapplicable at short distances,
further analysis is based on another method, based on an asymptotic expansion of the full $4d$ neutrino propagator
\eqref{J(X)_def}. The expansion is derived in Appendix \ref{sec:SBLasymptoticsOfNeutrinoPropagator}
in terms of the dimensionless variables $\delta_j$, $\eta_j$, and $\Delta_j$ defined by%
\footnote{In the Appendix, the index $j$ is omitted to shorten the formulas.
          Notice that $\delta_j$ and $\eta_j$ weakly depend on $m_j$ if, as will be eventually assumed,
          $\vec{\kP}^2 \sim \kP_0^2 \gg m_j^2$; in this approximation, the index $j$ can be ignored, and the approximations for $\vert\delta_j\vert$, $\eta_j$, and $\vert\Delta_j\vert$
          in Eqs.~\eqref{SmallParameters} refer just to this case.}
\begin{subequations}
\label{SmallParameters}
\begin{align}
\label{delta_j}
\delta_j = &\ \sig^2\frac{\rho_{\mu\nu}p_j^{\mu}X^\nu}{\rho_{\mu\nu}p_j^{\mu}p_j^{\nu}}, \quad  \vert\delta_j\vert \simeq \frac{\sig^2\vert\vec{X}\vert}{\vert\vec{\kP}\vert}, \\
\label{eta_j}
\eta_j   = &\ \frac{\sig^2}{E_j^2} \simeq \frac{\sig^2}{\vert\vec{\kP}\vert^2}, \\
\label{Delta_j}
\Delta_j = &\ \frac{{\kP}p_j-m_j^2}{\rho_{\mu\nu}p_j^{\mu}p_j^{\nu}}, \quad \vert\Delta_j\vert \lesssim \frac{2\sig}{\sqrt{\rho^{\mu\nu}l_{\mu}l_{\nu}}\vert\vec{\kP}\vert},
\end{align}
\end{subequations}
which are all assumed to be small (see Appendix~\ref{sec:Jg}), and under the important additional constraint
\begin{equation}
\label{sigX_0>>1}
{\sig}X_0 \gg 1.
\end{equation}
Here we defined
\[
p_j=(E_j,\vec{\kP}), \quad E_j=\sqrt{\vec{\kP}^2+m_j^2}.
\]
The 4-vector $p_j$ is a convenient intermediate notation for now, but, as will soon become clear, it can be given the meaning
of the most probable 4-momentum of $\nu_j$ in the sense that the $j$-th contribution to the amplitude \eqref{Amplitude_2} is maximal at $\kP_0=E_j$.
The Lorentz invariant positive-definite function
\begin{equation}
\label{SigDef}
\sig \equiv \sigSBL = \vert\kR\vert^{1/8}
\end{equation}
(where $\vert\kR\vert \equiv \det(\kR) = \vert\wRe\vert^{-1}$), characterizes the scale of the neutrino energy-momentum uncertainty
and generally depends on the momenta, masses, and momentum dispersions (spreads)
of the external (in and out) wave packets. The dimensionless tensor $\rho$ is defined by
\begin{equation}
\label{rhoDef}
\rho^{\mu\nu} = \vert\kR\vert^{-1/4}\kR^{\mu\nu}.
\end{equation}
It integrally describe the shapes and asymmetries of the space-time regions of overlap of the external wave packets in the vertices of the macrodiagram \ref{fig:Macrograph_Class_A}.
The conditions of smallness of Lorentz scalars \eqref{delta_j} and \eqref{Delta_j} are briefly summarized in Appendix \ref{sec:Jg},
and restriction \eqref{sigX_0>>1} is discussed in Appendix \ref{sec:Jg-Jv}. All this will be covered in more detail below. 

In this article, we will mainly study the leading-order approximation by putting $\delta_j=\eta_j=\Delta_j=0$ in
Eq.~\eqref{J_g_final2_inside_calc}, which gives
\begin{equation}
\label{final_simple_integral_approximation}
i\int\frac{d^4q\,\varPhi(q)e^{-iqX}}{q^2-m_j^2+i\epsilon}
=8\pi^2\sqrt{\frac{\pi\vert\kR\vert}{\mathcal{G}_j}}\exp\left(\Omega_j\right).
\end{equation}
The following notations are used here:
\begin{gather*}
\begin{aligned}
\Omega_j = &\ \Omega_c-\frac{1}{4\mathcal{G}_j}\left({\kP}p_j-m_j^2-2i\kR_{\mu\nu}p_j^{\mu}X^{\nu}\right)^2, \\
\Omega_c = &\ -i\kP X+\kR^{\mu\nu}\left(\frac{1}{4}Y_\mu Y_\nu-X_\mu X_\nu\right)-\frac{1}{4}F_0,    \\
F_0      = &\ \wRe_s^{\mu\nu}q_{s\mu}q_{s\nu}+\wRe_d^{\mu\nu}q_{d\mu}q_{d\nu},
\quad
\mathcal{G}_j = \kR_{\mu\nu}p_j^{\mu}p_j^{\nu},
\end{aligned}
\end{gather*}
The tensor function \eqref{tenzor_function} now takes the form
\begin{multline}
\label{krivaya_G}
\mathbb{G}^j_{\nu\nu'\mu'\mu}\left(\{\vec{p}_{\varkappa},x_{\varkappa}\}\right) \,
= \frac{1}{2\pi^2}\sqrt{\frac{{\pi\vert\kR\vert}}{\mathcal{G}_j}}\,
  \vert\mathbb{V}_d(\kP)\mathbb{V}_s(\kP)\vert \\
  \times\Delta_{\nu\nu'}(\kP-p_{\beta})(\hat\kP+m_j)\Delta_{\mu'\mu}(\kP+p_{\alpha}) \\
  \times\exp\left(-\varOmega_j-i\Theta\right),
\end{multline}
where $\mathbb{V}_{s,d}$ are the overlap integrals defined by Eq.~\eqref{OverlapVolumes_def}, 
which, in the framework of the CRGP model, have exactly the same form as in the LBL case:
\begin{gather*}
\begin{aligned}
\mathbb{V}_{s,d}(q) = &\ (2\pi)^4\widetilde{\delta}_{s,d}\left(q{\mp}q_{s,d}\right)  \\
                      &\ \times\exp\left[-\mathfrak{S}_{s,d} \pm i\left(q \mp q_{s,d}\right)X_{s,d}\right],
\end{aligned}
\intertext{where}
\begin{aligned}
\widetilde{\delta}_{s,d}(K)
= &\ \int\frac{d^4x}{(2\pi)^4}\exp\left(iKx-\mathfrak{R}_{s,d}^{\mu\nu}x_{\mu}x_{\nu}\right) \\
= &\ \dfrac{\exp\left(-\dfrac{1}{4}\widetilde{\Re}_{s,d}^{\mu\nu}K_{\mu}K_{\nu}\right)}{(4\pi)^2\sqrt{\vert\Re_{s,d}\vert}}. 
\end{aligned}
\end{gather*}
The function $\varOmega_j$ (not to be confused with the function $\Omega_j$ used above) is given by
\begin{multline}
\label{varOmega_j}
\varOmega_j = i{\kP}X+\frac{1}{\mathcal{G}_j}\left[\frac{1}{2}\left(\kP p_j-m_j^2\right)-i\kR_{\mu\nu}X^{\mu}p_j^\nu\right]^2 \\
              +\kR_{\mu\nu}X^{\mu}X^{\nu}.
\end{multline}
It is convenient to enter the intermediate notation
\begin{equation*}
\mathcal{M} = \frac{g^4}{64}\mathcal{J}_d^{*\nu}\Delta_{\nu\nu'}\overline{u}(\vec{p}_\beta)O^{\nu'}{\hat\kP}O^{\mu'}v(\vec{p}_\alpha)\Delta_{\mu'\mu}\mathcal{J}_s^{\mu},
\end{equation*}
with which and considering that $O^{\mu}O^{\nu}=0$ the amplitude \eqref{Amplitude_2} can be rewritten as
\begin{multline*}
\mathcal{A}_{\beta\alpha}
= \frac{\mathcal{M}\sqrt{\vert\kR\vert}\,\vert\mathbb{V}_s(\kP)\mathbb{V}_d(\kP)\vert}{2\pi^{3/2}\mathcal{N}} \\
  \times\sum_j\frac{V_{\beta j}V^*_{\alpha j}}{\sqrt{\mathcal{G}_j}}\exp\left(-\varOmega_j-i\Theta\right).
\end{multline*}
Let us now investigate the real part of the function $\varOmega_j$ in more detail.
As is seen from Eq.~\eqref{varOmega_j}, it splits into two parts with very different physical content:
\begin{gather}
\label{ReOmega_j}
\text{Re}\,\varOmega_j \equiv R_{1j}+R_{2j}, \\
\begin{aligned}
R_{1j} = &\ \frac{1}{4\mathcal{G}_j}\left({\kP}p_j-m_j^2\right)^2, \nonumber \\
R_{2j} = &\ \kR_{\mu\nu}\left(X^{\mu}X^{\nu}-\frac{1}{\mathcal{G}_j}\kR_{\mu'\nu'}X^{\mu}p_j^{\nu}X^{\mu'}p_j^{\nu'}\right). \nonumber
\end{aligned}
\end{gather}

\subsection{Off-shell control}
\label{sec:Off-ShellControl}

To clarify the physical meaning of $R_{1j}$, we rewrite this function as follows:
\begin{align}
\label{R1j_approx_1}
R_{1j}    =   &\ \frac{1}{4\mathcal{G}_j}E_j^2\left(E_j-\kP_0\right)^2 \nonumber \\
       \simeq &\ \frac{1}{4\sig^2\rho_{j}}
                 \left(\vert\vec{\kP}\vert-\kP_0+\frac{m_j^2}{2\vert\vec{\kP}\vert}\right)^2,
\end{align}
where
\[
\rho_{j} = \rho_{00}-2\rho_{0m}v_{jm}+\rho_{mn}v_{jm}v_{jn} > 0,
\]
$\vec{v}_j=\vec{\kP}/E_j$ is the most probable velocity of the $j$-th neutrino (see below),
of course different from the ``virtual velocity'' $\vec{v}=\vec{\kP}/\kP_0$;
the Lorentz scalar $\sig$ and tensor $\rho$ are defined by Eqs.~\eqref{SigDef} and \eqref{rhoDef}, respectively.

Let us digress for a while to clarify the meaning of the function $\sig$ by considering a toy model (hereinafter referred to as \ToyModel\ model) in which neutrinos are produced
in two-particle decays ($\pi_{\mu_2}$ for certainty) and, for one reason or another, contribution to the tensor $\kR$ from the detector vertex can be ignored,
so that $\kR=\Re_s=T_\pi+T_\mu$. Then, neglecting for further simplification the neutrino mass and assuming energy-momentum conservation, we find:
\[
\sig^{(\TM)} = \left[\frac{1}{2}\sigma_{\pi}\sigma_{\mu}\left(\sigma_{\pi}^2+\sigma_{\mu}^2\right)\left(\frac{m_{\pi}}{m_{\mu}}-\frac{m_{\mu}}{m_{\pi}}\right)\right]^{1/4},
\]
where $m_{\pi,\mu}$ ($\sigma_{\pi,\mu}$) are the masses (momentum spreads) of the pion and muon wave packets, respectively.
If, e.g., $\sigma_{\mu}=\sigma_{\pi}$ then $\sig^{(\TM)} \approx 0.87\sigma_{\pi}$.
The extreme simplicity of this example should not, however, be misleading, because for a process involving three packets already, 
$\sig$ (as well as $\mathcal{G}_j$, see below) is an invariant function of external momenta, not a constant.

\begin{figure}[hbt]
\centering\includegraphics[width=\linewidth]{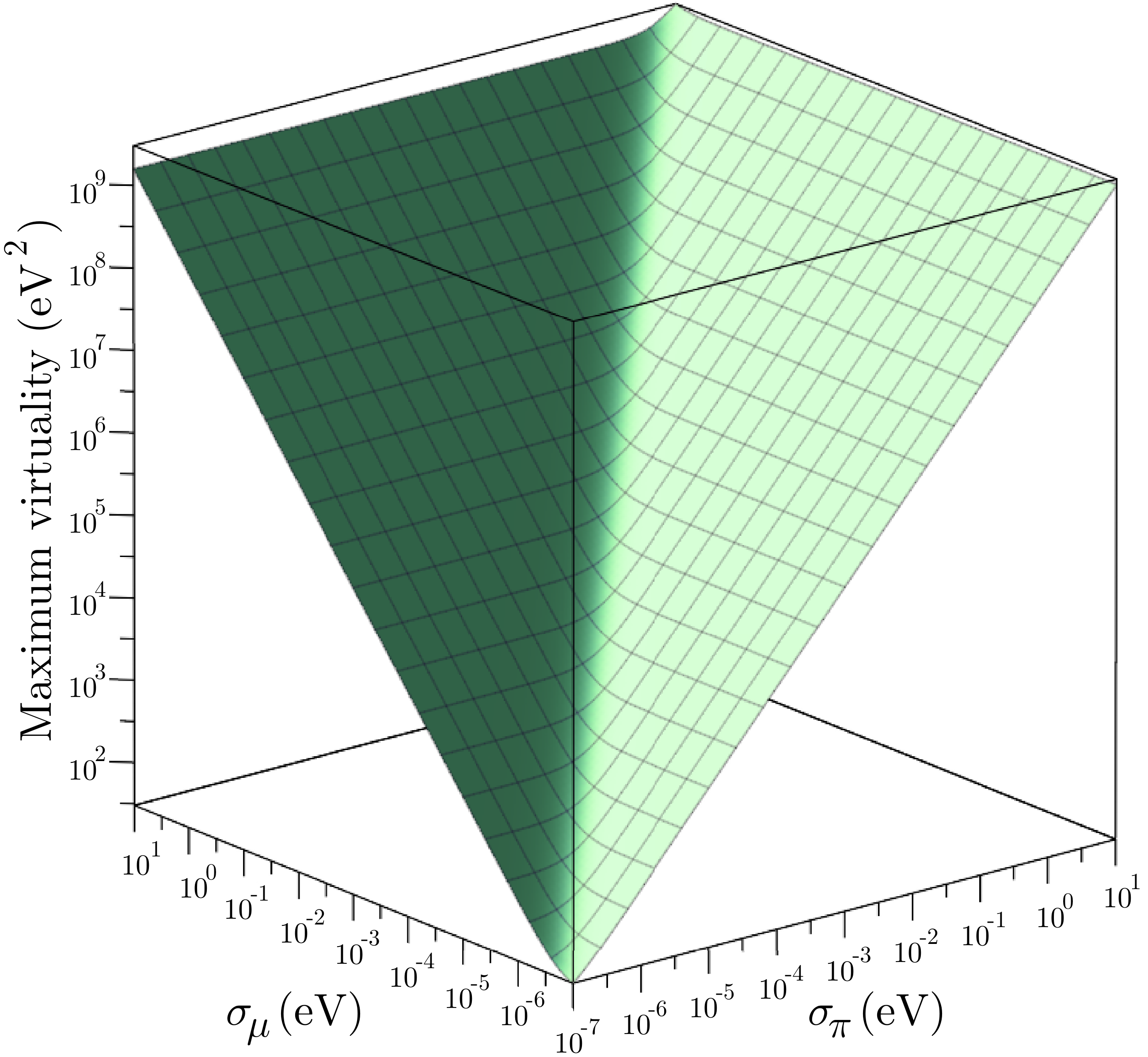}
\caption{Maximum permissible neutrino virtuality vs.\ $\sigma_\pi$ and $\sigma_\mu$ estimated in the \ToyModel\ model.
         Calculation is made assuming energy-momentum conservation and neglecting the neutrino mass.
         \label{MaxVirtuality}
        }
\end{figure}

Now, returning to the general case, we will restrict ourselves to the ultra-relativistic approximation:
\[
\kP_0^2 \sim \vec{\kP}^2 \gg m_j^2.
\]
In fact, this condition is not quite trivial for deeply virtual neutrinos, since the components of the 4-vector $\kP$ are formally mutually independent.
However, it is a direct consequence, provided by the natural conditions for the energy and momentum transfers in the vertices of the macrodiagram \ref{fig:Macrograph_Class_A},
\[
\left(q_s^0\right)^2 \sim \vec{q}_s^2 \gg m_j^2
\enskip\text{and}\enskip 
\left(q_d^0\right)^2 \sim \vec{q}_d^2 \gg m_j^2
\quad (\forall j),
\]
and by the presence of the multipliers $\widetilde\delta_s(q-q_s)$ and $\widetilde\delta_d(q+q_d)$ (smeared $\delta$-functions) in the amplitude.%
\footnote{They also ensure that $\vert\vec{q}_s\vert \sim \vert\vec{q}_d\vert$ and $q_s^0 \sim -q_d^0$.}
It is in this sense that we will call virtual neutrinos ultrarelativistic. In this approximation Eq.~\eqref{R1j_approx_1} can be rewritten as
\begin{equation*}
R_{1j} \simeq \frac{1}{\rho_{j}}\left(\frac{\kP^2-m_j^2}{4\sig\vert\vec{\kP}\vert}\right)^2.
\end{equation*}
From this relationship follows that $0 \le R_{1j} \lesssim 1$ when
\begin{equation}
\label{VirtualityInequalities}
0 \le \vert\kP^2-m_j^2\vert \lesssim 4\vert\vec{\kP}\vert\sig\sqrt{\rho_{j}} \approx 4\sqrt{\mathcal{G}_j}.
\end{equation}
Thus, the factor $\exp\left(-R_{1j}\right)$ controls the magnitude of the neutrino virtuality:
it allows a significant escape of the virtual neutrinos from theis mass shell, but suppresses the contributions
to the amplitude from external 4-momentum configurations that give \emph{too} much virtuality $\vert{\kP^2}-m_j^2\vert$.
For a numerical illustration, consider again the \ToyModel\ model. In the exact energy-momentum conservation limit, we find:
\begin{equation*}
\begin{aligned}
\mathcal{G}_j^{(\TM)}   =   &\ \frac{1}{8}\bigg\{\frac{1}{m_\mu^2}\left[m_\pi^2-(m_\mu+m_j)^2\right] \\ 
                            &\ \times\left[m_\pi^2-(m_\mu-m_j)^2\right]\sigma_\mu^2                  \\
                            &\ +2\left(m_\pi^2-m_\mu^2+m_j^2\right)\sigma_\pi^2\bigg\}               \\
                     \simeq &\ \frac{m_\pi^2-m_\mu^2}{8}\left[2\sigma_\pi^2+\left(\frac{m_\pi^2}{m_\mu^2}-1\right)\sigma_\mu^2\right].
\end{aligned}
\end{equation*}
It is apparent that the neutrino mass is completely negligible here and $\mathcal{G}_j^{(\TM)} \ggg m_j^4$.
The dependence of the maximum virtuality on $\sigma_\pi$ and $\sigma_\mu$
is shown in Fig.~\ref{MaxVirtuality}.
It can be seen that although the on-shell neutrinos make the maximum contribution to the amplitude,
the neutrino virtuality can be potentially very large if the dispersions of the momenta of the external wave packets
($\sigma_\pi$ and $\sigma_\mu$ in this example) are large enough.
It is noteworthy that constraints \eqref{VirtualityInequalities} (as well as \eqref{VelocityInequalities} below)
are actually independent of neutrino mass, provided that $\vec{\kP}^2 \sim\kP_0^2 \gg m_j^2$.

\begin{figure}[hbt]
\centering\includegraphics[width=\linewidth]{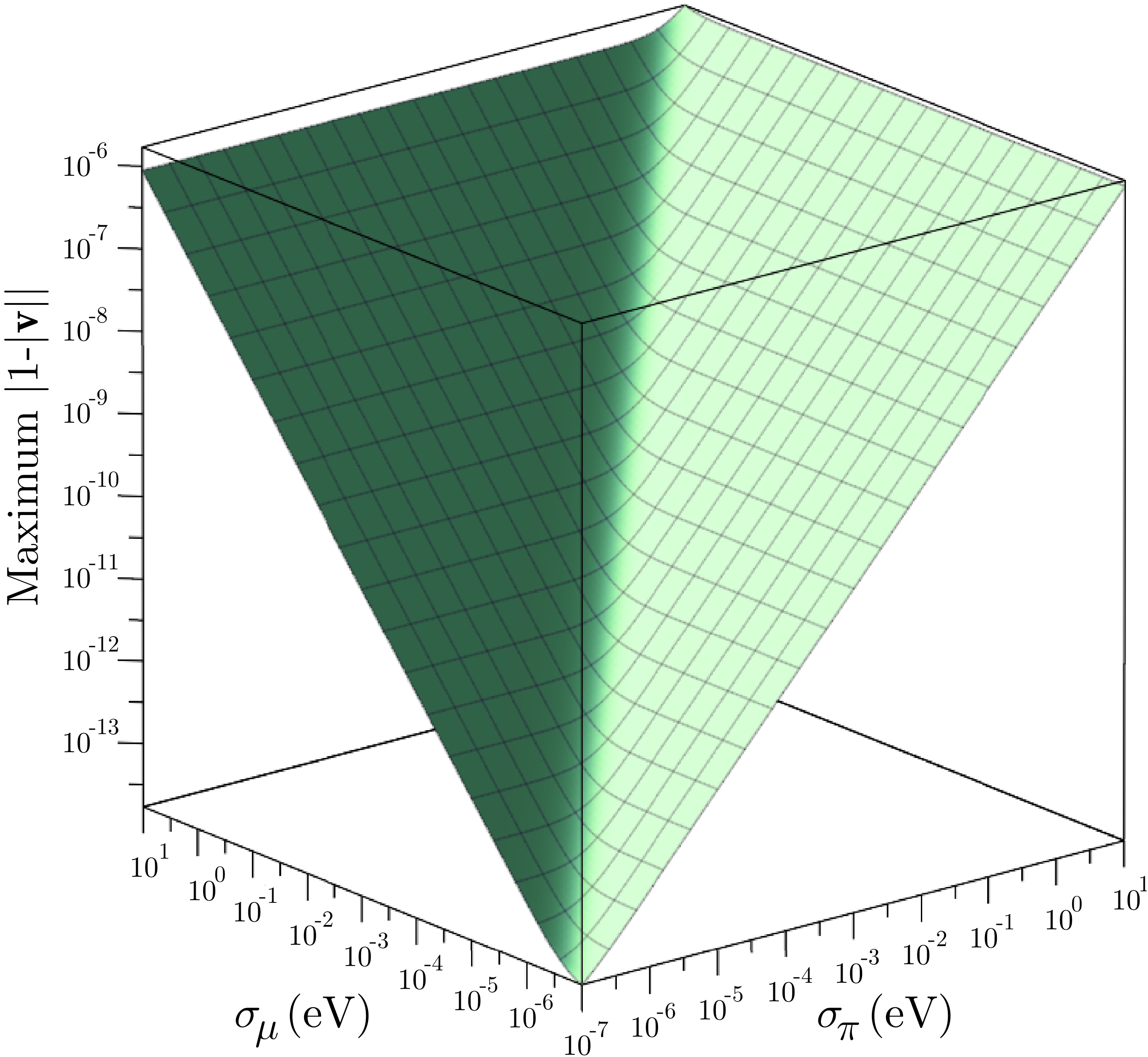}
\caption{Maximum permissible deviation of the virtual neutrino velocity from the speed of light
         vs.\ $\sigma_\pi$ and $\sigma_\mu$ estimated in the \ToyModel\ model.
         Calculation is made assuming energy-momentum conservation and neglecting the neutrino mass.
         The virtual neutrino energy in Eqs.~\protect\eqref{VelocityInequalities} is taken equal to
         the real neutrino energy in the pion rest frame, that is, $\kP_0=(m_\pi^2-m_\mu^2)/(2m_\pi)\approx 29.8$~MeV.
         \label{BoundsForVelocity}
         }
\end{figure}

Although the virtual neutrinos can behave as a bradyonic, luxonic, or tachyonic particle, the virtual velocity value
$\vert\vec{v}\vert=\vert\vec{\kP}/\kP_0\vert$ always remains close to the speed of light,
\begin{equation}
\label{VelocityInequalities}
\vert 1-\vert\vec{v}\vert \vert \lesssim \frac{2\sig\sqrt{\rho_{j}}}{\kP_0}
\simeq \frac{2\sqrt{\mathcal{G}_j}}{\kP_0^2} \ll 1,
\end{equation}
and the most probable velocity of $\nu_j$ (determined by the condition $R_{1j}=0$) is, as expected, 
\begin{equation}
\label{VelocityMostProbable}
\vert\vec{v}_j\vert = \frac{\vert\vec{\kP}\vert}{E_j} = \left(1+\frac{m_j^2}{\vert\vec{\kP}\vert^2}\right)^{-1/2} \! \simeq 1-\frac{m_j^2}{2\vert\vec{\kP}\vert^2}.
\end{equation}
Figure \ref{BoundsForVelocity} shows the maximum value of $\vert 1-\vert\vec{v}\vert \vert$ calculated in the \ToyModel\ model.
For comparison: the velocity of the real $\nu_j$ from $\pi_{\mu_2}$ decay at rest is equal to
\[
1-\frac{2m_\pi^2m_j^2}{(m_\pi^2-m_\mu^2)^2} \approx 1-5.6\times10^{-18}\left(\frac{m_j}{0.1~\text{eV}}\right)^2.
\]

\subsection{Quasiclassical motion}
\label{sec:QuasiclassicalMotion}

Consider now the second term in Eq.~\eqref{ReOmega_j}. The factor $\exp\left(-R_{2j}\right) $ suppresses the corresponding ($j$-th)
term in the amplitude when neutrino deviates from its classical world line. To prove this, let us rewrite $R_{2j}$ as follows:
\begin{subequations}
\label{trajectory_suppression_term}
\begin{align}
\label{trajectory_suppression_term_a}
R_{2j} = &\ \frac{1}{\mathcal{G}_j}\!\left(\kR_{\mu\nu}\kR_{\lambda\rho}-\kR_{\mu\lambda}\kR_{\nu\rho}\right)p_j^{\lambda}p_j^{\rho}X^{\mu}X^{\nu} \\
\label{trajectory_suppression_term_b}
       = &\ \frac{1}{\mathcal{G}_j}\kR_{\mu\nu}\kR_{\lambda\rho}X^{\mu}p_j^{\rho}\left(X^{\nu}p_j^{\lambda}-X^{\lambda}p_j^{\nu}\right).
\end{align}
\end{subequations}
It is easy to prove that $R_{2j}>0$. In the neutrino rest frame (checkmarked) Eq.~\eqref{trajectory_suppression_term_a} converts to
\begin{align*}
R_{2j} = &\ \frac{1}{\check{\kR}_{00}}\left(\check{\kR}_{00}\check{\kR}_{\mu\nu}-\check{\kR}_{0\mu}\check{\kR}_{0\nu}\right)\check{X}^{\mu}\check{X}^{\nu} \\
       = &\ \frac{1}{\check{\kR}_{00}}\left(\check{\kR}_{00}\check{\kR}_{kn}-\check{\kR}_{0k}\check{\kR}_{0n}\right)\check{X}_k\check{X}_n.
\end{align*}
By rotating the coordinate system to align the $z$-axis along $\vec{\check{X}}$, we find
\[
R_{2j} = \frac{\vert\vec{\check{X}}\vert^2}{\check{\kR}_{00}}\left(\check{\kR}_{00}\check{\kR}_{33}-\check{\kR}_{03}^2\right).
\]
This fuction is positive because the expression in parentheses and the denominator are the principal
minors of the symmetric positive-definite matrix $\vert\vert\kR_{\mu\nu}\vert\vert$. 

From Eq.~\eqref{trajectory_suppression_term_b} follows that $R_{2j}$ vanishes along the classical trajectory
$X^\mu=u_j^\mu\tau$, where $\tau$ is the proper time, $u_j=p_j/m_j=\left(E_j/m_j,\vec{\kP}/m_j\right)$
is the neutrino 4-velocity, and it is assumed that $m_j>0$.
Below we will need another useful representation of the function $R_{2j}$, which also allows us to extend
the above conclusion to massless neutrinos. After some manipulations, expression \eqref{trajectory_suppression_term_b}
can be rewritten as
\begin{equation}
\label{R_2j}
R_{2j} = \frac{D_{j\,kn}\left(X_k-v_{jk}X_0\right)\left(X_n-v_{jn}X_0\right)}{\kR_{00}-2\kR_{0n}v_{jn}+\kR_{kn}v_{jk}v_{jn}},
\end{equation}
where
\begin{multline}
\label{D_j}
D_{j\,kn} = \kR_{00}\kR_{kn}-\kR_{0k}\kR_{0n}                                      \\
           +\left(\kR_{0k}\kR_{nm}-2\kR_{0m}\kR_{kn}+\kR_{0n}\kR_{km}\right)v_{jm} \\
           +\left(\kR_{lm}\kR_{kn}-\kR_{ln}\kR_{km}\right)v_{jm}v_{jl}.
\end{multline}
Although the relativistic invariance and positivity of expression~\eqref{R_2j} are not seen explicitly,
they certainly take place, since it is an identical rewrite of Eq.~\eqref{trajectory_suppression_term_b}.
Evidently $R_{2j}=0$ on the classical trajectory $\vec{X}=\vec{v}_jX_0$, where $\vec{v}_j=\vec{\kP}/E_j$.
It is easy to verify that the function $R_{2j}$ is invariant under the group of uniform rectilinear motions
(displacements along the classical word lines)
\[
g_j = \left\{X_0 \mapsto X_0+\theta, \vec{X}\mapsto\vec{X}+\vec{v}_j\theta;~\vert\theta\vert<\infty\right\}.
\]
This means that the virtual neutrino behaves like a quasi-classical on-shell particle or, more precisely,
like the asymmetric (``true Gaussian'') wave packet studied in Ref.~\cite{Naumov:2014jpa}.
In other words, there is a duality between the description of the virtual neutrino motion by the causal propagator
and by an effective wave packet (see also Ref.~\cite{Naumov:2013bea} and Sect.~\ref{sec:EvolutionOfNeutrinoWavePacket} below).
One more important fact should be pointed out: as proved in Ref.~\cite{Naumov:2014jpa}, the true Gaussian wave packet
does not coincide in any approximation with CRGP, which describes the neutrino state in the LBL regime
\cite{Naumov:2010um,Naumov:2020yyv}.
Thus, the neutrino propagation at short distances is profoundly different from that at long distances.

\section{Microscopic probability}
\label{sec:Probability}

The following important identity
\begin{eqnarray}
\label{SquaredOverlapIntegral}
\vert\mathbb{V}_{s,d}(q)\vert^2 = (2\pi)^4\delta_{s,d}(q{\mp}q_{s,d})\mathrm{V}_{s,d},
\end{eqnarray}
valid for any 4-vector $q$, was proved in Ref.~\cite{Naumov:2020yyv}. Here
\begin{equation}
\label{delta}
\delta_{s,d}(K) = \frac{\exp\left(-\dfrac{1}{2}\,\wRe_{s,d}^{\mu\nu}K_{\mu}K_{\nu}\right)}{(2\pi)^2\sqrt{\vert\Re_{s,d}\vert}},
\end{equation}
\begin{equation}
\label{mathrmVsd}
\mathrm{V}_{s,d}
 = \int d^4x \prod\limits_{\varkappa{\in}S,D}\left\vert\psi_{\varkappa}\left(\vec{p}_{\varkappa},x_{\varkappa}-x\right)\right\vert^2,
\end{equation}
and $\psi_{\varkappa}\left(\vec{p}_{\varkappa},x_{\varkappa}\right)$ is the coordinate-dependent part of the 
wave function of the packet $\varkappa$ in the coordinate representation \eqref{psi_coord}
with $p=p_{\varkappa}$, $x=x_{\varkappa}$ and $m=m_{\varkappa}$.
The functions $\delta_{s,d}(K)$, of course, do not coincide with the functions $\widetilde{\delta}_{s,d}(K)$ used so far,
but have the same plane-wave limit \eqref{delta_s,d_PWL} as $\widetilde{\delta}_{s,d}(K)$ and similar properties.
The physical meaning of the functions $\mathrm{V}_s$ and $\mathrm{V}_d$ is obvious from the
integral representation \eqref{mathrmVsd}; it implies that they should be treated as $4d$ (space-time) overlap volumes
of incoming and outgoing wave packets in the source and detector, respectively.
It can be proved that they take maximum values when the classical world lines of the packets intersect at the impact points.

Using identity \eqref{SquaredOverlapIntegral} and taking into account 
the explicit form of the normalization factor $\mathcal{N}$ \cite{Naumov:2010um,Naumov:2020yyv},
\begin{gather*}
\label{NormalizationFactor}
\mathcal{N}^2 = \langle\mathrm{in}\vert\mathrm{in}\rangle\langle\mathrm{out}\vert\mathrm{out}\rangle
              = \prod\limits_{\varkappa{\in}S+D}2E_{\varkappa}\mathrm{V}_{\varkappa}(\vec{p}_{\varkappa}),
\intertext{where}
\mathrm{V}_{\varkappa}\left(\vec{p}_{\varkappa}\right)
              = \int d^3{x}\big\vert\psi_{\varkappa}(\vec{p}_{\varkappa},x)\big\vert^2
\end{gather*}
is the effective spatial volume of the wave packet $\varkappa$,
one can derive the following formula for the modulus squared transition amplitude
(neutrino detection probability):
\begin{multline}
\label{MicroscopicProbability_1}
\vert\mathcal{A}_{\beta\alpha}\vert^2
= \frac{64\pi^5\vert\kR\vert\delta_s(\kP-q_s)\delta_d(\kP+q_d)\mathrm{V}_s\mathrm{V}_d}
  {\prod\limits_{\varkappa{\in}S}2E_{\varkappa}\mathrm{V}_{\varkappa}\left(\vec{p}_{\varkappa}\right)
   \prod\limits_{\varkappa{\in}D}2E_{\varkappa}\mathrm{V}_{\varkappa}\left(\vec{p}_{\varkappa}\right)} \\
   \times\left\vert\sum_j\frac{\mathcal{M}V_{\beta j}V^*_{\alpha j}}{\sqrt{\mathcal{G}_j}}
   \exp\left(-\varOmega_j\right)\right\vert^2.
\end{multline}
At first glance, this expression does not show the ISL-like behavior (that is $\vert\mathcal{A}_{\beta\alpha}\vert^2\propto1/\vert\vec{X}\vert^2$).
However, since the exponential factor
in the right-hand side of Eq.~\eqref{MicroscopicProbability_1} suppresses the configurations
of the impact points $X_{s,d}$ that deviate significantly from the classical trajectories $\vec{X}_d-\vec{X}_s=\vec{v}_j(X_d^0-X_s^0)$,
we can expect ISL to hold at least approximately. In order to see this explicitly, 
let us first recast Eq.~\eqref{MicroscopicProbability_1} to a form that is the Lorentz-invariant $4d$ analogue of the
$1d$ integral representation proposed by Cardall \cite{Cardall:1999ze}.
For this purpose, consider the identity
\begin{multline*}
(2\pi)^4{\delta}_s\left(q-q_s\right){\delta}_d\left(q+q_d\right)f(q)
= \frac{1}{\sqrt{\vert\Re_s\vert\vert\Re_d\vert}} \\
  \times\exp\left[-\frac{1}{2}\left(F_0-2Y^{\mu}q_{\mu}+\widetilde{\Re}^{\mu\nu}q_{\mu}q_{\nu}\right)\right]f(q),
\end{multline*}
in which $f(q)$ is an arbitrary smooth function of $q$. By integrating this identity with respect to $q$ and using the saddle-point approximation, we obtain
\begin{multline*}
(2\pi)^2\int d^4q{\delta}_s\left(q-q_s\right){\delta}_d\left(q+q_d\right)f(q) \\
=\frac{f(\kP)}{\sqrt{\vert\Re_s\vert\vert\Re_d\vert\vert\widetilde{\Re}\vert}}
\exp\left[-\frac{1}{2}(F_0-\kR^{\mu\nu}Y_{\mu}Y_{\nu})\right].
\end{multline*}
Then, substituting
\begin{align*}
F_0-\kR^{\mu\nu}Y_{\mu}Y_{\nu} = &\  \wRe_s^{\mu\nu}(\kP-q_s)_\mu(\kP-q_s)_{\nu} \\ 
                                 &\ +\wRe_d^{\mu\nu}(\kP+q_d)_\mu(\kP+q_d)_{\nu}
\end{align*}
and using definition \eqref{delta} we arrive at the approximate relation
\begin{multline}
\label{CardallTrick4D}
(2\pi)^2\sqrt{\vert\kR\vert}\,\delta_s\left(\kP-q_s\right)\delta_d\left(\kP+q_d\right)f(\kP) \\
=\int d^4q\delta_s\left(q-q_s\right)\delta_d\left(q+q_d\right)f(q),
\end{multline}
valid with the same accuracy with which the amplitude itself was derived (that is, with the accuracy of the saddle point approximation). 
Using this result, we can rewrite Eq.~\eqref{MicroscopicProbability_1} in the desired form:
\begin{multline}
\label{MicroscopicProbability_6}
\vert\mathcal{A}_{\beta\alpha}\vert^2 
= \int d^4q \frac{16\pi^3\sqrt{{\vert\kR\vert}}\delta_s(q-q_s)\delta_d(q+q_d)}
     {\prod\limits_{\varkappa{\in}S}2E_{\varkappa}\mathrm{V}_{\varkappa}\left(\vec{p}_{\varkappa}\right)  \!
      \prod\limits_{\varkappa{\in}D}2E_{\varkappa}\mathrm{V}_{\varkappa}\left(\vec{p}_{\varkappa}\right)} \\
      \times\mathrm{V}_s\mathrm{V}_d\sum_{ij}
      \frac{V_{{\alpha}i}V_{{\beta }j}\mathcal{M}
      \left(V_{{\beta }i}V_{{\alpha}j}\mathcal{M}\right)^*}{\sqrt{\mathcal{G}_j\mathcal{G}_j}}            \\
      \times\exp\left[i\left(q_{i}-q_{j}\right)X+i\left(\omega_i-\omega_j\right)\right]                   \\
      \times\exp\left[-\left(R_{1i}+R_{1j}+R_{2i}+R_{2j}\right)\right],
\end{multline}
where
\begin{equation*}
\omega_j = \frac{\left(E_j-q_0\right)\left(\kR_{0k}+\kR_{kn}v_j^n\right)\left(X^k-v_j^{k}X_0\right)}{\kR_{\mu\nu}v_j^{\mu}v_j^{\nu}}.
\end{equation*}
and $v_j=(1,\vec{q}/E_j)$. Of course, it is implicitly assumed that we replaced $\kP_0 \longmapsto q_0$ and
$\vec{\kP} \longmapsto \vec{q}$ in the functions $R_{1i}$, $R_{1j}$, $R_{2i}$, $R_{2j}$, $\omega_i$, and $\omega_j$.
Representation \eqref{MicroscopicProbability_6} provides the basis for further study. Its main difference from the corresponding LBL asymptotics
is in the last exponential multiplier, which takes into account the non-zero virtuality and deviation from the classical trajectories.
Below we will convert it to a $1d$ integral in order to obtain an expression for the event count rate as close as possible
to the corresponding LBL result obtained in Ref.~\cite{Naumov:2010um} for the on-shell regime and, in particular, to reproduce the inverse-square law.

\subsection{Account for virtuality}
\label{sec:Virtuality}

To simplify expression \eqref{MicroscopicProbability_6}, we take advantage of the fact that the function
$\exp{\left(-R_{1i}-R_{1j}\right)}$ has a sharp peak at its saddle point. In the ultrarelativistic limit it
it has the obvious meaning of the half-sum of the energies of $\nu_i$ and $\nu_j$ and in general case it is equal to
\begin{align}
\label{q_{ij}^0}
q_{ij}^0   =   &\  \frac{E_i^3\mathcal{G}_j+E_j^3\mathcal{G}_i}{E_i^2\mathcal{G}_j+E_j^2\mathcal{G}_i}                                  \\
           =   &\  \frac{E_i+E_j}{2}\left[1+\frac{\left(E_i-E_j\right)\left(E_i^2\mathcal{G}_j-E_j^2\mathcal{G}_i\right)}{\left(E_i+E_j\right)
                   \left(E_i^2\mathcal{G}_j+E_j^2\mathcal{G}_i\right)}\right]                                                 \nonumber \\
        \simeq &\  \vert\vec{q}\vert\bigg[1+\frac{m_i^2+m_j^2}{4\vert\vec{q}\vert^2}-\frac{m_i^2m_j^2}{8\vert\vec{q}\vert^4} 
                  -\frac{\left(m_i^2-m_j^2\right)^2}{16\vert\vec{q}\vert^4}                                                   \nonumber \\
               &\  \qquad\times\frac{\kR_{00}-\kR_{mn}l_ml_n}{\kR^{\mu\nu}l_{\mu}l_{\nu}}\bigg].                              \nonumber 
\end{align}
Quite apparently, the latter approximation (not used further) is valid only in the region
$\vert\vec{q}\vert^2\sim\vert\vec{q}_s\vert^2 \sim \vert\vec{q}_d\vert^2 \gg m_{i,j}^2$,
which, however, gives the main contribution to the integral \eqref{MicroscopicProbability_6}.
It is notable that the difference with the usual on-shell neutrino energy appears here only in the fourth order
with respect to $m_{i,j}/\vert\vec{q}\vert$.
Given Eq.~\eqref{q_{ij}^0}, we can replace $q_0 \longmapsto q_{ij}^0$ everywhere except the exponent
of $-\left(R_{1i}+R_{1j}\right)$. Integrating the latter in $q_0$, we obtain
\begin{align}
\label{module_integral}
I_0 = &\ \int\limits_{-\infty}^{\infty}dq_0\exp\left[-\left(R_{1i}+R_{1j}\right)\right] \nonumber \\
    = &\ 2\sqrt{\frac{\pi\mathcal{G}_i\mathcal{G}_j}{E_i^2\mathcal{G}_j+E_j^2\mathcal{G}_i}}
          \exp\left[-\frac{E_i^2E_j^2\left(E_j-E_i\right)^2}{4\left(E_i^2\mathcal{G}_j+E_j^2\mathcal{G}_i\right)}\right].
\end{align}
This is the exact formula. Formal expansion of the right-hand side of Eq.~\eqref{module_integral} in powers of $m_{i,j}^2$ yields:
\begin{multline}
\label{module_integral_approx}
I_0 \simeq \sqrt{2\pi\kR^{\mu\nu}l_{\mu}l_{\nu}}\Bigg[1+\frac{\kR_{0k}l_k-\kR_{kn}l_kl_n}{\kR^{\mu\nu}l_{\mu}l_{\nu}}\\
           \times\frac{m_i^2+m_j^2}{4\vert\vec{q}\vert^2}
          +\frac{\left(m_i^2-m_j^2\right)^2}{32\kR^{\mu\nu}l_{\mu}l_{\nu}\vert\vec{q}\vert^2}\Bigg].
\end{multline}
Of course, this expansion is also valid only in the region $\vert\vec{q}\vert^2\sim\vert\vec{q}_s\vert^2\sim\vert\vec{q}_d\vert^2 \gg m_{i,j}^2$.
The last term in square brackets can compete with the second one if
\[
\sig^2 \cdot O(1) \lesssim \frac{1}{8} \frac{\left(m_i^2-m_j^2\right)^2}{m_i^2+m_j^2} \lesssim 10^{-4}~\text{eV}^2, 
\]
but both these terms are negligible under experimentally interesting conditions.
The common multiplier $\propto\sig$ in Eq.~\eqref{module_integral_approx} is however crucial:
it shows that the non-zero neutrino virtuality, along with other factors to be discussed below,
is responsible for the flavor transition and survival probabilities.
This is an important distinguishing feature of the neutrino propagation at short distances between
a source and a detector.

\subsection{Account for quasi-classical motion}
\label{sec:QuasiclassicalMotion_Account}

Next, we can integrate over the variables $q_1$ and $q_2$, given that the virtual neutrinos move almost
quasi-classicaly  and so the main contribution to the integral comes from the region where $\vert{q}_1\vert\ll\vert{q}_3\vert$
and $\vert{q}_2\vert\ll\vert{q}_3\vert$ in the coordinate frame where $X_1=X_2=0$ (and hence $\vec{l}=(0,0,1)$).
Since all the factors in the integrand, except the exponent $\exp\left(-R_{2j}-R_{2i}\right)$, slowly evolve in the region
$\vert{q}_{1,2}\vert\ll\vert{q}_3\vert$, we can safely put $q_1=q_2=0$ in these factors. 
The remaining integral can be evaluated by the formal expansion of $R_{2i}$ and $R_{2i}$, represented according to Eq.~\eqref{R_2j}, in powers
of $q_1/q_3$ and $q_2/q_3$ to the second order. As a result of simple but time-consuming calculations, we obtain:%
\footnote{Recall that all ingredients in Eq.~\eqref{angle_integral} are written in the coordinate system, where $\vec{l}=(0,0,1)$.
          For the convenience of further calculations, all tensor convolutions, except those contained in the coefficients $\Zeta_n$,
          are written in a form invariant to rotations of the coordinate axes.}
\begin{gather}
\begin{aligned}
\label{angle_integral}
I_{12} = &\ \int\limits_{-\infty}^{\infty}dq_1 \int\limits_{-\infty}^{\infty}dq_2 \exp\left(-R_{2i}-R_{2j}\right)             \\
       = &\ \frac{{\pi}q_3^2}{2X_0^2}\sqrt{\frac{\kR^{\mu\nu}l_{\mu}l_{\nu}}{\vert\kR\vert\wRe^{\mu\nu}l_{\mu}l_{\nu}}}\,Z(X) \\
		 &\ \times\exp\bigg\{-\frac{\mathcal{Q}^4\left(v_{i3}-v_{j3}\right)^2}{2\kR^{\mu\nu}l_{\mu}l_{\nu}}X_0^2              \\
         &\ -\frac{2}{\wRe^{\mu\nu}l_{\mu}l_{\nu}}\left[X_3-\frac{(v_{i3}+v_{j3})}{2}X_0\right]^2\bigg\},
\end{aligned} \\
\begin{aligned}
\mathcal{Q}^4 = &\ \left(\kR^{00}\kR^{\mu\nu}-\kR^{0\mu}\kR^{0\nu}\right)l_{\mu}l_{\nu} \\
              = &\ \left(\kR_{00}\kR_{kn}-\kR_{0k}\kR_{0n}\right)l_{k}l_{n},
\end{aligned} \\
\label{ISLVseries}
Z(X) = 1+\sum_{n\ge1}\Zeta_n\left(\frac{X_3-X_0}{X_0}\right)^n.
\end{gather}
In this calculation, we neglected the neutrino masses in the pre-exponential multiplier \eqref{ISLVseries},
because taking them into account would exceed the accuracy.%
\footnote{It is, however, of interest that series \eqref{ISLVseries} contains all degrees of the skew symmetric
          corrections $(X_3-X_0)/X_0$, since they are invariant with respect to the $PT$ transformation.}
The same is obviously true for the term proportional to $\left(v_{i3}-v_{j3}\right)^2$ in the exponent; it could be significant only
at very large times satisfying the condition 
\[
\bigg[\frac{{\sig}X_0\left(m_i^2-m_j^2\right)}{2q_3^2}\bigg]^2 \gtrsim 1,
\]
which, as will soon be shown, strongly contradicts the restrictions determining the applicability of the formalism under discussion.
Therefore, this term should certainly be neglected in further calculations.
The leading term in the exponent expectedly shows that the main contributions to the modulus squared amplitude \eqref{MicroscopicProbability_1}
give the classical trajectories $X_3=v_{j3}X_0$ of the effective wave packets of virtual neutrinos $\nu_j$.
The deviations from the classical trajectories is suppressed by the universal Lorentz-invariant multiplier
$1/\wRe^{\mu\nu}l_{\mu}l_{\nu}=O\left(\sig^2\right)$ -- perfectly the same as in the LBL mode~\cite{Naumov:2010um}. 

\subsection{ISL violation at supershort baselines}
\label{sec:ISLVatSSBL}

The main result of this exercise is, of course, a mere reproduction in QFT of the classical ISL at \emph{short times},
$1/\sigSBL \ll T_0 \ll \vert\vec{q}\vert/\sigSBL^2$, which is the SBL analogue to the Grimus-Stokinger theorem,
applicable to asymptotically \emph{long distances}, $L \gg \vert\vec{q}\vert/\sigLBL^2$,
and also deviations from ISL at $T_0 \approx L \lesssim 1/\sigSBL$, defined by the series \eqref{ISLVseries}.
The dimensionless coefficients $\Zeta_n$ of this series can, in principle, be calculated for any finite $n$,
but they are too cumbersome to reproduce here, and, furthermore, they do not seem to be of practical interest since measurements
at the supershort distances $L\lesssim 1/\sig \approx 0.2~\mu\text{m}\,\left(1\,\text{eV}/\sig\right)$ are not feasible.

As a propaedeutic example, we write out only the coefficient for the leading correction:%
\footnote{\label{Rotation} Recall that the function $\Zeta_1$ is not an explicit rotation invariant.
          Here and in similar cases below, it is implicitly assumed that the tensors are properly transformed at the transition
          to the rotated coordinate system, but we do not rename or label them so as not to complicate already cumbersome formulas.
          We hope that this will not lead to misunderstandings, because it is always clear from the context which coordinate system
          we are talking about.}
\begin{gather*}
\begin{multlined}
\Zeta_1 =   \frac{1}{\alpha_{11}\alpha_{22}-\alpha_{12}^2}\bigg[-\frac{1}{2}\alpha_{33}\left(\alpha_{11}+\alpha_{22}\right) \\
          + \alpha_{11}\beta_{22}+\alpha_{22}\beta_{11}-\alpha_{12}\left(\beta_{12}+\beta_{21}\right)                       \\
		  +2\sum_{s=1,2}\frac{\left(\alpha_{3\bar{s}}\alpha_{ss}-\alpha_{3s}\alpha_{s\bar{s}}\right)
		    \left(\kR_{0\bar{s}}-\kR_{3\bar{s}}\right)}{\kR_{00}-2\kR_{03}+\kR_{33}}\bigg],
\end{multlined}
\intertext{were}
\begin{aligned}
  \alpha_{kn} = &\ \left(\kR_{00}-2\kR_{03}+\kR_{33}\right)\kR_{kn}                              \\
                &\ -\left(\kR_{0k}-\kR_{3k}\right)\left(\kR_{0n}-\kR_{3n}\right),                \\
   \beta_{kn} = &\ \left(\kR_{0k}-\kR_{3k}\right)\kR_{3n}+\left(\kR_{03}-\kR_{33}\right)\kR_{kn} \\
                &\ -2\left(\kR_{0n}-\kR_{3n}\right)\kR_{3k}, 
\end{aligned}
\end{gather*}
and $\bar{s}=3-s$. 
It follows from dimensional considerations that all the functions $\Zeta_n$ do not contain large or small dimensional parameters,
and the smallness of the ISLV corrections at supershort baselines is determined by the common exponential factor
in Eq.~\eqref{angle_integral}. Thus, the corrections 
\[
\propto\left\vert\frac{X_3-X_0}{X_0}\right\vert^n \sim \frac{1}{\left({\sig}X_0\right)^n}
\]
are small under condition \eqref{sigX_0>>1}.
As explained in Appendix \ref{sec:Jg-Jv}, under this condition we must all the more neglect the exponentially small
contribution $J_v$ defined by Eq.~\eqref{J_v(X)_inside_calc} in Appendix \ref{sec:Jv}.

\subsection{Evolution of neutrino wave packet}
\label{sec:EvolutionOfNeutrinoWavePacket}

Let us summarize at a qualitative level the behavior of the effective neutrino wave packet (ENWP)
over a wide range of distances between the source and the detector.
In the vicinity of the impact point $X_s$, up to the ``mesoscopic'' distances $L \sim 1/\sig$ there is still no the duality
``propagator $\longleftrightarrow$ wave packet'' and the neutrino propagator has the form $J_g(X)+J_v(X)$
(see Appendix \ref{sec:SBLasymptoticsOfNeutrinoPropagator}).
At short but macroscopic distances, $1/\sigSBL \ll L \ll L_{\text{tr}} = \vert\vec{q}\vert/\sigSBL^2$, the contribution $J_v(X)$ disappears
and a ``virtual'' packet with off-mass-shell 4-momentum is formed, which moves almost along the classical world line with a velocity satisfying
the condition \eqref{VelocityInequalities}; the most probable 4-momentum at this stage is however on shell and the most probable velocity
is given by Eq.~\eqref{VelocityMostProbable}.
The packet evolves with time and distance and at long distances, $L \gg L_{\text{tr}}' = \vert\vec{q}\vert/\sigLBL^2$, turns into the on-shell ENWP. 
The ``transition lengths'' $L_{\text{tr}}$ and $L_{\text{tr}}'$ have the same order of magnitude, but may considerably differ from each other.
In any case, there is a certain transition region where the virtual ENWP converts into the real (on-shell) one.
Needless to say this region is described by neither SBL nor LBL asymptotics.

\subsection{Modulus squared amplitude}  
\label{sec:FinalExpression}

Combining Eqs.~\eqref{module_integral} and \eqref{angle_integral} together, we obtain the explicitly rotation invariant expression:
\begin{gather}
\label{MicroscopicProbability_integrated}
\begin{multlined}
\vert\mathcal{A}_{\beta\alpha}\vert^2
= \sum_{ij} \int\frac{d\vert\vec{q}\vert\vert\vec{q}\vert^2}{16\pi^3X_0^2}
  \sqrt{\frac{\kR^{\mu\nu}l_{\mu}l_{\nu}}{\pi\wRe^{\mu\nu}l_{\mu}l_{\nu}}}\,\mathrm{V}_s\mathrm{V}_d  \\
  \times\frac{(2\pi)^4\delta_s\left(q_{ij}-q_s\right)(2\pi)^4\delta_d\left(q_{ij}+q_d\right)}
 {\prod\limits_{\varkappa{\in}S}2E_{\varkappa}\mathrm{V}_{\varkappa}\left(\vec{p}_{\varkappa}\right)
  \prod\limits_{\varkappa{\in}D}2E_{\varkappa}\mathrm{V}_{\varkappa}\left(\vec{p}_{\varkappa}\right)} \\
  \times\frac{\vert\mathcal{M}\vert^2\VVVV}{\sqrt{E_i^2\mathcal{G}_j+E_j^2\mathcal{G}_i}}\exp\left(-\Xi_{ij}\right),
\end{multlined}\\ 
\begin{multlined} 
\Xi_{ij} = \frac{2}{\wRe^{\mu\nu}l_{\mu}l_{\nu}}\left\vert\vec{X}-\frac{\vec{v}_i+\vec{v}_j}{2}X_0\right\vert^2  \\
		  +i\left(E_j-E_i\right)X_0+\frac{\mathcal{Q}^4}{2\kR^{\mu\nu}l_{\mu}l_{\nu}}\left(v_j-v_i\right)^2X_0^2 \\
          +\frac{E_i^2E_j^2\left(E_j-E_i\right)}{E_i^2\mathcal{G}_j+E_j^2\mathcal{G}_i}
           \left\{\frac{1}{4}\left(E_j-E_i\right)\right. \\
           \left.\vphantom{\frac{E_j-E_i}{4}}
          +i\kR_{{\mu}n}\left[v_i^\mu\left(X^n-v_i^n X_0\right)+v_j^\mu\left(X^n-v_j^n X_0\right)\right]\right\}, \nonumber
\end{multlined}
\end{gather}
$l=(1,\vec{l})$, $\vec{l}=\vec{X}/\vert\vec{X}\vert$, $q_{ij}=(q_{ij}^0,\vert\vec{q}\vert\vec{l})$, where $q_{ij}^0$ is defined by Eq.~\eqref{q_{ij}^0}.
The integration limits in the integral over $\vert\vec{q}\vert$ are not specified, since the main contribution is determined by a narrow region
where $\vert\vec{q}\vert\sim\vert\vec{q}_s\vert\simeq\vert\vec{q}_d\vert$ allowed by the factor
\begin{equation*}
\label{deltadelta}
\delta_s\left(q_{ij}-q_s\right)\delta_d\left(q_{ij}+q_d\right).
\end{equation*}
It is useful to identically rewrite this factor in the form
\begin{equation}
\label{q_ijtoq}
\delta_s\left(q-q_s\right)\delta_d\left(q+q_d\right)\exp(-\varTheta_{ij}),
\end{equation}
where $q=\left(\vert\vec{q}\vert,\vec{q}\right)$ ($\vec{q}=\vert\vec{q}\vert\vec{l}$) represents the 4-momentum of the (fictitious) massless neutrino.
This expedient allows us to factor the functions $\delta_{s,d}$ out from the sums in $i$ and $j$.
Using definition \eqref{delta} we find
\begin{multline}
\label{varTheta_ij_General}
\varTheta_{ij} = \left(q_{ij}^0-\vert\vec{q}\vert\right)\left[\wRe_s^{0\mu}\left(q-q_s\right)_\mu+\wRe_d^{0\mu}\left(q+q_d\right)_\mu\right] \\
                 +\frac{1}{2}\left(q_{ij}^0-\vert\vec{q}\vert\right)^2\left(\wRe_s^{00}+\wRe_d^{00}\right).
\end{multline}
In the ultrarelativistic approximation, we arrive at the following result:
\begin{multline}
\label{varTheta_ij_UR_SBL}
\varTheta_{ij} \simeq \frac{\vert\vec{q}\vert}{4}\left[\wRe_s^{0\mu}\left(q-q_s\right)_\mu+\wRe_d^{0\mu}\left(q+q_d\right)_\mu\right] \\
                      \times\bigg[\frac{m_i^2+m_j^2}{\vert\vec{q}\vert^2}
                     -\frac{\left(m_i^2+m_j^2\right)^2}{8\vert\vec{q}\vert^4}
                     -\frac{\left(m_i^2-m_j^2\right)^2}{8\vert\vec{q}\vert^4} \\
                      \times\frac{\kR^{0\mu}l_{\mu}+3\left(\kR_{0n}l_{n}-\kR_{kn}l_{k}l_{n}\right)}{\kR^{\mu\nu}l_{\mu}l_{\nu}}\bigg] \\
                     +\frac{\left(m_i^2-m_j^2\right)^2}{32\vert\vec{q}\vert^4}\left(\wRe^{00}_s+\wRe^{00}_d\right).
\end{multline}
Although the expression in the first square brackets in Eq.~\eqref{varTheta_ij_UR_SBL} can formally be of arbitrary value, the factor $\delta_s(q-q_s)\delta_d(q+q_d)$ in Eq.~\eqref{q_ijtoq}
guarantees that in the main approximation,
\begin{multline}
\label{varTheta_ij_UR_MainApprox_SBL}
\varTheta_{ij} \simeq \frac{m_i^2+m_j^2}{4\vert\vec{q}\vert}\left[\wRe_s^{0\mu}\left(q-q_s\right)_\mu\right. \\
                                                           +\left.\wRe_d^{0\mu}\left(q+q_d\right)_\mu\right],
\end{multline}
$\vert\varTheta_{ij}\vert=O\left(m_{i,j}^2/\sig\vert\vec{q}\vert\right)$ inside the integral over $\vert\vec{q}\vert$.
The function $\varTheta_{ij}$ does not have a constant sign and does not disappear in the diagonal terms (with $i=j$) of the modulus squared amplitude,
i.e., the factors $\exp\left(-\varTheta_{ij}\right)$ could potentially play a role in both transition and survival probabilities,
but only for heavy (beyond SM) neutrinos and extremely small $\sig$. Otherwise, they can be ignored.
It should be recorded that the corresponding function in the LBL mode has (in the same approximation) a very different form:
\begin{multline}
\label{varTheta_ij_UR_MainApprox_LBL}
\varTheta_{ij}^{(\text{LBL})} \simeq
\frac{m_i^2+m_j^2}{4q_0}\left[\wRe_s^{{\mu}k}\left(q'-q_s\right)_\mu\right. \\
                       +\left.\wRe_d^{{\mu}k}\left(q'+q_d\right)_\mu\right]l^k,
\end{multline}
where $q'=\left(q_0,q_0\vec{l}\right)=q_0l$.
But of course, the functions \eqref{varTheta_ij_UR_MainApprox_SBL} and \eqref{varTheta_ij_UR_MainApprox_LBL} are of the same order of magnitude.
It is easy to show~\cite{Naumov:2010um} that $\vert{\mathcal{M}}\vert=\vert{M_sM_d}\vert$, where 
\begin{align*}
M_s     =   &\ \frac{g^2}{8}\overline{u}_-(\vec{q})\mathcal{J}_s^{\mu}\Delta_{\mu\mu'}O^{\mu'}u(\vec{p}_\alpha), \\
     \simeq &\ -i\frac{G_F}{\sqrt{2}}\,\overline{u}_-(\vec{q})\mathcal{J}_s^{\mu}O_{\mu}v(\vec{p}_\alpha),       \\
M_d^*   =   &\ \frac{g^2}{8}\overline{v}(\vec{p}_\beta)O^{\mu'}\Delta_{\mu'\mu}\mathcal{J}_d^{*\mu}u_-(\vec{q}), \\
	 \simeq &\ -i\frac{G_F}{\sqrt{2}}\,\overline{u}(\vec{p}_\beta)\mathcal{J}_d^{*\mu}O_{\mu}u_-(\vec{q}),
\end{align*}
$G_F$ is the Fermi coupling constant), and $u_-(\vec{q})$ is the Dirac bispinor for the left-handed \emph{massless} neutrino;
no need to justify the reasons for this trick.
It should be readily apparent that $M_s$ and $M_d^*$ represent the matrix elements for the neutrino production and absorption
in the reactions $I_s \to F_s'\ell_\alpha^+\nu$ and $\nu I_d \to F_d'\ell_\beta^-$, respectively.
As a consequence, we can rewrite Eq.~\eqref{MicroscopicProbability_integrated} in the form
\begin{multline}
\label{MicroscopicProbability_integrated2}
\vert\mathcal{A}_{\beta\alpha}\vert^2
=   \int\frac{d\vert\vec{q}\vert\vert\vec{q}\vert^2}{16\pi^3X_0^2}
        \sqrt{\frac{\kR^{\mu\nu}l_{\mu}l_{\nu}}{\pi\wRe^{\mu\nu}l_{\mu}l_{\nu}}}\,\mathrm{V}_s\mathrm{V}_d \\
  \times\frac{(2\pi)^8\delta_s(q-q_s)\vert{M}_s\vert^2\delta_d(q+q_d)\vert{M}_d\vert^2}
       {\prod\limits_{\varkappa{\in}S}2E_{\varkappa}\mathrm{V}_{\varkappa}\left(\vec{p}_{\varkappa}\right)
        \prod\limits_{\varkappa{\in}D}2E_{\varkappa}\mathrm{V}_{\varkappa}\left(\vec{p}_{\varkappa}\right)} \\
  \times\sum_{ij}\frac{\VVVV}{\sqrt{E_i^2\mathcal{G}_j+E_j^2\mathcal{G}_i}}
        \exp\left(-\Xi_{ij}-\varTheta_{ij}\right).
\end{multline}
Here and in what follows, $q=(\vert\vec{q}\vert,\vec{q})$ is the 4-momentum of the massless neutrino.

\section{Count rate}
\label{sec:CountRate}

In order to obtain observable quantities, the generic result \eqref{MicroscopicProbability_integrated2} must
be averaged over the unmeasurable and/or unused variables on which the in and out wave-packet states depend.
The averaging procedure can be implemented only by taking into account the physical conditions
of a real experimental environment. This makes further analysis model-dependent.

Here we will use the same model as in Refs.~\cite{Naumov:2010um,Naumov:2020yyv}.
The model supposes that the statistical distributions of the incoming wave packets $a{\in}I_{s,d}$ on the mean momenta,
spin projections and space-time coordinates in the source and detector devices can be described by single-particle
distribution functions $f_a(\boldsymbol{p}_a,s_a,x_a)$, $x_a=(x_a^0,\vec{x}_a)$. Each function $f_a$ is normalized
to the total number, $N_a(x_a^0)$, of the packets $a$ at time $x_a^0$:
\begin{equation*}
\sum\limits_{s_a}\int\frac{d\vec{x}_ad\vec{p}_a}{(2\pi)^3}f_a(\vec{p}_a,s_a,x_a)=N_a(x_a^0), \enskip a{\in}I_{s,d}.
\end{equation*}
To proceed further, we redefine the terms ``source'' and ``detector'', which so far have been used to refer to
the vertices of the macrodiagram shown in Fig.~\ref{fig:Macrograph_Class_A}.
We will use these terms and notations $\mathcal{S}$ and $\mathcal{D}$ both for the corresponding devices
and for the supports of the product of the distribution functions $f_a$ in space and time, namely,
\begin{equation}
\label{mathcalS,D}
\mathcal{S} = \stackrel[\{x_a\}]{}{\mathrm{supp}}\prod\limits_{a{\in}I_s}f_a
\enskip\text{and}\enskip
\mathcal{D} = \stackrel[\{x_a\}]{}{\mathrm{supp}}\prod\limits_{a{\in}I_d}f_a,
\end{equation}
which are assumed finite and mutually non-intersecting in space. 
We also assume that the spatial dimensions of $\mathcal{S}$ and $\mathcal{D}$ are very large compared with the effective
dimensions ($\sim\sigma_{\varkappa}^{-1}$) of the wave packets $\varkappa$ moving inside $\mathcal{S}$ and $\mathcal{D}$.
But we will not use the generally accepted assumption that the dimensions $\mathcal{S}$ and $\mathcal{D}$ are small
compared to the effective distance between them, because this assumption is not suitable for such experiments as 
GALLEX~\cite{Hampel:1997fc}, SAGE \cite{Abdurashitov:1998ne,Abdurashitov:2005tb}, and BEST \cite{Barinov:2021asz,Barinov:2022wfh},
where the source is surrounded by a detector environment.
We neglect the background events caused by outgoing particles produced in $\mathcal{S}$ and falling in $\mathcal{D}$;
this issue would not deserve special mention in the case of long baseline experiments.
For certainty, we consider only such experiments in which the momenta of all or some types of particles arising
in $\mathcal{D}$ are measured.
To avoid non-principal complications we suppose a detection efficiency of 100\%.
Under all these assumptions, the macroscopically averaged probability \eqref{MicroscopicProbability_integrated2}
(representing in this instance the number of events registered in $\mathcal{D}$) reads
\begin{multline}
\label{AveragedProbability_Gen}
\langle\!\langle\vert\mathcal{A}_{\beta\alpha}\vert^2\rangle\!\rangle
= \sum  \limits_{\text{spins}} 
        \int        \prod\limits_{a{\in}I_s}\frac{d\vec{x}_a  d\vec{p}_a f_a(\vec{p}_a,s_a,x_a)}{(2\pi)^32E_a\mathrm{V}_a}  \\
  \times\int\left(\,\prod\limits_{b{\in}F_s}\frac{d\vec{x}_b  d\vec{p}_b }{(2\pi)^32E_b\mathrm{V}_b}\right)\mathrm{V}_s     \\
  \times\int        \prod\limits_{a{\in}I_d}\frac{d\vec{x}_a  d\vec{p}_a f_a(\vec{p}_a,s_a,x_a)}{(2\pi)^3 2E_a\mathrm{V}_a} \\
  \times\int\left(\,\prod\limits_{b{\in}F_d}\frac{d\vec{x}_b [d\vec{p}_b]}{(2\pi)^32E_b\mathrm{V}_b}\right)\mathrm{V}_d     \\
  \times\int\frac{d\vert\vec{q}\vert\vert\vec{q}\vert^2}{16\pi^3X_0^2}
        \sqrt{\frac{\kR^{\mu\nu}l_{\mu}l_{\nu}}{\pi\wRe^{\mu\nu}l_{\mu}l_{\nu}}}                                            \\
  \times(2\pi)^4\delta_s(q-q_s)\vert{M}_s\vert^2(2\pi)^4\delta_d(q+q_d)\vert{M}_d\vert^2                                    \\
  \times\sum_{ij}\frac{\VVVV}{\sqrt{E_i^2\mathcal{G}_j+E_j^2\mathcal{G}_i}}
        \exp\left(-\Xi_{ij}-\varTheta_{ij}\right).
\end{multline}
The symbol $\sum_{\text{spins}}$ denotes the averaging over the spin projections $s_a$ of the packets $a {\in} I_{s,d}$
and summation over the spin projections $s_b$ of the packets $b {\in} F_{s,d}$, supposing that the projections $s_b$ are not measured.
The symbol $[d\vec{p}_b]$ indicates no integration in $\vec{p}_b$ (i.e.\ $\int[d\vec{p}_b]{\equiv}d\vec{p}_b$), and hence that these
momenta are measured. If, however, the momenta of particles of certain types $b' {\in} F_d$ are not measured, the brackets must be removed for $b=b'$.
Therefore, it should be clearly apparent that the expression \eqref{AveragedProbability_Gen} means the number ($N_{\beta\alpha}$) of
detected particles $b$ with 3-momenta between $\vec{p}_b$ and $\vec{p}_b+d\vec{p}_b$ (or any momenta) in $\mathcal{D}$.
As in the LBL case, the indices $\alpha$ and $\beta$ indicate that neutrino exchange generally leads to the lepton number violation
(when $\alpha\ne\beta$). The final state leptons $\ell_\alpha$ and $\ell_\beta$ provide the experimental signature of such a violation.

\subsection{Averaging over spatial variables}
\label{sec:SpaceAveraging}

An approximate multidimensional integration over the spatial variables $\vec{x}_{\varkappa}$ in 
Eq.~\eqref{AveragedProbability_Gen} can be performed explicitly using the integral representation
\eqref{mathrmVsd} for the $4d$ overlap volumes
\[
\mathrm{V}_s = \mathrm{V}_s\left(\{\vec{p}_{\varkappa},x_{\varkappa}\}\right)
\enskip \text{and} \enskip
\mathrm{V}_d = \mathrm{V}_d\left(\{\vec{p}_{\varkappa},x_{\varkappa}\}\right)
\] 
and naturally assuming that the distribution functions $f_a$ and the factor $\exp\left(-\Xi_{ij}\right)$ change significantly
only on macroscopic scales (comparable with the dimensions of $\mathcal{S}$ and $\mathcal{D}$);
most often, in real experiments, spatial variations of the distribution functions can be neglected altogether.
On the other hand, the integrand
\[
\mathscr{P}_{s,d} \equiv \prod\limits_{\varkappa{\in}S,D}\left\vert\psi_{\varkappa}\left(\vec{p}_{\varkappa},x_{\varkappa}-x\right)\right\vert^2
\]
in Eq.~\eqref{mathrmVsd} varies on a very short scale; in fact it differs from zero only when the classical
world lines of \emph{all} wave packets $\varkappa$ pass through a narrow (microscopic or at least mesoscopic) neighborhood
of the integration variable (let the latter be $x$ for $\mathrm{V}_s$ and $y$ for $\mathrm{V}_d$).
Hence, neglecting the possible (microscopic) edge effects in $\mathcal{S}$ and $\mathcal{D}$, it is safe to substitute
$x_{\varkappa}$ by $x$ ($y$) for $\varkappa{\in}S$ ($\varkappa{\in}D$) in the mentioned slowly-varying factors.
Then, as is seen from Eq.~\eqref{X_sd} defining the impact points, $X_s=x$ and $X_d=y$.
The remaining integrals over the variables $\vec{x}_{\varkappa}$ yield the factor
$\prod_{\varkappa{\in}S{\oplus}D}\mathrm{V}_{\varkappa}$, which cancels the product of the corresponding (inverse) factor in the integrand.

To formalize the procedure described above, let us prove that for an arbitrary 4-vector $x$ there is always a
configuration of spatially separated points $\{x_{\varkappa}\}$ for which $\mathscr{P}_{s,d }=1$.
Let, for certainty, $\varkappa{\in}S$.
Because of the invariance of $\vert\psi_{\varkappa}(\vec{p}_{\varkappa},x_{\varkappa}-x)\vert$ with respect to shifts along
the world lines of the centers of the wave packet $\varkappa$, we can replace the space-time coordinates $x_{\varkappa}$
by $\tilde{x}_{\varkappa}$, where
\[
\tilde{x}_{\varkappa}^0=X_s^0 \enskip \text{and} \enskip 
\tilde{\vec{x}}_{\varkappa} = \vec{x}_{\varkappa}+\left(X_s^0-x_{\varkappa}^0\right)\vec{v}_{\varkappa}.
\]
It is apparent that $\mathscr{P}_s=1$ as $\tilde{x}_{\varkappa}=x$, i.e., at the same moment in time, but at non-matching
spatial coordinates $\vec{x}_{\varkappa} = \vec{x}+\left(x_{\varkappa}^0-x^0\right)\vec{v}_{\varkappa}$,
which are (by construction) remote from each other.
Of course, $x=X_s$ at $\tilde{x}_{\varkappa}=x$ (recall that $X_s$ does not change when $x_{\varkappa}$ is substituted
by $\tilde{x}_{\varkappa}$).
It is implicitly assumed here that no more than one packet is at rest, but it is clear that the contribution
to the integral from the configurations in which two or more wave packets have zero velocities is insignificant
-- such configurations form a set of measures zero.

In consequence of this we can rewrite Eq.~\eqref{AveragedProbability_Gen} as follows:
\begin{multline}
\label{AveragedProbability_Simplified}
N_{\beta\alpha}
= \sum\limits_{\text{spins}}\int d\vec{x} \int d\vec{y} \int d\mathfrak{P}_s \int d\mathfrak{P}_d  \\
  \times\int\frac{d\vert\vec{q}\vert\vert\vec{q}\vert^2}{16\pi^3\left\vert\vec{y}-\vec{x}\right\vert^2}
        \sqrt{\frac{\kR^{\mu\nu}l_{\mu}l_{\nu}}{\pi\wRe^{\mu\nu}l_{\mu}l_{\nu}}}                   \\
  \times\sum_{ij}\frac{\VVVV}{\sqrt{E_i^2\mathcal{G}_j+E_j^2\mathcal{G}_i}}
        \exp\left(-\Xi_{ij}-\varTheta_{ij}\right).
\end{multline}
Here we made the substitution
\[
\frac{1}{X_0^2} \simeq \frac{1}{\vert\vec{y}-\vec{x}\vert^2}\left[1+O\left(\frac{1}{\sig\vert\vec{y}-\vec{x}\vert}\right)\right],
\]
allowed by condition \eqref{sigX_0>>1}, and introduced the differential forms
\begin{equation}
\label{mathfrakP}
\begin{aligned}
d\mathfrak{P}_s = &\ \prod\limits_{a{\in}I_s}\frac{ d\vec{p}_a f_a(\vec{p}_a,s_a,x)}{(2\pi)^32E_a}
                     \prod\limits_{b{\in}F_s}\frac{ d\vec{p}_b }{(2\pi)^32E_b} \\
                  &\ \times (2\pi)^4{\delta}_s(q-q_s)\vert{M}_s\vert^2,        \\
d\mathfrak{P}_d = &\ \prod\limits_{a{\in}I_d}\frac{ d\vec{p}_a f_a(\vec{p}_a,s_a,y)}{(2\pi)^32E_a}
                     \prod\limits_{b{\in}F_d}\frac{[d\vec{p}_b]}{(2\pi)^32E_b} \\
                  &\ \times (2\pi)^4{\delta}_d(q+q_d)\vert{M}_d\vert^2.
\end{aligned}
\end{equation}

\subsection{Averaging over time intervals}
\label{sec:TimeAveraging}

The next step is to average the number of events \eqref{AveragedProbability_Simplified} over the $\mathcal{S}$ and $\mathcal{D}$
running time intervals. This is not the most general setting of the problem and, moreover, it very much depends on the specific conditions
of an experiment. But it typical for the current SBL experiments and is useful to clarify the physical meaning of the obtained results and
compare them with the results derived for on-shell regime (long baselines, or LBL) in the same terms.
For these aims, we will use the simple model for the particle distribution functions \eqref{AveragedProbability_Simplified}
proposed in Refs.~\cite{Naumov:2009zza,Naumov:2010um}, in which they do not depend on time (or very slowly evolve with time)
during the source operating period $\tau_s=x^0_2-x^0_1$ and detector exposure period $\tau_d=y^0_2-y^0_1$, that is
\begin{equation}
\label{DistributionFunctionsModel}
\begin{aligned}
f_a(\vec{p}_a,s_a;x) = &\ \theta\left(x^0-x^0_1\right)\theta\left(x^0_2-x^0\right) \\
                       &\ \times\overline{f}_a(\vec{p}_a,s_a;\vec{x})
                          \enskip\text{for}\enskip a{\in}I_{s},                    \\
f_a(\vec{p}_a,s_a;y) = &\ \theta\left(y^0-y^0_1\right)\theta\left(y^0_2-y^0\right) \\
                       &\ \times\overline{f}_a(\vec{p}_a,s_a;\vec{y})
                          \enskip\text{for}\enskip a{\in}I_{d}.
\end{aligned}
\end{equation}
In the case of a detector, the step functions in Eq.~\eqref{DistributionFunctionsModel} can be considered as ``hardware'' or ``software''
trigger conditions. 
In the context of present study (short distances between $\mathcal{S}$ and $\mathcal{D}$), the steady-state operation periods for $\mathcal{S}$
and $\mathcal{D}$ can be long, as in the SBL reactor antineutrino experiments and experiments with artificial radioactive neutrino sources, or very short,
as in accelerator-based SBL neutrino-tagging experiments (which, however, requires more specific analysis). In either case it is assumed that the time intervals
required to turn the devices $\mathcal{S}$ and $\mathcal{D}$ on and off are negligible compared to $\tau_s$ and $\tau_d$, respectively.
In the general case, we do not impose any synchronization conditions on the operation periods of $\mathcal{S}$ and $\mathcal{D}$.

As a result of this simplification, the integration over the source and detector operation times reduces to calculation of the following integral:
\begin{equation*}
\int\limits_{x_1^0}^{x_2^0}dx_0\int\limits_{y_1^0}^{y_2^0}dy_0\exp\left(-\Xi_{ij}\right).
\end{equation*}
Using the general formula for the double Gaussian integral
\begin{multline*}
\int\limits_{x_1^0}^{x_2^0}dx_0\int\limits_{y_1^0}^{y_2^0}dy_0
  \exp\left[-a^2\left(y_0-x_0\right)^2-b\left(y_0-x_0\right)\right]                         \\ 
= \frac{\sqrt\pi}{2a^2}\exp\left(\frac{b^2}{4a^2}\right)\sum\limits_{l,l'=1}^2(-1)^{l+l'+1} \\ 
  \times\Ierf\left[a\left(x_l^0-y_{l'}^0\right)-\frac{b}{2a}\right],
\end{multline*}
where
\begin{equation*}
\label{IerfDef}
\Ierf(z) = \int\limits_0^zdz'\text{erf}(z')+\frac{1}{\sqrt{\pi}}=z\,\text{erf}(z)+\frac{e^{-z^2}}{\sqrt{\pi}},
\end{equation*}
the massless neutrino approximation in the pre-exponential factors, and leading order on ${\Delta}m_{ij}^2=m_i^2-m_j^2$ in the exponent,
we can rewrite Eq.~\eqref{AveragedProbability_Simplified} in the following relatively compact form
(see Appendix~\ref{sec:N_alphabeta_Exact} for a more precise formula):
\begin{gather}
\begin{multlined}
\label{AveragedProbability_Simplified_mod4}
\frac{N_{\beta\alpha}}{\tau_d} = \sum\limits_{\text{spins}}\int{d}\vec{x}\int{d}\vec{y}\int{d}\mathfrak{P}_s\int{d}\mathfrak{P}_d \\
                   \times\int{d}\vert\vec{q}\vert\frac{\mathcal{P}_{\alpha\beta}\left(\vert\vec{q}\vert,\vert\vec{y}-\vec{x}\vert\right)}{4(2\pi)^3\vert\vec{y}-\vec{x}\vert^2},
\end{multlined} \\
\begin{multlined}
\label{oscill_f}
\mathcal{P}_{\alpha\beta}\left(\vert\vec{q}\vert,\vert\vec{y}-\vec{x}\vert\right) = \sum_{ij}\VVVV \\
   \times\exp\left(i\varphi_{ij}-\mathcal{A}_{ij}^2-\mathcal{C}_{ij}^2-\varTheta_{ij}\right)S_{ij},
\end{multlined} \\
\begin{multlined}
\label{Instrumental}
S_{ij} = \frac{\exp(-\mathcal{B}_{ij}^2)}{4\mathfrak{D}\tau_d}\sum\limits_{l,l'=1}^2 (-1)^{l+l'+1} \\
         \times\Ierf\left[2\mathfrak{D}\left(x_l^0-y_{l'}^0+\vert\vec{y}-\vec{x}\vert\right){+}i\mathcal{B}_{ij}\right], 
\end{multlined}
\end{gather}
and $\mathfrak{D} = 1/\sqrt{2\wRe^{\mu\nu}l_{\mu}l_{\nu}}$.
The unit vector $\vec{l}$ is directed along the vector $\vec{y}-\vec{x}$.
But, given that all structures (tensor convolutions) in the above formula are rotation invariants, one can orient $\vec{l}$ in any fixed direction,
say, along the geometric centers of $\mathcal{S}$ and $\mathcal{D}$, provided they have appropriate symmetries.
The differential forms $\mathfrak{P}_{s,d}$ in Eq.~\eqref{AveragedProbability_Simplified_mod4} are defined by Eqs.~\eqref{mathfrakP},
in which the distribution functions $f_a$ should be substituted by $\overline{f}_a$, see Eqs.~\eqref{DistributionFunctionsModel}.
Therefore, according to Eqs.~\eqref{mathcalS,D}, integration in $\vec{x}$ and $\vec{y}$ is carried out over the volumes of $\mathcal{S}$ and $\mathcal{D}$.

With the same arguments as in the LBL regime \cite{Naumov:2010um}, the function \eqref{oscill_f} can be treated as a QFT generalization of
the QM neutrino flavor transition probability.
The ingredients involved in Eq.~\eqref{oscill_f} are gathered in Table~\ref{Tab:Ingredients}, which also contains the corresponding structures
arising in the LBL asymptotics.
\begin{table*}[!htbp]
\centering
\caption{Ingredients of Eq.~\protect\eqref{AveragedProbability_Simplified_mod4} in leading order for off-shell (short distances) and on-shell (long distances) regimes;
         the latter are taken from Ref.~\cite{Naumov:2010um}. Here $L=\vert\vec{y}-\vec{x}\vert$ and ${\Delta}m_{ij}^2=m_i^2-m_j^2$.
         Last column shows the order of magnitude (OoM) of the corresponding quantity. Evidently, $E_\nu=\vert\vec{q}\vert$ in the given approximation. 
        }
\begin{tabularx}{\textwidth}{@{\extracolsep\fill}llll}  
\hline                                                                                                           \noalign{\smallskip}
Quantity               & Off-shell regime  & On-shell regime  & OoM                                           \\ \noalign{\smallskip}
\hline                                                                                                           \noalign{\smallskip}
  $\varphi_{ij}$       & $\dfrac{{\Delta}m_{ij}^2L}{2\vert\vec{q}\vert}$
                       & $\dfrac{{\Delta}m_{ij}^2L}{2E_\nu}$
                       & $\dfrac{\vert{\Delta}m_{ij}^2\vert{L}}{E_\nu}$                                       \\ \noalign{\smallskip}
  $\mathcal{A}_{ij}^2$ & $\left(\dfrac{{\Delta}m_{ij}^2L}{2\vert\vec{q}\vert^2}\right)^2\dfrac{\mathcal{Q}^4}{2\kR^{\mu\nu}l_{\mu}l_{\nu}}$ 
                       & $\left(\dfrac{{\Delta}m_{ij}^2L}{2E_\nu^2}\right)^2\dfrac{1}{2\wRe^{\mu\nu}l_{\mu}l_{\nu}}$ 
                       & $\left(\dfrac{{\Delta}m_{ij}^2}{E_\nu^2}{\sig}L\right)^2$                            \\ \noalign{\smallskip}
  $\mathcal{B}_{ij}$   & $\dfrac{{\Delta}m_{ij}^2}{4\vert\vec{q}\vert}\sqrt{\dfrac{\wRe^{\mu\nu}l_{\mu}l_{\nu}}{2}}\dfrac{\kR^{0\mu}l_{\mu}}{\kR^{\mu\nu}l_{\mu}l_{\nu}}$
                       & $\dfrac{{\Delta}m_{ij}^2}{4E_\nu}\sqrt{\dfrac{\wRe^{\mu\nu}l_{\mu}l_{\nu}}{2}}\,\dfrac{Y_kl_k}{Y^{\mu}l_{\mu}}$
                       & $\dfrac{\vert{\Delta}m_{ij}^2\vert}{{\sig}E_\nu}$                                    \\ \noalign{\smallskip}
  $\mathcal{C}_{ij}^2$ & $\left(\dfrac{{\Delta}m_{ij}^2}{2\vert\vec{q}\vert}\right)^2\dfrac{1}{8\kR^{\mu\nu}l_{\mu}l_{\nu}}$
                       & $0$
                       & $\left(\dfrac{{\Delta}m_{ij}^2}{{\sig}E_\nu}\right)^2$                               \\ \noalign{\smallskip}
\hline
\end{tabularx}
\label{Tab:Ingredients}
\end{table*}
As is seen, the functions $\mathcal{A}_{ij}$ and $\mathcal{B}_{ij}$, as well as the function $\varTheta_{ij}$ we discussed above (cf.\ Eqs.~\eqref{varTheta_ij_UR_MainApprox_SBL}
and \eqref{varTheta_ij_UR_MainApprox_LBL}) relating to the SBL and LBL modes have a certain similarities, but not identical, and do not coincide with each other in any approximation. 
The positivity of $\mathcal{A}_{ij}^2$ in the SBL mode is obvious, given that 
$\mathcal{Q}^4 = \kR_{00}\kR_{33}-\left(\kR_{03}\right)^2$
in the coordinate system, where $\vec{l}=(0,0,1)$, and this expression is a principal minor of the positive-definite matrix $\vert\vert\kR_{\mu\nu}\vert\vert$
(see footnote \ref{Rotation}).
The suppression factor $\exp\left(-\mathcal{A}_{ij}^2\right)$ is potentially important for very long $L$, but this does not apply to the SBL mode.
The function $\mathcal{C}_{ij}^2$ is absent in the LBL mode, but has the same order of magnitude as $\mathcal{B}_{ij}^2$ and is also
negligibly small for the SM neutrinos.
Although the suppression factors in Eq.~\eqref{oscill_f} are probably not important from the experimental viewpoint, they are of interest
as a reflection of the process (obscured by the averaging of various kinds) responsible for the creation and evolution of ENWP,
and shortly discussed in Sect.~\ref{sec:EvolutionOfNeutrinoWavePacket}.

Remarkably the oscillation phase $\varphi_{ij}$ is the same in both modes and, moreover, the same as in the QM approach,
although it is clear that at short distances this phase is also small for the standard $3\nu$ oscillations.
The higher-order corrections to $\varphi_{ij}$ are presented in Appendix~\ref{sec:N_alphabeta_Exact}, but
they are of potential interest only if there is mixing of standard neutrinos with hypothetical heavy sterile neutrinos.

\subsection{Instrumental factor}
\label{sec:S}

\begin{figure}[htb]
\centering\includegraphics[width=\linewidth]{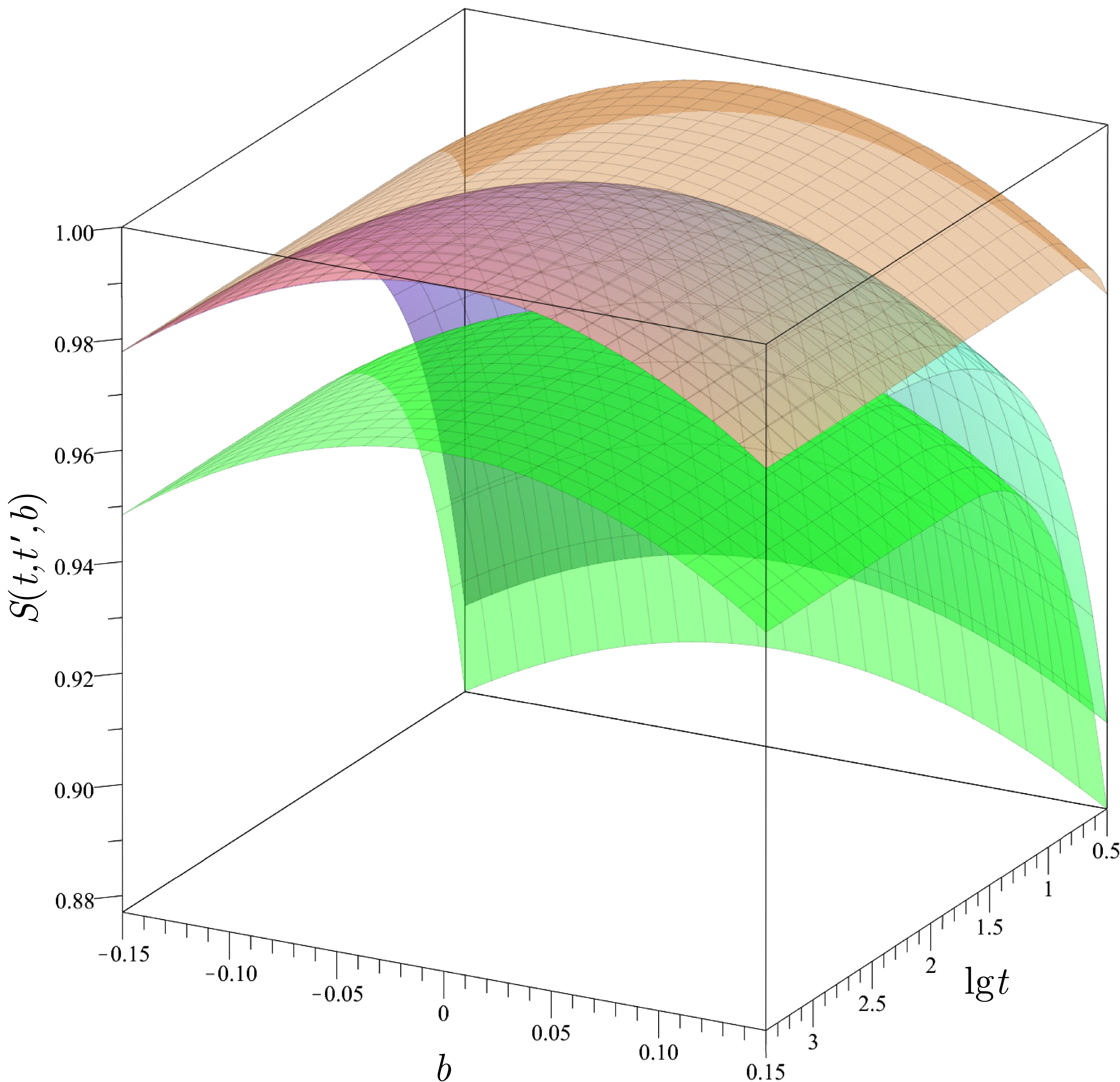}
\caption{Instrumental factor \eqref{S} at $t'=0.97t$, $t'=t$ and $t'=1.30t$ vs.\ $\lg t$ and $b$ shown by the lower, middle and upper surfaces,
         respectively. The middle and upper surfaces merge at very large $t$.
         \label{fig:S}
        }
\end{figure}

Within the approximations made, the function $S_{ij}$ (hereinafter referred to as decoherence function or instrumental factor)
is the same as in the LBL mode, except for the difference in the explicit form of the function $\mathcal{B}_{ij}$
(see Table~\ref{Tab:Ingredients}), on which $S_{ij}$ depends.
The general properties and several examples of the decoherence function have been studied in detail in Ref.~\cite{Naumov:2020yyv}.
Here we briefly discuss an additional simple example most relevant to SBL-experiments.
Namely, we consider the case when the periods $\tau_s$ and $\tau_d$ are large compared to $\max\vert\vec{y}-\vec{x}\vert$. Then
\[
x^0_1 = -\frac{\tau_s}{2},
\quad
x^0_2 =  \frac{\tau_s}{2},
\quad
y^0_1 = -\frac{\tau_d}{2},
\quad
y^0_2 =  \frac{\tau_d}{2},
\]
and, as shown in Ref.~\cite{Naumov:2020yyv}, the decoherence function is expressed through the universal (independent of indices $i,j$)
real-valued function
\begin{multline}
\label{S}
S(t,t',b) = \frac{\exp\left(-b^2\right)}{2t'}\text{Re}\left[\Ierf\left(t+t'+ib\right)\right. \\ 
           -\left.\Ierf\left(t-t'+ib\right)\right]
\end{multline}
of three independent dimensionless real variables. In these terms,
\begin{equation}
\label{S_ij}
S_{ij} = S(\mathfrak{D}\tau_s,\mathfrak{D}\tau_d,\mathcal{B}_{ij}).
\end{equation}
Here are the simplest properties of function \eqref{S}:
\begin{itemize}
\item $0 < S \le 1$ for all values of the arguments and decreases with increasing $b$;
\item at $t \ge t' \gg 1$ the function $S$ weakly depends on $t$ and $t'$ and approaches its asymptotic value $\exp(-b^2)$ with increasing $t$,
      so that $S_{ij} \approx 1$ for $i,j=1,2,3$ (SM neutrinos) and $S_{ii} = 1$ for any $i$ (since $\mathcal{B}_{ii}=0$);
\item $S$ decreases rapidly as the ratio $t/t'$ increases.
\end{itemize}
For illustration, Fig.~\ref{fig:S} displays three particular cases of the function $S(t,t',b)$ for $t'=0.97t$, $t'=t$, and $t'=1.30t$.
The interval of values of $b$ used in the figure is, of course, unrealistically wide and is chosen only for the convenience of visualization. 
The figure shows that $S$ depends very weakly on $b$ at $\vert b \vert \ll1$ (which is the case in the real experimental circumstances),
but the dependence on the parameters $t$ and $t'$ is very significant and should be taken into account when processing the data 
and deriving the count rate from the measured number of events.

\subsection{Flavor transition probability}
\label{sec:FlavorTransitionProbability}

The function \eqref{oscill_f} matches the QM flavor transition probability only if $S_{ij}=1$ and
$\mathcal{A}_{ij}=\mathcal{B}_{ij}=\mathcal{C}_{ij}=\varTheta_{ij}=0$, which is not the case.
So it really can be interpreted as a QFT refinement of the standard QM result.
However, as in the LBL mode, the probabilistic interpretation of the function $\mathcal{P}_{\alpha\beta}$ can be true only in attenuated form,
since the instrumental factor and the functions $\mathcal{A}_{ij}$, $\mathcal{B}_{ij}$, $\mathcal{C}_{ij}$, and $\varTheta_{ij}$ involve
the tensor componets $\wRe^{\mu\nu}$ and $\kR^{\mu\nu}$ that depend on the momenta, masses, and momentum spreads of all external wave packets
and are in general different for the reactions \eqref{Macroprocess_A} with different leptonic pairs $(\ell_\alpha,\ell_\beta)$ in the final states.
The distinctions are determined by kinematics of the subprocesses in $\mathcal{S}$ and $\mathcal{D}$ (in particular, by the reaction thresholds)
and specific sets of participating in and out particles. It is also appropriate to recall that the function $\varTheta_{ii}$ is not sign-definite
and $\varTheta_{ii} \ne 0$. As a result, the unitarity relations
\[
\sum_{\alpha}\mathcal{P}_{\alpha\beta}=\sum_{\beta}\mathcal{P}_{\alpha\beta}=1
\]
are generally do not hold, or hold only approximately even within the kinematic boundaries for the corresponding subprocesses.
 
By applying the aforementioned asymptotics $S=\exp(-b^2)$ (presumably valid for a wide class of possible SBL experiments),
and using the same arguments as in Refs.~\cite{Naumov:2009zza,Naumov:2010um,Naumov:2020yyv}, we arrive at formally the same expression
as in the LBL case, represented by Eq.~\eqref{EventNumber}, where
\begin{multline}
\label{oscill_final}
\mathcal{P}_{\alpha\beta}\left(\vert\vec{q}\vert,\vert\vec{y}-\vec{x}\vert\right) =  \sum_{ij}\VVVV \\
\times\exp\left(i\varphi_{ij}-\mathcal{A}_{ij}^2-\mathcal{B}_{ij}^2-\mathcal{C}_{ij}^2-\varTheta_{ij}\right)
\end{multline}
with the exponent very close to unit for the SM neutrinos. In this extremely simple but quite realistic case, it follows
from Eq.~\eqref{oscill_final} and unitarity of the mixing matrix that
\begin{equation}
\label{NoOscillations!}
\mathcal{P}_{\alpha\beta} \simeq \delta_{\alpha\beta}.
\end{equation}
Thus, as was to be expected, the flavor transitions are strongly forbidden at short distances and/or high energies.
The same conclusion follows from the simple QM formula for the flavor transition probability.
It would seem that this leaves no room for experimental verification of the theory presented here.
However, there is a subtle point: the commonplace result \eqref{NoOscillations!} is only valid when $\sig$ is large enough, or
better to say not too small;
in fact, we have explicitly used this assumption in deriving the results discussed, as well as in some approximations and simplifications.

On the other hand, in the plane-wave limit, $\sig\to0$ and thus $\mathcal{B}_{ij}^2\to+\infty$ and $\mathcal{C}_{ij}^2\to+\infty$ at $i \ne j$, 
which (together with the ultrarelativistic approximation) leads to the following result%
\footnote{The derivation of this limit is not entirely trivial because of the factor $\exp\left(-\varTheta_{ij}\right)$ in Eq.~\eqref{oscill_final},
          but becomes more transparent if we rewrite Eq.~\eqref{ProbabilityInPWL} in the more accurate form
          \begin{equation*}
          \delta_s(q-q_s)\delta_d(q+q_d)\mathcal{P}_{\alpha\beta}
          \, \stackrel{\text{PWL}}{\longrightarrow} \,\delta(q-q_s)\delta(q+q_d)\sum_i\vert{V_{{\alpha}i}}\vert^2\vert{V_{{\beta}i}}\vert^2.
          \end{equation*}
         }
\begin{equation}
\label{ProbabilityInPWL}
\mathcal{P}_{\alpha\beta} \, \stackrel{\text{PWL}}{\longrightarrow} \, \sum_i\vert{V_{{\alpha}i}}\vert^2\vert{V_{{\beta}i}}\vert^2.
\end{equation}
which is completely different from Eq.~\eqref{NoOscillations!}.
As is well known, the oscillationless flavor transitions are ruled out by a huge number of (terrestrial) neutrino experiments,
which within our formalism indicates that $\sig>0$. 
At nonzero (and not very small!) $\sig$ we have to take into account the power corrections for the neutrino propagator,
derived in Appendix~\ref{sec:Jg}, which could lead to a potentially observable consequence not predicted by the QM formalism
and basically distinguishable from the ISLV effect \cite{Naumov:2013bea} predicted for the LBL regime. 
The absence of additional masking factors (decoherence, damping, and oscillations themselves) would have helped to detect the presumably small effect.

\section{ISL violation at short baselines}
\label{sec:ISLV}

The previous discussion was based on using the leading order of the asymptotic expansion of the neutrino propagator
in small parameters \eqref{SmallParameters}. 
Taking into account the next-to-leading order terms in the expression for number of events
\eqref{EventNumber} would require getting rid of several approximations made in deriving this formula.
In particular, the degree of approximation used in Eqs.~\eqref{CardallTrick4D} and \eqref{angle_integral} should be improved.
We will postpone this issue for the future, and here we will confine our consideration to the potential effect from the first sub-leading
term $\propto\delta^2$ in series \eqref{Re+Im}.

Within the approximations made so far, we must neglect in the series \eqref{Re+Im} the terms of order
$O(\eta)$, $O(\Delta)$ and higher, as well as the neutrino mass. Moreover, we can safely put in Eq.~\eqref{Re+Im}
$p \simeq q \simeq {\vert\vec{q}\vert}l$ and $X \simeq X_0l \simeq {\vert\vec{X}\vert}l$.
Then the series \eqref{Re+Im} in the leading order becomes extremely simple:
\[
\mathcal{I} \simeq -g_1\delta^2.
\]
Here
\[
g_1 \simeq F_{200} = \theta_2-2\omega_2+\omega_1^2,
\quad
\delta \simeq \sig^2\frac{\vert\vec{X}\vert}{\vert\vec{q}\vert},
\]
and the functions $\theta_2$, $\omega_1$, and $\omega_2$ are defined by Eqs.~\eqref{rot_inv_coeffs}
in which we must put $\vec{v}=\vec{l}$.
As a result, the inverse-square law for the differential neutrino flux $d\mathfrak{F}_\nu$, which appears in the SBL version of
Eq.~\eqref{EventNumber}, is violated under appropriate conditions (cf.\ the corresponding results of Refs. \cite{Ioannisian:1998ch}
and \cite{Beuthe:2001rc}):
\begin{subequations}
\label{TheISLviolatinFactorSBL}
\begin{align}
d\mathfrak{F}_\nu 
\approxprop &\ \frac{1}{\vert\vec{y}-\vec{x}\vert^2}\exp\left(-\frac{\sigEff^4\vert\vec{y}-\vec{x}\vert^2}{\vert\vec{q}\vert^2}\right) \\
\approx     &\ \frac{1}{\vert\vec{y}-\vec{x}\vert^2}\left(1-\frac{\sigEff^4\vert\vec{y}-\vec{x}\vert^2}{\vert\vec{q}\vert^2}\right),
\end{align}
\end{subequations}
where 
\begin{align*}
\sigEff^4 \equiv &\ 2F_{200}\sig^4 = 2F_{200}\sqrt{\vert\kR\vert} \\
             =   &\ 2\left[\Tr\vec{\kR}^2-2\vec{l}^T\vec{\kR}^2\vec{l}+\left(\vec{l}^T\vec{\kR}\vec{l}\right)^2\right].
\end{align*}
In Appendix \ref{sec:PositivityOfF_200} it is proved (without neglecting the neutrino mass) that $F_{200}>0$.
Therefore $\sigEff^4>0$.
This means that the number of neutrino events should \emph{decrease} compared to what would be expected in the absence
of the ISL violation.
This effect is ultimately associated with incomplete overlap in space and time of the noncollinearly interacting incoming and
outgoing wave packets involved in the processes of neutrino production and absorption, which must lead to a \emph{decrease}
of the probability of their interactions. Therefore, the positivity of $\sigEff^4$ is natural.

A more intuitive, though more speculative, interpretation of the ISLV effect relates to the aforementioned duality
``propagator $\longleftrightarrow$ wave packet''. The effective neutrino wave packet moves quasi-classically,
i.e., its center only approximately follows a straight line connecting the neutrino production and absorption points.
According to Eq.~\eqref{Positivity2}
\[
\sigEff^4 = 2\left(\kR_{11}^2+2\kR_{12}^2+\kR_{22}^2\right)
\]
in the coordinate system, where $\vec{l}=(0,0,1)$, i.e.\ (within our approximations) $\sigEff$ is determined
only by the spatial components of the tensor $\kR$, transverse to the vector $\vec{l}$.
Thus, the motion of the neutrino wave packet can be roughly thought as classical motion along $\vec{l}$ superimposed on
the quantum jitter perpendicular to $\vec{l}$. 
The resulting path length, $\bar{L}$, is longer than the length of the classical rectilinear trajectory.
Therefore the number of events ($\propto 1/\bar{L}^2$) must be smaller than for the classical motion.
The formalism automatically accounts for this effect and gives the expected reduction of the number of events. 

The neutrino virtuality also plays some part in the ISLV effect, and series \eqref{J_g_final2_inside_calc} accounts
for it with the terms $\propto\Delta^c$. But we neglected these terms (as well as the terms $\propto\eta^b$) because
taking them into account would definitely exceed the accuracy of our estimate.

\begin{figure}[htb]
\centering\includegraphics[width=\linewidth]{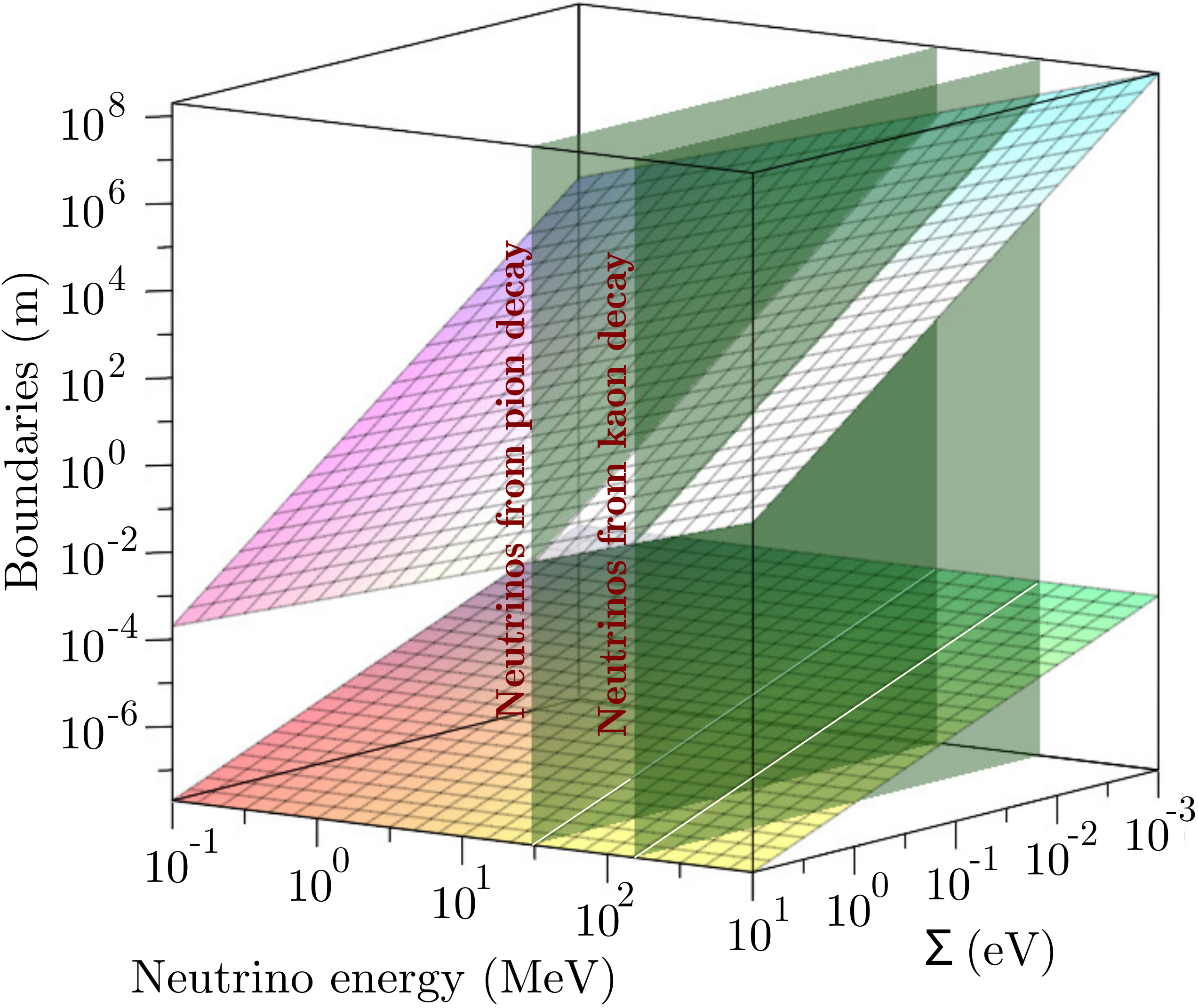}
\caption{The upper and lower limits of applicability of the formalism vs.\ neutrino energy and
         scale parameter $\sig$ (inclined planes).
         The vertical planes show the energies of neutrinos originated from two-particle pion and kaon decay at rest.
         \label{fig:Boundaries}
        }
\end{figure}

Regardless of the interpretation, the function $\sigEff$ in Eqs.~\eqref{TheISLviolatinFactorSBL} depicts a cumulative effect
from overlap of the external wave-packet states $\varkappa$ in the source and detector, depending (throught the tensor $\kR$)
on their momenta $\vec{p}_{\varkappa}$, masses $m_{\varkappa}$, and momentum spreads $\sigma_{\varkappa}$.
With suitable values of these quantities, the ISLV correction can be measurable at macroscopic distances satisfying the conditions
of applicability of the approximations used in the formalism. The basic constraints can be formulated as the inequalities
\begin{equation}
\label{TheConditions}
\begin{gathered}
2 \times 10^{-5}\left(\frac{1\,\text{eV}}{\sig}\right)~\text{cm}
\ll \vert\vec{y}-\vec{x}\vert \\ 
\ll 20\left(\frac{1\,\text{eV}}{\sig}\right)^2\left(\frac{\vert\vec{q}\vert}{1\,\text{MeV}}\right)~\text{cm},
\end{gathered}
\end{equation}
satisfied at sufficiently high neutrino energies and at short distances, exceeding the ``mesoscopic scale'' $\sim1/\sig$.
They are virtually independent of the neutrino masses, provided that the latter are negligible compared to the neutrino energy.
The constraints \eqref{TheConditions} are illustrated in Fig.~\ref{fig:Boundaries} for representative intervals of neutrino energy
and values of the scale parameter $\sig$.
The energies of the neutrinos from $\pi_{\mu_2}$ and $K_{\mu_2}$ decays in the meson rest frame are also shown (by vertical planes)
to epitomize acceptable baselines in possible dedicated experiments.

Last but not least, the QFT approach predicts the exact opposite dependence of the ISLV effect on the distance
$L=\vert\vec{y}-\vec{x}\vert$ for the SBL and LBL asymptotics: the corresponding corrections are given by the series in powers
of $\delta_i^2 \approxprop L^2$ and $1/L^2$, respectively, and the lowest order corrections are inversely proportional
to each other to within a dimensionless positive factor.
This feature makes it possible, at least in principle, to distinguish experimentally these two asymptotics from each other,
as well as from the predictions of other approaches, in particular from the result of Ref.~\cite{Beuthe:2001rc}, which predicts 
the absence of even an approximate fulfillment of ISL in the entire region $L\lesssim\vert\vec{q}\vert/\sigSBL^2$.
Eventually, it is hoped that the transition region, which has not yet been explored within the formalism under discussion,
will also become the subject of experimental study.
A promising source of (anti)neutrinos could be, e.g., stopped-kaon decays, which provide a relatively wide range
of applicability of the formalism (see Fig.~\ref{fig:Boundaries}).

\section{Summary}
\label{sec:Conclusion}

In the framework of the covariant perturbative QFT approach with wave packets as asymptotically free initial and final states,
the wave-packet modified neutrino propagator is expressed as an initial segment of the triple Taylor series in powers of
dimensionless rotation-invariant variables \eqref{SmallParameters}.
This asymptotic expansion is applicable at high energies and short but macroscopic space-time distances between the vertices
of the corresponding Feynman macrodiagram of a fairly general form (see Fig.~\ref{fig:Macrograph_Class_A}).
It is also worth noting that the obtained asymptotic expansion applies not exclusively to the neutrino propagator.

Using this result, it is shown that in the short-baseline regime, high-energy neutrinos are deeply virtual and behave like quasi-classical
particles, that is, propagate approximately along classical trajectories connecting the points of their production and absorption.
In the leading-order approximation, this results in the classical inverse-square dependence of the modulus squared amplitude.
But there are two kinds of corrections to the classical behavior: at supershort distances and times, $\vert\vec{X}\vert \sim X_0$,
which, however, exceed the characteristic scale $\sim 1/\sig$, relative corrections are described by a series in powers of
$\left(\vert\vec{X}\vert-X_0\right)/X_0\sim1/({\sig}X_0)$, while at $1/\sig\ll\vert\vec{X}\vert\sim X_0\ll\vert\vec{q}\vert/\sig^2$ --
by an expansion in powers of $\vert\vec{X}\vert^2$, which follows from the above expansion of the neutrino propagator. 
As in the LBL regime, the main ISL violating correction is negative, i.e., it leads to a decrease in the neutrino event rate.
But the functional dependence of the ISLV correction is radically different from that in the LBL mode, where the correction
is given by an expansion in inverse powers of $\vert\vec{X}\vert^2$.
This makes it potentially possible to distinguish between the SBL and LBL asymptotics and thereby test the theory
in dedicated experiments with variable baseline or with several identical detectors at different baselines.

An expression for the time-averaged number of neutrino-induced events is derived and compared with the corresponding result
for the long-baseline mode. It is, in particular, shown that although the corresponding formulas have the same structure,
the decoherence factors 
are different and do not reduce to each other even in the lowest-order approximation.
This results from the fact that the space-time behavior of the neutrino propagator and, hence, the effective neutrino wave packet
in the two modes is strongly different, and there is a transition region, $\vert\vec{X}\vert\sim(0.1-10)\vert\vec{q}\vert/\sig^2$,
in which the neutrino wave packet transformes from deeply virtual (with the 4-momentum independent of the neutrino mass)
into the more habitual on-mass-shell packet.
Negligible decoherence and flavor transition probabilities in the SBL regime should simplify the experimental study of the ISLV effect,
and the specific dependence of the predicted count rate on distance and energy could potentially help distinguish its
reduction due to ISLV from that caused by possible mixing of active and sterile neutrinos. 

\begin{appendices} 

\section{SBL asymptotics of neutrino propagator}
\label{sec:SBLasymptoticsOfNeutrinoPropagator}

The aim of this Appendix is to represent the integral%
\footnote{For simplicity, hereafter we omit the neutrino mass eigenfield index.}
\begin{equation}
\label{J(X)_def}
J(X) = \int\frac{d^4q\,\varPhi(q)e^{-iqX}}{q^2-m^2+i\epsilon},
\end{equation}
where $\varPhi(q)$ is defined by Eq.~\eqref{Phi_just_gauss},
through certain asymptotic expansions over several Lorentz-invariant and rotation-invariant dimensionless quantities,
guaranteed to be small in absolute value. 
As will be shown, the resulting representation is applicable at relatively short macroscopic distances
$\vert\vec{X}\vert$ and times $X_0$, sufficiently high (not necessary ultrarelativistic) neutrino energies $\kP_0$,
and not too high neutrino virtualities $\kP^2-m^2$, where $\kP=(\kP_0,\vec{\kP})$ is the saddle point of $\varPhi(q)$.
The exact meaning of these restrictions will become clear from what follows.
The results obtained below are quite general, in the sense that they do not apply only to neutrinos.

\subsection{Transformation of integral $J(X)$}
\label{sec:J}

Performing the $4d$ Fourier transform
\begin{equation} 
\label{4dFourierTransform}
\varPhi(q) = \int d^4x\,\varphi(x)e^{iqx}
\end{equation}
and substituting Eq.~\eqref{4dFourierTransform} to integral \eqref{J(X)_def} we get
\begin{multline} 
\label{chet_mer_int_q_0_posled}
J(X) = \int d^4x\,\varphi(x)\int d^3q\,e^{-i\vec{q}\left(\vec{x}-\vec{X}\right)} \\
       \times\int\frac{dq_0\,e^{iq_0\left(x_0-X_0\right)}}{q_0^2-\vec{q}^2-m^2+i\epsilon}.
\end{multline}
Considering the two possibilities $x_0 \gtrless X_0$ and using the residue theorem, we obtain
\begin{equation*}
\int\frac{dq_0\,e^{iq_0\left(x_0-X_0\right)}}{q_0^2-\vec{q}^2-m^2+i\epsilon}
= -\frac{i\pi}{E_{\vec{q}}}e^{-iE_{\vec{q}}\vert{x_0-X_0}\vert},
\end{equation*}
where $E_{\vec{q}}=\sqrt{\vec{q}^2+m^2}$.
By substituting this result into Eq~\eqref{chet_mer_int_q_0_posled} and changing the order of integration,
we can rewrite integral \eqref{chet_mer_int_q_0_posled} as
\begin{multline} 
\label{nash_int_int_po_d^3x_posled}
J(X) = -i\pi\int dx_0\int\frac{d^3q}{E_{\vec{q}}}e^{i\left(\vec{qX}-E_{\vec{q}}\vert{x_0-X_0}\vert\right)} \\
         \times \int d^3x\,\varphi(x)e^{-i\vec{qx}}.
\end{multline}
Using Eq.~\eqref{4dFourierTransform} it can be shown that
\begin{equation}\nonumber
\int d^3x\,\varphi(x) e^{-i\vec{qx}}=\frac{1}{2\pi}\int dq_0\varPhi(q)e^{-iq_0x_0}.
\end{equation}
Applying this identity and shifting the integration variable from $x_0$ to $X_0-x_0$,
we rewrite Eq.~\eqref{nash_int_int_po_d^3x_posled} as
\begin{multline*}
J(X) = -i\int\frac{d^4q}{2E_{\vec{q}}}e^{-i\left(q_0X_0-\vec{qX}\right)}\varPhi(q) \\
          \times\int\limits_{-\infty}^{\infty} dx_0\,e^{i\left(q_0 x_0-E_{\vec{q}}\vert{x}_0\vert\right)}.
\end{multline*}
It is convenient to split this integral into two parts,
\begin{equation}
\label{splitting_into_two_parts}
J(X) = J_+(X)+J_-(X),
\end{equation}
where
\begin{gather*}
J_\pm(X) = -i\int\frac{d^4q}{2E_{\vec{q}}}e^{-i\left(q_0X_0-\vec{qX}\right)}\varPhi(q)I_\pm(q), \\
I_-(q)   = \int\limits_{-\infty}^0 dx_0 \, e^{i\left(q_0+E_{\vec{q}}\right)x_0},                \\
I_+(q)   = \int\limits_0^{ \infty} dx_0 \, e^{i\left(q_0-E_{\vec{q}}\right)x_0}.
\end{gather*}
Now, representing $\varPhi(q)$ as a $1d$ Fourier integral
\begin{equation}
\label{integral_representation_of_Phi}
\varPhi(q)=\int\limits_{-\infty}^{\infty}dy_0\,\varphi(\vec{q},y_0)e^{-iq_0y_0},
\end{equation}
we have
\begin{multline*}
J_+(X) = -i\int\frac{d^3 q}{2E_{\vec{q}}}e^{i\vec{qX}}\int\limits_{-\infty}^{\infty}dy_0\varphi(\vec{q}, y_0) \\
            \times\int\limits_0^\infty dx_0 e^{-iE_{\vec{q}}x_0}\int\limits_{-\infty}^{\infty}dq_0\,e^{iq_0(x_0-y_0-X_0)}.
\end{multline*}
It is readily seen that
\begin{multline*}
\int\limits_0^\infty dx_0\,e^{-iE_{\vec{q}}x_0}\int\limits_{-\infty}^{\infty}dq_0\,e^{iq_0(x_0-y_0-X_0)} \\
= 2\pi e^{-iE_{\vec{q}}\left(y_0+X_0\right)}\theta\left(y_0+X_0\right),
\end{multline*}
where $\theta$ is the Heaviside step function. Therefore
\begin{multline}
\label{J_+_s_varphi}
J_+(X) = -2i\pi\int\frac{d^3 q}{2E_{\vec{q}}}e^{i\vec{qX}} \\
            \times\int\limits_{-X_0}^{\infty}dy_0\varphi(\vec{q}, y_0)e^{-iE_{\vec{q}}(y_0+X_0)}.
\end{multline}
In exactly the same way, we get
\begin{multline}\label{J_-_s_varphi}
J_-(X) = -2i\pi\int\frac{d^3q}{2E_{\vec{q}}}e^{i\vec{qX}+iE_{\vec{q}}X_0} \\
            \times\int\limits_{-\infty}^{-X_0}dy_0\varphi(\vec{q},y_0)e^{iE_{\vec{q}}y_0}.
\end{multline}
To move on, we must use the explicit form of $\varPhi(q)$ given by Eqs.~\eqref{Phi_just_gauss} and \eqref{F(q)_def}.
Let us rewrite expression \eqref{F(q)_def} as follows:
\begin{equation} \label{F_j(q_0}
F(q) = F_0-2Y_{\mu}q^{\mu}+\widetilde{\Re}_{\mu\nu}q^{\mu}q^{\nu}.
\end{equation}
Now, applying Eqs.~\eqref{integral_representation_of_Phi} and \eqref{F_j(q_0},
we obtain an explicit expression for the function $\varphi(\vec{q},y_0)$:
\begin{multline}
\label{phi_vec_q_y_0}
\varphi(\vec{q},y_0) = \frac{1}{2\pi}\int\limits_{-\infty}^{\infty}dq_0\varPhi(\vec{q},q_0)e^{iq_0 y_0} \\
                     = \frac{1}{\sqrt{\pi\wRe_{00}}}
                       \exp\bigg[-\frac{1}{4}\left(\wRe_{kn}q_kq_n+2Y_kq_k+F_0\right) \\
                      +\frac{1}{4\wRe_{00}}\left(Y_0+\wRe_{0k}q_k+2iy_0\right)^2\bigg].
\end{multline}
Combining Eq.~\eqref{phi_vec_q_y_0} with Eqs.~\eqref{J_+_s_varphi} and \eqref{J_-_s_varphi} we arrive at
\begin{multline}
\label{J_pm_intermediate}
J_\pm(X) = -i\pi\int\frac{d^3q}{2E_{\vec{q}}}e^{i\left(\vec{qX}{\mp}E_{\vec{q}}X_0\right)}\varPhi(\vec{q},{\pm}E_{\vec{q}}) \\ 
             \times\left[1\pm\erf\left(\frac{X_0}{\sqrt{\wRe_{00}}}+\frac{i}{2}\Upsilon_\pm(\vec{q})\sqrt{\wRe_{00}}\right)\right],
\end{multline}
where
\begin{equation*}
\Upsilon_\pm(\vec{q}) = \frac{Y_0+\wRe_{0k}q_k}{\wRe_{00}}{\mp}E_{\vec{q}}.
\end{equation*}
Given Eqs.~\eqref{Phi_just_gauss}, \eqref{splitting_into_two_parts}, and \eqref{J_pm_intermediate}, we obtain:
\begin{equation}
\label{Partitioning}
J(X) = J_g(X)+J_v(X),
\end{equation}
where
\begin{equation}
\label{J_g_x}
J_g(X) = -i\pi\int\frac{d^3q}{E_{\vec{q}}}e^{i\left(\vec{qX}-E_{\vec{q}}X_0\right)}\varPhi(\vec{q},E_{\vec{q}})
\end{equation}
and
\begin{multline}
\label{J_v_x}
J_v(X) = i\pi\exp\left[-\frac{1}{4}F_0-\frac{1}{\wRe_{00}}\left(X_0+\frac{i}{2}Y_0\right)^2\right] \\
          \times\int\frac{d^3 q}{2E_{\vec{q}}}\exp\left[-\left(\frac{\wRe_{kn}}{4}
         -\frac{\wRe_{k0}\wRe_{n0}}{4\wRe_{00}}\right)q_kq_n\right.                                \\
          \left.-\left(\frac{Y_k}{2}+\frac{i\wRe_{0k}(2X_0-Y_0)}{2\wRe_{00}}-iX_k\right)q_k\right] \\
          \times\left[w\left(\Psi_+\right)-w\left(\Psi_-\right)\right].
\end{multline}
Here the following notation is used:
\begin{gather}
\label{faddeeva_func}
w(x) = e^{-x^2}\erfc(-ix), \\ 
\Psi_{\pm} = i\frac{X_0}{\sqrt{\wRe_{00}}}-\frac{1}{2}\Upsilon_{\pm}\sqrt{\wRe_{00}}. \nonumber
\end{gather}
Although the functions $J_g(X)$ and $J_v(X)$ individually are not exact relativistic invariants, the partitioning \eqref{Partitioning}
is convenient because, as will be shown in Sect.~\ref{sec:Jg-Jv}, under some simple condition valid in the given (laboratory)
frame of reference, the second term in Eq.~\eqref{Partitioning} can be neglected. This is sufficient for our purposes.

\subsection{Fedosov's lemma}
\label{sec:Fedosov}

Below we will make extensive use of the important lemma proved by Fedosov~\cite[p.~79]{Fedosov:1996}, which states:

Let $\varphi(x) = xAx+ax+a_0$ 
be a quadratic function on $\mathbb{R}^n$, where $A$ is a complex symmetric matrix with a positive definite real part,
$a$ and $a_0$ are a complex vector and a complex constant, respectively.
Then for any (smooth) function $f(x)$ the integral 
\[
I = \int d^nx f(x)\exp[-\varphi(x)]
\]
may be represented in the form
\begin{gather*}
I = \sqrt{\frac{\pi^n}{\vert{A}\vert}}\exp[-\varphi(z_0)]\left.e^{D_\tau}T_N(\tau_0,\tau)\right\vert_{\tau=0}+I_N, \\
\begin{multlined}
\vert{I_N}\vert\le\int{d\tau}\left\vert{f}(\tau+\tau_0)-T_N(\tau_0,\tau)\right\vert \\
                  \times\exp[-\text{Re}\,\varphi(\tau+\tau_0)],
\end{multlined} \\
D_\tau = \frac{1}{4}\partial_{\tau}A^{-1}\partial_{\tau}
\equiv\frac{1}{4}\sum_{r,s=1}^n\frac{\partial}{\partial{\tau_r}}\left(A^{-1}\right)_{rs}
\frac{\partial}{\partial{\tau_s}},
\end{gather*}
where $T_N(\tau_0,\tau)$ is the $N$-th order Taylor polynomial of $f(x)$ at $x=\tau_0=\text{Re}\,z_0$,
and $z_0\in\mathbb{C}^n$ is a stationary point of $\varphi(z)$.

Note that the stationary point $z_0$ (solution of the linear system $\partial\varphi/\partial x_r=0$) is given by 
$z_0 = -\frac{1}{2}A^{-1}a$ and thus $\varphi(z_0) = -\frac{1}{4}aA^{-1}a+a_0$.
If $f(x)$ is analytic at $z_0$ then $f(\tau+\tau_0)-T_N(\tau_0,\tau)\to0$ as $N\to\infty$ and
\begin{equation}
\label{FedosovLemma-lim}
I = \sqrt{\frac{\pi^n}{\vert{A}\vert}}\exp[-\varphi(z_0)]\left.e^{D_{\tau}}f(\tau_0+\tau)\right\vert_{\tau=0}.
\end{equation}
We will also need a corollary of this lemma: Let
\begin{gather*}
\varphi(x;t) = xAx+(a+ibt)x+a_0+ic_0t, \\
f(x;t) = f_0(x)\exp[-ig(x)t],
\end{gather*}
where $A$, $a$, and $a_0$ have the same meaning as above, $b\in\mathbb{R}^n$, and $t\in\mathbb{R}^1$.
It has been proven \cite{Naumov:2014jpa} that
\begin{multline}
\label{FedosovLemma-lim-Int}
\int\limits_{-\infty}^{\infty}dt \int d^nx\ f(x;t)\exp\left[-\varphi(x;t)\right]           \\ 
= 2\sqrt{\frac{\pi^{n+1}}{\vert{A}\vert bA^{-1}b}}\exp\left(\frac{1}{4}aA^{-1}a-a_0\right) \\
   \times\exp\left[-\frac{1}{bA^{-1}b}\left(\frac{aA^{-1}b}{2}-c_0\right)^2\right]         \\
   \times e^{D_\tau}f_0(\tau_0+\tau)\exp\bigg\{\frac{g(\tau_0+\tau)}{bA^{-1}b}             \\
   \times \left[aA^{-1}b-2c_0-g(\tau_0+\tau)\right]\bigg\}\Bigg\vert_{\tau=0}.
\end{multline}
The extention of Fedosov's lemma and its corollaries to Minkowski signature is trivial: as long as the integration variables are independent,
it is only necessary to take care of the correct position of the Lorentz indices in the tensor and vector convolutions.

\subsection{Integral $J_g(X)$}
\label{sec:Jg}

The integral~\eqref{J_g_x} can be identically represented as
\begin{align*}
J_g(X) = &\ -i\int\limits_{-\infty}^{\infty}dy_0  \int\frac{d^4q}{2E_{\vec{q}}}\varPhi(\vec{q},q_0) \\
         &\   \times\exp\left[-iqX+iy_0\left(q_0-E_{\vec{q}}\right)\right]   \\
       = &\ -i\int\limits_{-\infty}^{\infty}dy_0\int\frac{d^4q}{2E_{\vec{p}}}f(q;y_0)\exp\left[-\frac{1}{4}F(q)\right. \\
         &\ \left.+i\left(y_0\frac{qp-m^2}{E_{\vec{p}}}-qX\right)\right],
\end{align*}
where
\begin{gather*}
f(q;y_0) = \frac{E_{\vec{p}}}{E_{\vec{q}}}\exp\left[-\frac{iy_0(E_{\vec{q}}E_{\vec{p}}-m^2-\vec{qp})}{E_{\vec{p}}}\right], \\
      p  = (E_{\vec{p}},\vec{p}) = \left(\sqrt{\vec{\kP}^2+m^2},\vec{\kP}\right), 
\quad
\kP^\mu  = \kR^{\mu\nu}Y_\nu,
\end{gather*} 
The 4-vector $\kP$ is the stationary point and $\kR$ is the tensor such that $\kR_{\mu\lambda}\widetilde{\Re}^{\lambda\nu}=\delta_\mu^\nu$.
Here it is also appropriate to recall that $p^0 \equiv E_{\vec{p}} \ne \kP^0 \equiv \kR^{0\mu}Y_\mu$ and $q^0 \ne E_{\vec{q}}$.

The seeming complication (representing $J_g$ as a $5D$ integral) allows us to straightforwardly apply Eq.~\eqref{FedosovLemma-lim-Int},
which, after some transformations, gives
\begin{equation}
\label{GenFedosovLemmaForAmplitude}
J_g(X) = \left.-8i\pi^2\sqrt{\frac{\pi\vert\kR\vert}{\mathcal{G}}}e^\Omega
           e^{D_{\vec\tau}}e^{\Omega_{\vec\tau}}\right\vert_{\vec\tau=0}.
\end{equation}
Here 
\begin{gather*}
\begin{aligned}
  \Omega              = &\ \Omega_c-\frac{1}{4\mathcal{G}}\left({\kP}p-m^2-2i\kR^{\mu\nu}p_{\mu}X_\nu\right)^2, \\
  \Omega_{\vec{\tau}} = &\ \frac{W_{\vec{\tau}}}{4\mathcal{G}}\left[-4i\kR^{\mu\nu}X_{\mu}p_{\nu}-W_{\vec{\tau}}
                                                                              +2\left({\kP}p-m^2\right)\right], \\
  W_{\vec{\tau}}      = &\ E_{\vec{p}}\left(E_{\vec{p}+\vec{\tau}}-E_{\vec{p}}\right)-\vec{p}\vec{\tau}, 
  \quad
  D_{\vec\tau}        =    \partial_{\vec\tau}\vec{\kR}\partial_{\vec\tau},
\end{aligned}                                                                                                   \\
 \vec{\kR}            =    \vert\vert\kR_{kn}\vert\vert,
 \quad
 \mathcal{G}          =    \kR_{\mu\nu}p^{\mu}p^{\nu},
\end{gather*}
and other notations are explained in the main text. 
Note that the differential operator $D_{\vec\tau}$ is a $3\times3$ matrix, since the function $f(q;y_0)$ does not depend on $q_0$.
A direct application of the operator $e^{D_{\vec\tau}}=\sum_n D_{\vec\tau}^n/n!$ in Eq.~\eqref{GenFedosovLemmaForAmplitude}
shows that the function $J_g(X)$ can be represented by the initial segment of the triple asymptotic series
\begin{multline}
\label{J_g_final}
J_g(X) = -8i\pi^2\sqrt{\frac{\pi\vert\kR\vert}{\mathcal{G}}}\exp(\Omega) \\ 
\times\sum\limits_{a\ge0}\sum\limits_{b\ge0}\sum\limits_{c\ge0}f_{abc}(i\delta)^a\eta^b\Delta^c
\end{multline}
in powers of three dimensionless real variables
\begin{equation*}
\delta=\sig^2 \frac{\rho_{\mu\nu}p^{\mu}X^\nu}{\rho^{\mu\nu}p_{\mu}p_{\nu}},
\enskip
\eta=\frac{\sig^2}{E_{\vec{p}}^2},
\enskip \text{and} \enskip
\Delta=\frac{{\kP}p-m^2}{\rho^{\mu\nu}p_{\mu}p_{\nu}}.
\end{equation*}
Here $\sig$ is the energy-dimension scalar defined by Eq.~\eqref{SigDef} and assumed small compared to the masses of in and out wave packets
at the source and detector vertices and $\rho$ is the tensor with dimensionless space-time components defined by Eq.~\eqref{rhoDef}.
We emphasize that $\delta$ and $\Delta$ are Lorentz scalars, while $\eta$ is a rotation invariant.
Therefore, the dimensionless coefficient functions $f_{abc}$ must also be invariant to rotation.
Under the natural conditions that follow from the formulation of the problem and from the formalism discussed in the main text, we see that
\[
\vert\delta\vert \simeq {\sig^2\vert\vec{X}\vert}/{\vert\vec{\kP}\vert}
\]
as $\vert\vec{X}\vert \simeq X_0$ (quasi-classical trajectories + ultrarelativism) and
\[
\vert\Delta\vert  \simeq  \frac{\vert\kP^2-m^2\vert}{2\rho^{\mu\nu}l_{\mu}l_{\nu}\vert\vec{\kP}\vert^2}
                 \lesssim \frac{2\sig}{\sqrt{\rho^{\mu\nu}l_{\mu}l_{\nu}}\vert\vec{\kP}\vert}
\]
as $\vert\vec{\kP}\vert \simeq \kP_0$ (suppression of too high virtualities + ultrarelativism + approximate energy-momentum conservation).
From definition \eqref{rhoDef} and dimensional considerations, it follows that $\rho_{\mu\nu} = O(1)$. 
Taking all this into account, we can conclude that
\[
\vert\delta\vert \ll 1, \enskip \eta \ll 1, \enskip \text{and} \enskip \vert\Delta\vert \ll 1
\]
provided that the conditions
\begin{subequations}
\label{TheApplicabilityConditions}
\begin{equation}
\label{TheApplicabilityCondition_a}
\sig\vert\vec{X}\vert \ll \vert\vec{\kP}\vert/\sig
\end{equation}
and
\begin{equation}
\label{TheApplicabilityCondition_b}
\sig \ll \vert\vec{\kP}\vert
\end{equation}
\end{subequations}
are both fulfilled.
These conditions determine the range of applicability of expansion \eqref{J_g_final} and, consequently, of the entire formalism under discussion.
While  condition \eqref{TheApplicabilityCondition_b} is trivial and is fulfilled with a margin in the modern neutrino experiments,
the condition \eqref{TheApplicabilityCondition_a} means that the formalism is only applicable at sufficiently short distances and/or high neutrino energies.
The area of applicability increases with decreasing $\sig$.
Several lower-order coefficients $f_{abc}$ are:
\begin{align*}
f_{000} = &\ 1,                                                                                      \\
f_{001} = &\  \frac{1}{2}\theta_1-\frac{1}{2}\omega_1,                                               \\
f_{002} = &\  \frac{1}{8}\theta_1^2+\frac{1}{4}\theta_2-\frac{1}{2}\omega_2+\frac{3}{8}\omega_1^2
                                                                       -\frac{1}{4}\omega_1\theta_1, \\
f_{010} = &\ 3\omega_1-\theta_1-\frac{1}{\varrho}\left(\frac{1}{2}\theta_2
                                                              -\omega_2+\frac{1}{4}\theta_1^2\right. \\
          &\  \left.-\frac{1}{2}\omega_1\theta_1+\frac{3}{4}\omega_1^2\right),                       \\
f_{011} = &\ -\frac{3}{2}\theta_2+9\omega_2-\frac{3}{4}\theta_1^2+\frac{9}{2}\omega_1\theta_1
                                                                             -\frac{45}{4}\omega_1^2 \\
          &\ -\frac{1}{\varrho}\left(\theta_3-3\omega_3
                              +\frac{3}{4}\theta_1\theta_2-\frac{3}{4}\omega_1\theta_2\right.        \\
          &\  \left.-\frac{3}{2}\theta_1\omega_2+\frac{9}{2}\omega_1\omega_2+\frac{1}{8}\theta_1^3
                                                               -\frac{3}{8}\omega_1\theta_1^2\right. \\
		  &\ \left.+\frac{9}{8}\omega_1^2\theta_1-\frac{15}{8}\omega_1^3\right),                     \\
f_{020} = &\ -30\omega_2+\frac{105}{2}\omega_1^2+\frac{3}{2}\theta_1^2+3\theta_2-15\omega_1\theta_1  \\
          &\ -\frac{1}{\varrho}\left(36\omega_3-3\theta_1\theta_2-4\theta_3
                                                                       -\frac{1}{2}\theta_1^3\right. \\
          &\  \left.+9\omega_1\theta_2+\frac{105}{2}\omega_1^3-90\omega_1\omega_2
                                                               +\frac{9}{2}\omega_1\theta_1^2\right. \\
          &\  \left. -\frac{45}{2}\omega_1^2\theta_1+18\theta_1\omega_2 \right)                      \\
          &\ -\frac{1}{\varrho^2}\left(\frac{45}{4}\omega_1^2\omega_2
                             -\frac{3}{8}\omega_1\theta_1^3-\frac{9}{16}\omega_1^2\theta_1^2\right.  \\
          &\  \left.+\frac{15}{8}\omega_1^3\theta_1-\frac{105}{32}\omega_1^4-\frac{9}{2}\omega_2^2
                                                                                  +9\omega_4 \right. \\
          &\  \left.-\frac{9}{8}\theta_2^2-\frac{9}{8}\omega_1^2\theta_2-3\theta_1\theta_3
		                                                 +\frac{9}{4}\omega_1\theta_1\theta_2\right. \\
          &\  \left. +\frac{9}{8}\theta_1^2\theta_2-\frac{9}{32}\theta_1^4
		                   +\frac{9}{4}\theta_1^2\omega_2-\frac{9}{2}\omega_1\theta_1\omega_2\right. \\
          &\  \left.-\frac{1}{2}\omega_2\theta_2-9\omega_1\omega_3\right),                           \\
f_{100} = &\  \omega_1-\theta_1,                                                                     \\
f_{101} = &\ -\theta_2+2\omega_2-\frac{3}{2}\omega_1^2+\omega_1\theta_1-\frac{1}{2}\theta_1^2,       \\
f_{110} = &\ 3\theta_2-18\omega_2+\frac{3}{2}\theta_1^2-9\omega_1\theta_1+\frac{45}{2}\omega_1^2     \\
          &\ +\frac{1}{\varrho}\left(2\theta_3-6\omega_3+\frac{3}{2}\theta_1\theta_2
		                                                      -\frac{3}{4}\omega_1\theta_1^2 \right. \\
          &\  \left.-\frac{3}{2}\omega_1\theta_2-3\theta_1\omega_2+9\omega_1\omega_2
		                                                       +\frac{9}{4}\omega_1^2\theta_1\right. \\
		  &\ \left.-6\omega_3+\frac{1}{4}\theta_1^3-\frac{15}{4}\omega_1^3\right),                   \\
f_{200} = &\ \theta_2+\frac{1}{2}\theta_1^2-2\omega_2+\frac{3}{2}\omega_1^2-\omega_1\theta_1.
\end{align*}
Here
\begin{equation}
\label{rot_inv_coeffs}
\begin{gathered}
\varrho   =   \frac{1}{E_{\vec{p}}^2}\rho^{\mu\nu}p_{\mu}p_{\nu} 
          =   \rho^{\mu\nu}v_{\mu}v_{\nu}
       \simeq \rho^{\mu\nu}l_{\mu}l_{\nu}, \\
\theta_s = \Tr\,{\vec\rho}^s,
\enskip
\omega_s = \vec{v}^T{\vec\rho}^s\vec{v}, \enskip s=0,1,\ldots, \\
v = (1,\vec{v}),
\enskip
\vec{v} = \frac{\vec{p}}{E_{\vec{p}}},
\enskip
{\vec\rho} = \vert\vert\rho_{kn}\vert\vert.
\end{gathered}
\end{equation}
Obviously, the approximate equality $\varrho\simeq\rho^{\mu\nu}l_{\mu}l_{\nu}$ is valid only for ultrarelativistic neutrinos.

The presented results were derived using the computer algebra systems Maple$^\text{\textregistered}$ and Maxima for
a finite number of terms of the asymptotic series \eqref{J_g_final}.
More specifically, the calculations were performed using the $K$-th order Taylor polynomial of the operator exponent $e^{D_{\vec\tau}}$.
Up to the order $K=8$ it is proved that $f_{abc}=0$ at $b>K-a$ and $c>K-b$ for $a \le K$. Hence
\[
\sum\limits_{a\ge0}\sum\limits_{b\ge0}\sum\limits_{c\ge0} \longmapsto
\sum\limits_{a=0}^K\sum\limits_{b=0}^{K-a}\sum\limits_{c=0}^{K-b}
\]
in Eqs.~\eqref{J_g_final}.
The method for deriving the coefficients $f_{abc}$ is reduced to the transformation of sums of products
of tensor and vector components resulting from differentiation of $\exp\left(\Omega_{\vec\tau}\right)$
to the identical combinations of the rotation invariants $\theta_s$, $\omega_s$, and $\varrho$.
The calculations are simplified by dimensional and symmetry considerations, but are still intricate.
Obviously, the required computer processing time grows exponentially as $K$ increases,
so that the calculation of high-order coefficients becomes rather time-consuming.
But for our purposes, it would be enough to make sure that the structure of series \eqref{J_g_final}
is correctly ``guessed'' even in the lowest (second) subleading order in $\delta$, yielding the ISL violation.
 
It is convenient to rewrite Eq.~\eqref{J_g_final} in the equivalent exponential form
\begin{multline}
\label{J_g_final2_inside_calc}
J_g(X) = -8i\pi^2\sqrt{\frac{\pi\vert\kR\vert}{\mathcal{G}}}\exp\left(\Omega\right) \\
            \times\exp\left[\sum\limits_{a\ge0}\sum\limits_{b\ge0}\sum\limits_{c\ge0}F_{abc}(i\delta)^a\eta^b\Delta^c\right].
\end{multline}
This representation is a bit more suitable
since it facilitates calculation of the modulus squared transition amplitude.
The corresponding coefficients $F_{abc}$ can be obtained by re-expanding the power series \eqref{J_g_final}:
\begin{align*}
F_{000} = &\ 0,                                                                                      \\
F_{001} = &\  \frac{1}{2}\theta_1-\frac{1}{2}\omega_1,                                               \\
F_{002} = &\  \frac{1}{4}\theta_2-\frac{1}{2}\omega_2+\frac{1}{4}\omega_1^2,                         \\
F_{010} = &\ 3\omega_1-\theta_1-\frac{1}{\varrho}\left(\frac{1}{2}\theta_2
                                                              -\omega_2+\frac{1}{4}\theta_1^2\right. \\
          &\  \left.-\frac{1}{2}\omega_1\theta_1+\frac{3}{4}\omega_1^2\right),                       \\
F_{011} = &\ -\frac{3}{2}\theta_2+9\omega_2-\frac{1}{4}\theta_1^2+\frac{5}{2}\omega_1\theta_1
                                                                             -\frac{39}{4}\omega_1^2 \\
          &\ -\frac{1}{\varrho}\left(\theta_3-3\omega_3
                              +\frac{1}{2}\theta_1\theta_2-\frac{1}{2}\omega_1\theta_2\right.        \\
          &\  \left.-\theta_1\omega_2+4\omega_1\omega_2+\frac{1}{2}\omega_1^2\theta_1
                                                                      -\frac{3}{2}\omega_1^3\right), \\
F_{020} = &\ -30\omega_2+48\omega_1^2+\theta_1^2+3\theta_2-12\omega_1\theta_1                        \\
          &\ -\frac{1}{\varrho}\left(36\omega_3-\frac{5}{2}\theta_1\theta_2-4\theta_3
                                                                       -\frac{1}{4}\theta_1^3\right. \\
          &\  \left.+\frac{15}{2}\omega_1\theta_2+\frac{201}{4}\omega_1^3-87\omega_1\omega_2 \right. \\
          &\  \left. +{\frac{13}{4}}\omega_1\theta_1^2-\frac{81}{4}\omega_1^2\theta_1
		                                                                 +17\theta_1\omega_2 \right) \\
          &\ -\frac{1}{\varrho^2}\left(\frac{21}{2}\omega_1^2\omega_2
                             -\frac{1}{2}\omega_1\theta_1^3-\frac{1}{4}\omega_1^2\theta_1^2\right.   \\
          &\  \left.+\frac{3}{2}\omega_1^3\theta_1-3\omega_1^4-4\omega_2^2
                                                                        +9\omega_4-\theta_2^2\right. \\
          &\  \left.-\frac{3}{4}\omega_1^2\theta_2-3\theta_1\theta_3+2\omega_1\theta_1\theta_2\right.\\
          &\  \left. +\frac{5}{4}\theta_1^2\theta_2
                        -\frac{1}{4}\theta_1^4+2\theta_1^2\omega_2-4\omega_1\theta_1\omega_2\right.  \\
          &\  \left.-9\omega_1\omega_3\right),                                                       \\
F_{100} = &\  \omega_1-\theta_1,                                                                     \\
F_{101} = &\ -\theta_2+2\omega_2-\omega_1^2,                                                         \\
F_{110} = &\ 3\theta_2-18\omega_2+\frac{1}{2}\theta_1^2-5\omega_1\theta_1+\frac{39}{2}\omega_1^2     \\
          &\ +\frac{1}{\varrho}\left(2\theta_3-6\omega_3+\theta_1\theta_2-\omega_1\theta_2 \right.   \\
          &\  \left.-2\theta_1\omega_2+8\omega_1\omega_2+\omega_1^2\theta_1-3\omega_1^3\right),      \\
F_{200} = &\  \theta_2-2\omega_2+\omega_1^2.
\end{align*}

It is also instructive to represent the triple power series in the exponent in Eq.~\eqref{J_g_final2_inside_calc} as
\begin{equation}
\label{Re+Im}
\sum\limits_{a\ge0}(-1)^a\left(g_a+ig'_a\delta\right)\delta^{2a} \equiv \mathcal{I}.
\end{equation}
The explicit form of the real coefficient functions $g_a$ and $g'_a$ trivially follows from the expressions for
the coefficients $F_{abc}$.
A simple analysis shows that $\text{Im}\,\mathcal{I}$ is negligibly small under all experimental circumstances,
but $\text{Re}\,\mathcal{I}$ provide corrections of the order of $O(\eta)$ and $O(\Delta$) to the decoherence factors and,
more importantly, potentially measurable ISL-violating correction given by even powers of $\delta$ (see main text).

\subsection{Positivity of $F_{200}$}
\label{sec:PositivityOfF_200}

Let us prove that the coefficient function $F_{200}$ is positive (we will need this fact in Sect.~\ref{sec:ISLV}).
The proof can be fulfilled by an orthogonal diagonalization of the symmetric positive definite $3\times3$ matrix $\vec\rho$:
\[
\vec\rho = \vec{O}^T\vec{r}\,\vec{O},
\quad
\vec{r} = \diag\left(r_1,r_2,r_3\right),
\quad 
\vec{v}=\vec{O}\boldsymbol{u}.   
\]
Here $\boldsymbol{u}=(u_1,u_2,u_3)^T$, $r_k$ are the \emph{positive} eigenvalues of $\vec\rho$,
$\vec{O}$ is an orthogonal matrix of rotation, and $\vec{v}$ is defined in Eq.~\eqref{rot_inv_coeffs}.
In this representation, using definitions \eqref{rot_inv_coeffs} we find:
\begin{equation*}    
F_{200} = \sum_kr_k^2\left(1-u_k^2\right)^2+2\sum_k\eta_k^{ln}r_lr_nu_l^2u_n^2 > 0,
\end{equation*}
where $\eta_k^{ln}=1$ for a cyclic permutation of $(1,2,3)$ and $\eta_k^{ln}=0$ otherwise.
It is pertinent to note here that the coefficient function $f_{200}$ in Eq.~\eqref{J_g_final}
is also positive, sinse 
\[
f_{200}=F_{200}+\frac{1}{2}\left(\theta_1-\omega_1\right)^2.
\]
A less formal way to prove the same is to rewrite $F_{200}$ in the coordinate system whose
$z$ axis is directed along the vector $\vec{v}$. In this system 
\begin{multline}
\label{Positivity2}
F_{200} =    \left[\rho_{33}\left(1-\v^2\right)\right]^2
           +2\left(\rho_{23}^2+\rho_{31}^2\right)\left(1-\v^2\right) \\
           + \rho_{11}^2+2\rho_{12}^2+\rho_{22}^2 > 0,
\end{multline}
where $\v \equiv \vert\vec{v}\vert=\vert\vec{v}'\vert$, $\v \le 1$. For massless neutrino
$F_{200} = \rho_{11}^2+2\rho_{12}^2+\rho_{22}^2$,
i.e.\ $F_{200}$ is determined only by the transverse spatial components of the tensor $\rho$.

\subsection{Integral $J_v(X)$}
\label{sec:Jv}

Applying Fedosov's lemma to Eq.~\eqref{J_v_x}, we obtain:
\begin{gather}
\label{FedosovRorJv}
\begin{aligned} 
J_v(X) = &\ 4i\pi^2\sqrt{\pi\vert\vec{\kR}\vert}\exp\left(\Omega_c\right) \\
         &\  \times e^{D_{\vec\tau}}\left.\frac{\left[w_{-}(\vec\tau)-w_{+}(\vec\tau)\right]}{E_{\vec{p}+\vec{\tau}}}\right\vert_{\vec{\tau}=0},
\end{aligned}                                                                                                         \\  \nonumber
w_{\pm}(\vec\tau)      = w\left(i\frac{X_0}{\sqrt{\wRe_{00}}}-\frac{1}{2}\sqrt{\wRe_{00}}\,\Gamma_{\pm}(\vec\tau)\right), \nonumber \\
\Gamma_{\pm}(\vec\tau) = \kP_0\pm E_{\vec{p}+\vec{\tau}}-\frac{\wRe_{0k}\tau_k}{\wRe_{00}},
\quad
D_{\vec\tau}           = \partial_{\vec\tau}\vec{\kR}\partial_{\vec\tau}.                                                 \nonumber
\end{gather}
We mention in passing that the $3d$ stationary point $\vec{\kP}$ and operator $D_{\vec\tau}$ are exactly the same as above.
From the known asymptotic expansion of the complementary error function function of complex variable (see, e.g., Ref.~\cite[p.~164]{Olver:2010}) it follows that
\begin{gather*}
w(iz)=\frac{1}{z\sqrt\pi}\sum\limits_{n=0}^\infty(-1)^n\frac{(2n-1)!!}{\left(2z^2\right)^n}, \\
z\to\infty, \quad \vert\text{arg}(z)\vert < \frac{3\pi}{4}.
\end{gather*}
Combining this and Fedosov's expansions, the expression \eqref{FedosovRorJv} can be represented by the formal series
\begin{align}
\label{J_v(X)_inside_calc}
J_v(X) = &\ \frac{4i\pi^2}{E_{\vec{p}}}\sqrt{\vert\vec{\kR}\vert}\exp\left(\Omega_c\right)\sum\limits_{a\ge0}\kappa^{2a}  \nonumber \\
         &\ \begin{multlined}
            \times\sum\limits_{b\ge0}
            \left[ \frac{1}{\kappa_-^{2b+1}}\left(\mathcal{A}^{-}_{ab}+\frac{i\mathcal{B}^{-}_{ab}\kappa}{\kappa_-}\right)\right.   \\
            \left.-\frac{1}{\kappa_+^{2b+1}}\left(\mathcal{A}^{+}_{ab}+\frac{i\mathcal{B}^{+}_{ab}\kappa}{\kappa_+}\right)\right],
            \end{multlined}
\end{align}
in powers of the real variable
\[
\kappa = \frac{1}{E_{\vec{p}}\sqrt{\wRe_{00}}} > 0 
\] 
and inverse powers of two complex variables
\begin{gather*}
\kappa_\pm = \frac{X_0}{\sqrt{\wRe_{00}}}+\frac{i}{2}\sqrt{\wRe_{00}}\,\left(\kP_0{\pm}E_{\vec{p}}\right).
\end{gather*}
The specific structure of this series was checked only to the second order of the Taylor polynomials of $w_\pm(\vec\tau)$ and $e^{D_{\vec\tau}}$,
which is enough for our purposes.
The lowest-order real dimensionless coefficients $\mathcal{A}^{\pm}_{ab}$ and $\mathcal{B}^{\pm}_{ab}$ are found to be
\begin{align*}
\mathcal{A}^{\pm}_{00} = &\  1,                                                                                                            \\
\mathcal{A}^{\pm}_{10} = &\  3\omega_1-\theta_1,                                                                                           \\
\mathcal{A}^{\pm}_{01} = &\ - \frac{1}{2}\left(1+\zeta_0+\omega_1\pm\omega^0_1\right),                                                     \\
\mathcal{A}^{\pm}_{11} = &\   \frac{3}{4}\omega_1\left(4\theta_1-2-15\omega_1\right)
                            + \frac{1}{2}\left(\theta_1-\theta_2\right)-\frac{1}{4}\theta_1^2                                              \\
                         &\ +6\omega_2+\frac{1}{2}\left(\theta_1-3\omega_1\right)\zeta_0-3(\omega^0_1)^2+\omega_0^0                        \\
                         &\   \pm3\left(2\omega^0_2-5 \omega_1\omega^0_1+\omega^0_1\theta_1\right),                                        \\
\mathcal{B}^{\pm}_{00} = &\   \omega^0_1\pm\frac{1}{2}\left(3\omega_1-\theta_1\right),                                                     \\
\mathcal{B}^{\pm}_{10} = &\  3\left(5\omega_1\omega^0_1-2\omega^0_2-\omega^0_1\theta_1\right) \pm \frac{3}{4}
                              \left[35\omega_1^2-20\omega_2\right.                                                                         \\
                         &\   \left.+\theta_1\left(3\theta_1-10\omega_1\right)+2\left(\theta_2-\theta_1^2\right)\right],                   \\
\mathcal{B}^{\pm}_{01} = &\   \frac{3}{2}\omega^0_1\left(\theta_1-\zeta_0\right)+3\omega^0_2-\frac{3}{2}\omega^0_1-9\omega_1\omega^0_1     \\
                         &\   \pm\frac{3}{4}\left[\theta_1\left(1+\zeta_0+\omega_1\right)-\omega_1\left(3+3\zeta_0+5\omega_1\right)\right. \\
                         &\   \left.+2\omega_0^0+2\omega_2-6(\omega^0_1)^2\right],
\end{align*}
where
\begin{gather*}
\omega_0^0 = \rho_{0k}\rho_{0k},
\quad
\omega^0_1 = \rho_{0k}V_k,
\quad 
\omega^0_2 = \rho_{0k}\rho_{kn}V_{n}, \\ 
\zeta_0    = \rho_{00}-\widetilde{\rho}_{00}^{\,-1},
\quad
\widetilde{\rho}^{\,\mu\nu} = \vert\wRe\vert^{-1/4}\wRe^{\mu\nu},  
\end{gather*}
and the functions $\omega_{1,2}$ and $\theta_{1,2}$ are defined by Eqs.\ \eqref{rot_inv_coeffs}.
The coefficients $\mathcal{A}^{\pm}_{ab}$ and $\mathcal{B}^{\pm}_{ab}$ were derived by the same method as described at the end of Sect.~\ref{sec:Jg}.
Obviously, $\kappa \sim \sig/\kP_0 \ll 1$, but the series in powers $1/\kappa_{\pm}$ can be considered asymptotic expansions only at
$\text{Re}\,\kappa_\pm \gg 1$, which, under the obvious presumption $\widetilde{\rho}_{00}=O(1)$, is equivalent to the condition \eqref{sigX_0>>1}.
Let us show that the relative contribution of $J_v(X)$ is completely negligible under this condition.

\subsection{Comparison of $J_g(X)$ and $J_v(X)$}
\label{sec:Jg-Jv}

The patterns of the asymptotic expansions \eqref{J_g_final} and \eqref{J_v(X)_inside_calc} allows us to estimate the relative contribution of $J_g(X)$ and $J_v(X)$
to the modulus square of the amplitude \eqref{Amplitude_2}, using the leading-order terms of these expansions (that is the condition ${\sig}X_0 \gg 1$)
and neglecting the neutrino virtuality.%
\footnote{Admissibility of the latter simplification is discussed in the main text and, for this estimate, reduces to the approximate
          equalities $\kP_0 \simeq E_p \simeq \vert\vec{\kP}\vert$.}
Then after simple calculations we get
\begin{equation*}
\left\vert\frac{J(X)}{J_g(X)}\right\vert^2 \!
\simeq 1-\frac{2\mathfrak{f}e^{-\Theta}\sqrt{\widetilde{\Re}_{00}}}{(1+\varpi)X_0}
   +\frac{\mathfrak{f}^2e^{-2\Theta}\widetilde{\Re}_{00}}{(1+\varpi)X_0^2},
\end{equation*}
where
\begin{gather*}
\Theta         =   \frac{1}{\mathcal{G}}\left(\kR^{\mu\nu}p_{\mu}X_{\nu}\right)^2 \simeq \rho^{\mu\nu}l_{\mu}l_{\nu}(\sig X_0)^2 \gg 1, \\
\mathfrak{f}   =   \frac{1}{2}\sqrt{\frac{\vert\vec{\kR}\vert\mathcal{G}}{\pi\vert\kR\vert{E_{\vec{p}}^2}}}
            \simeq \frac{1}{2}\sqrt{\frac{\vert\vec{\rho}\vert\rho^{\mu\nu}l_{\mu}l_{\nu}}{\pi\vert\rho\vert}} = O(1),
\end{gather*}
and, according to conditions \eqref{TheApplicabilityConditions},
\[
\varpi = \frac{X_0^2}{E_{\vec{p}}^2\widetilde{\Re}_{00}^2} 
       = \left(\frac{{\sig}X_0}{\widetilde{\rho}_{00}E_{\vec{p}}}\right)^2 \ll 1.
\]
Given the condition \eqref{sigX_0>>1}, the relative contribution of $J_v(X)$ to the modulus squared amplitude%
\footnote{Note that the main contribution is made by the interference term 
          $2\left[\text{Re}(J_g)\text{Im}(J_v)+\text{Re}(J_v)\text{Im}(J_g)\right]/\vert{J_g}\vert^2$.}
is exponentially small, being proportional to
\[\exp\left[-\rho^{\mu\nu}l_{\mu}l_{\nu}({\sig}X_0)^2\right]/{({\sig}X_0)},\]
that is, to a good approximation, \[J(X)=J_g(X)+J_v(X) \simeq J_g(X),\] where $J_g(X)$ is given by Eq.~\eqref{J_g_final2_inside_calc}.
In the main text, we restrict ourselves to this approximation, since for realistic values of the scale parameter $\sig$,
the condition \eqref{sigX_0>>1} is satisfied in all potentially interesting applications of the formalism.
It is, however, a straightforward though tedious task to get rid of this approximation in exotic circumstances when ${\sig}X_0 \gtrsim 1$,
provided that a small number of terms in expansion \eqref{J_v(X)_inside_calc} is sufficient.
In the context of the ISLV effect discussed in Sect.~\ref{sec:ISLV}, it is interesting to note that accounting for the contribution of $J_v$,
as a NLO correction would lead to an additional reduction in the expected number of neutrino-induced events,
i.e., $J_v$ does not compete with the main ISLV correction.

\section{Enhanced expression for $N_{\beta\alpha}$}
\label{sec:N_alphabeta_Exact}

Here we give an expression for the number of events obtained without using the ultrarelativistic approximation, but still neglecting
the (LO) ISLV corrections 
$\propto \vert\delta_i+\delta_j\vert^2 \propto \sig^4\vert\vec{y}-\vec{x}\vert^2/\vert\vec{q}\vert^2$
and corrections to the oscillation phase 
$\propto \vert\delta_i-\delta_j\vert \propto \vert{\Delta}m_{ij}^2\vert\sig^2\vert\vec{y}-\vec{x}\vert/\vert\vec{q}\vert^3$,
which follow from the expansion \eqref{Re+Im} and Eq.~\eqref{delta_j}.
So, with the reservations made,  the number of events can be written in the form
\begin{gather*}
\begin{multlined}
  N_{\beta\alpha}
        = \sum\limits_{\text{spins}} \int d\vec{x}\int d\vec{y}\,\int d\mathfrak{P}_s \int d\mathfrak{P}_d                          \\
          \times\int\frac{d\vert\vec{q}\vert\vert\vec{q}\vert^2}{2(2\pi)^3\mathfrak{D}\vert\vec{y}-\vec{x}\vert^2}\sum_{ij}{\VVVV}  \\
          \times\exp\left[i\varphi_{ij}-A_{ij}^2-\mathcal{B}_{ij}^2-\mathcal{C}_{ij}^2-\Theta_{ij}\right]                           \\
          \times\frac{\chi_{ij}^2}{\left(\v_i+\v_j\right)^2}
                \sqrt{\frac{\kR^{\mu\nu}l_{\mu}l_{\nu}}{2\left(E_i^2\mathcal G_j+E_j^2\mathcal G_i\right)}}                               \\
          \times\sum\limits_{l,l'=1}^2(-1)^{l+l'+1}
                \Ierf\bigg[2\mathfrak{D}\bigg(\frac{\v_i+\v_j}{2\chi_{ij}}\left(x_l^0-y_{l'}^0\right)                                     \\
               +\chi_{ij}\vert\vec{y}-\vec{x}\vert\bigg)-i\mathcal{B}_{ij}\bigg],
\end{multlined}                                                                                                                    
\end{gather*}
where the following notations are used:
\begin{gather*}
\begin{aligned}
  \varphi_{ij}
          = &\ \frac{2\chi_{ij}^2}{\v_i+\v_j} \left(E_i-E_j\right)\vert\vec{y}-\vec{x}\vert                                         \\
	        &\ \times\bigg\{1-\frac{\kR_{{\mu}k}l_k}{\kR_{\mu\nu}\left(v_i^{\mu}v_i^{\nu}+v_j^{\mu}v_j^{\nu}\right)}                \\
	        &\ \times\bigg[\frac{\v_i+\v_j}{2\chi_{ij}^2}\left(v_i^\mu+v_j^\mu\right)                                  
                    -\left(v_i^\mu\v_j+v_j^\mu\v_i\right)\bigg]\bigg\},                                                                   
\end{aligned}                                                                                                                       \\
\begin{aligned}
  \mathcal{A}_{ij}^2
          = &\ 4\mathfrak{D}^2\chi_{ij}^2T_{ij}\vert\vec{y}-\vec{x}\vert^2,                                                         \\
  \mathcal{B}_{ij}
          = &\ \frac{\left(E_i-E_j\right)}{2\left(\v_i+\v_j\right)}
               \frac{\chi_{ij}\kR_{0\mu}\left(v_i^\mu+v_j^\mu\right)}
                    {\mathfrak{D}\kR_{\mu\nu}\left(v_i^{\mu}v_i^{\nu}+v_j^{\mu}v_j^{\nu}\right)},                                   \\
  \mathcal{C}_{ij}^2
         = &\ \frac{(E_i-E_j)^2}{4\kR_{\mu\nu}\left(v_i^{\mu}v_i^{\nu}+v_j^{\mu}v_j^{\nu}\right)},                                  \\
  T_{ij} = &\ \left(\frac{\v_i-\v_j}{\v_i+\v_j}\right)^2\frac{\left(\kR^{00}\kR^{\mu\nu}-\kR^{0\mu}\kR^{0\nu}\right)l_{\mu}l_{\nu}}
                         {2\mathfrak{D}^2\kR^{\mu\nu}l_{\mu}l_{\nu}},
\end{aligned}                                                                                                                       \\
  \mathfrak{D}^2 = \frac{1}{2\wRe^{\mu\nu}l_{\mu}l_{\nu}},
  \enskip
  \v_{i,j} = \vert\vec{v}_{i,j}\vert,
  \enskip
  \chi_{ij}^2 = \frac{1}{1+T_{ij}}, \\
  l = (1,\vec{l}),
  \enskip
  v_{i,j} = (1,\v_{i,j}\vec{l}),
  \enskip
  \vec{l}=\frac{\vec{y}-\vec{x}}{\vert\vec{y}-\vec{x}\vert}.
\end{gather*}
All notations not listed here are explained in the main text.
This is a formal generalization of Eqs.~\eqref{AveragedProbability_Simplified_mod4} -- \eqref{Instrumental} (see also Table~\ref{Tab:Ingredients}),
which can be of utility, for example, when analyzing possible experimental implications of models with ``heavy'' neutrinos.
In practical terms, this result allows us to quantify the accuracy of the ultrarelativistic approximation.

We deliberately do not distinguish here the factors that could be treated as ``flavor transition probability'' and ``instrumental function''.
Recall again that since all structures in the formula are rotation invariants, the unit vector $\vec{l}$ can be oriented in any fixed direction
independent of the vector $\vec{y}-\vec{x}$, for example, along the $z$ axis of a given coordinate system;
then all tensors and vectors must be written in this very coordinate system.

\end{appendices}

\bibliography{references}

\end{document}